\documentclass[manuscript]{aastex} 

\usepackage{natbib}
\usepackage{mathtools}
\usepackage{amsmath}
\usepackage{textcomp}
\usepackage{rotating}
\usepackage{lscape}
\usepackage[dvips]{color}
\definecolor{red}{rgb}{1, 0, 0}
\definecolor{green}{rgb}{0, 1, 0}

\newcommand{\M}{\mathrm}
\newcommand{\tfwhm}{t_\mathrm{FWHM}}
\newcommand{\DF}{{\Delta_F}}

\newcommand{\Msun}{M_{\odot}}

\newcommand{\E}{\,\mathrm}

\newcommand{\erf}{\mathop{\rm erf}\nolimits}

\begin{document}
\title{Microlensing events from the 11-year observations of the Wendelstein Calar Alto Pixellensing Project}
%
\author{C.-H. Lee\altaffilmark{1,2}, A. Riffeser\altaffilmark{1,2}, S. Seitz\altaffilmark{1,2}, R. Bender\altaffilmark{1,2}, J. Koppenhoefer\altaffilmark{2,1}}

\altaffiltext{1}{University Observatory Munich, Scheinerstrasse 1, 81679 Munich, Germany}
\altaffiltext{2}{Max Planck Institute for Extraterrestrial Physics, Giessenbachstrasse, 85748 Garching, Germany}

\begin{abstract}
We present the results of the decade-long M31 observation from 
the Wendelstein Calar Alto Pixellensing Project (WeCAPP). WeCAPP has  
monitored M31 from 1997 till 2008 in both $R$- and $I$-filters, thus provides the 
longest baseline of all M31 microlensing surveys. 
The data are analyzed with the difference imaging analysis, which is most suitable
to study variability in crowded stellar fields. We extracted light curves 
based on each pixel, and devised selection criteria that are optimized to 
identify microlensing events. This leads to 10 new events, and sums up to
a total of 12 microlensing events from WeCAPP, 
for which we derive their timescales, flux excesses, and colors from their 
light curves. The color of the lensed stars fall between $(R-I)$ = 0.56 to 1.36, 
with a median of 1.0 mag, in agreement with our expectation that the 
sources are most likely bright, red stars at post main-sequence stage. 
The event FWHM timescales range from 0.5 to 14 days, with a median of 
3 days, in good agreement with predictions based on the model of Riffeser 
et al. (2006). 

\end{abstract}


\section{Introduction}
\label{sec.intro}
Almost four decades after revealing evidences for dark matter in 
galaxies \citep{1980ApJ...238..471R}, its nature is still unknown. 
Dark matter can be smoothly distributed, e.g. the weakly interacting
massive particles (WIMPs), or in compact form.
Since dark matter does not emit light, the best way to study it is 
through gravitational interaction.
\cite{{1986ApJ...304....1P}} was the first to conceive the idea of gravitational
microlensing as a method to detect massive compact halo dark matter (MACHO). 
Based on his calculation, the optical depth\footnote{We note that this value depends primarily on the lens population characteristics.}, i.e. 
at any time the probability of a source to be closer along the line of sight
to a foreground lens than the lens' Einstein angle therefore
to be magnified by more than 1.34,
is of order 10$^{-6}$ towards Magellanic Clouds.
This has motivated several campaigns to search for microlensing 
towards dense stellar fields, see e.g. the review by \cite{2010GReGr..42.2047M}. The first microlensing events were 
reported by the \textit{MACHO} \citep{1993Natur.365..621A}, \textit{EROS} \citep{1993Natur.365..623A}, 
and \textit{OGLE} \citep{1993AcA....43..289U} teams, with \textit{MOA} \citep{1999PThPS.133..233M} 
joining later on. In their 5.7-year survey data, the \textit{MACHO} team has announced 13\footnote{In \citep{2000ApJ...542..281A} they found 13 microlensing events with tight criteria; they also presented 17 events with loose criteria, which contain more low S/N events.} 
microlensing events towards the Large Magellanic Cloud \citep{2000ApJ...542..281A}, 
while a later analysis \citep{2005ApJ...633..906B} showed that 1 of them is a true variable star and 2 of them 
are likely to be variables from a simple likelihood analysis. This leaves 10 of
them to be plausible microlensing events, and gives 
a MACHO halo fraction of 16\% for MACHO masses between 0.1 - 1 M$_{\odot}$ \citep{2005ApJ...633..906B}. 
The \textit{EROS-1} and \textit{EROS-2} surveys resulted in a MACHO halo fraction of less than
8\% for MACHO mass $\sim$ 0.4 M$_{\odot}$, and ruled out MACHOs with masses between 0.6$\times$10$^{-7}$ and 15 M$_{\odot}$
as a major component of the Milky Way halo \citep{2007A&A...469..387T}.
Analyzing the \textit{OGLE} phase II and III data, 
Wyrzykowski et al. (2010,2011a,2011b) concluded that 
microlensing events towards the Large and Small Magellanic Cloud can be reconciled with  
self-lensing by stars alone, i.e. without requiring compact halo objects in the Milky Way halo.
Later on, \cite{2013MNRAS.428.2342B} studied the tidal streams between the Magellanic Clouds 
and used theoretical modeling to show that the microlensing signal can be reproduced 
by the stars in the stream, though their modeling requires further verification. \cite{2013MNRAS.428.2342B} outlined several observational tests to verify their theoretical modeling, e.g. the sources of the microlensing events are low-metallicity SMC stars, the sources have high velocities relative to the LMC disk stars, and the presence of a very faint stellar counterpart to the Magellanic Stream and Bridge (surface brightness $>$ 34 mag/arcsec$^2$ in V-band).

In contrast to this, \cite{2013MNRAS.435.1582C} recently re-analyzed the \textit{OGLE} and \textit{EROS} data 
and found that some of the \textit{OGLE} events can be attributed
to halo lensing. The inconclusive results can be partially attributed to the 
fact that, by monitoring the Magellanic Clouds we can only sample a small fraction of 
the Milky Way halo. In order to have a census of the Milky Way halo, we
need targets that are distributed along different lines-of-sights through the Milky Way halo.
Other than the Milky Way, we could monitor a nearby spiral galaxy, for which we can have a complete view of its dark halo.

An alternative dense stellar target for a microlensing search for MACHOs is M31, 
as proposed by several authors \citep{1992ApJ...399L..43C, 1993A&A...277....1B, 1994A&A...286..426J}. 
The advantage of M31 is twofold. First, we can have multiple lines-of-sight 
through a dark halo towards M31,
in contrast to the single line-of-sight towards the Magellanic Clouds due 
to our location in the Milky Way. 
The second advantage is that one does not only probe the Milky Way halo, 
but also that of M31. The structure of M31 is well known: with an inclination 
angle of 77$^o$ \citep{1987A&AS...69..311W}, we expect to see an asymmetry of the microlensing 
event rate between the near- and far-side of M31 disk. 
Such an asymmetry can however also originate from extinction along different sight-lines
caused by dust. This has indeed been observed in the density distribution of variables from 
the POINT-AGAPE survey \citep{2004ApJ...601..845A} and from the WeCAPP \citep{2006A&A...445..423F}. 
To be able to account for this dust effect, we have studied the dust properties of M31 
\citep{2009A&A...507..283M} and derived an extinction map across the disk of M31, which can be used for 
quantifying detection efficiencies. Microlensing in M31 differs 
from the Magellanic Clouds, because at the distance of M31 \citep[770 kpc,][]{1990ApJ...365..186F}, 
most of the sources for possible microlensing events are not resolved, 
and each pixel of a CCD image contains up to hundreds of stars. 
Instead of monitoring individual resolved stars, one has to monitor pixel light curves and their
variations towards dense stellar fields. 
The sources of microlensing events are usually not resolved before and
after the high magnification phases and their ``baseline fluxes'' are thus unknown.
In most cases the magnified source (at least at maximum magnification)
appears as a resolved object. However, in order to achieve exquisite photometry
in such crowded fields, we thus perform difference image analysis 
and search for microlensing events at the position of each 
individual pixel. In order to extract light curves at the position of 
each individual pixel, we use a PSF constructed from resolved sources in the
image to perform PSF photometry and obtain pixel light curves.

The theoretical aspect of pixel lensing has been laid down by
\cite{1996ApJ...470..201G} under small impact parameter assumption. 
\cite{2006ApJS..163..225R} reformulated this theory in a more general context.

The first microlensing events towards M31 were reported by the VATT/Columbia microlensing
survey. They put the idea of \cite{1992ApJ...399L..43C} into practice 
\citep{1996AJ....112.2872T} and presented 6 microlensing events discovered by the 
joint observations of the Vatican Advanced Technology Telescope (VATT) and the KPNO 4m 
telescope taken during 1994 and 1995 \citep{1996ApJ...473L..87C}. Their observations 
continued from 1997 to 1999, with the VATT and the 1.3m telescope at the MDM observatory.
With additional data from the Isaac Newton Telescope, they presented 4 probable 
microlensing events out of their 3 year data \citep{2004ApJ...612..877U}. At the same
time, \cite{1997A&A...324..843A} also launched the Andromeda Gravitational 
Amplification Pixel Experiment (AGAPE); they observed M31 with the 2m telescope
Bernard Lyot (TBL) in the French Pyrenees in 1994 and 1995, which led to the 
discovery of one bright, short microlensing event \citep{1999A&A...344L..49A}. 
Following AGAPE, the Pixel-lensing Observations with the Isaac Newton Telescope-Andromeda 
Galaxy Amplified Pixels Experiment (POINT-AGAPE) has monitored M31 from 1999 to 2001; their
first microlensing event was announced in \cite{2001ApJ...553L.137A}, accompanied by 
three more in \cite{2002ApJ...576L.121P,2003A&A...405...15P}; another 3 were reported by 
\cite{2003A&A...405..851C} with additional data from the 1.3m telescope at the MDM 
observatory in 1998-1999. 
The full POINT-AGAPE data were analyzed with 3 different
pipelines based at Cambridge, Z\"{u}rich, and London, where 
3 \citep{2005MNRAS.357...17B}, 6 \citep{2005A&A...443..911C}, and 10 
events \citep{2010MNRAS.404..604T} were reported by the different 
nodes, respectively. 

Using the same INT data, the Microlensing Exploration of the Galaxy and Andromeda 
(MEGA) survey has presented 14 events \citep{2006A&A...446..855D}.
At the same time, the Nainital Microlensing Survey employed the 1.04m Sampurnanand Telescope
in India to observe from September 1998 until February 2002. They have extracted 1 
microlensing event from the 4-year data \citep{2005A&A...433..787J}. More recently, 
the Pixel Lensing Andromeda collaboration (PLAN) carried out observations using the 
1.5m Loiano telescope located in Italy \citep{2007A&A...469..115C} and reported 
2 events in their data collected in 2007 \citep{2009ApJ...695..442C}. 
They further incorporated observations from the 2m Himalayan Chandra 
Telescope (HCT) taken in 2010 and reported another event \citep{2014ApJ...783...86C}. 
In the mean time, the Pan-STARRS 1 collaboration conducted a high-cadence, long-term 
Andromeda monitoring campaign (PAndromeda) utilizing its wide-field ($\sim$ 7 deg$^2$) 
camera. Based on its first year data, \cite{2012AJ....143...89L} have reported 6 events 
in the central 40''$\times$40'' area, with the promises to detect more events taking 
advantage of the larger survey area and higher cadence from Pan-STARRS 1. 

It is worth to note that previous M31 microlensing event identifications may suffer from contaminations by variables.
For example, \cite{1996ApJ...473L..87C} suspected that part of their events are 
contaminated by long-period red supergiant variables. 
Despite the efforts of various campaigns, the MACHO fraction at the mass 
range of 0.1-1$\Msun$ is still under debate 
\citep[see][for a detailed discussion]{2012JPhCS.354a2001C}. 
For example, the POINT-AGAPE 
collaboration has reported evidence for MACHO signal \citep{2005A&A...443..911C},
while the MEGA collaboration
\citep{2006A&A...446..855D} on the contrary concluded that their
events can be fully explained by self-lensing.

Due to the small number of reported events, the origin of M31 microlensing remains an open issue. 
In this study, we aim to increase the number of microlensing detections and suppress
contaminations by variables with long-term observations of the M31 bulge.
This paper is structured as follows. In section \ref{sec.obs} we present the observations of 
our long-term survey. Our data reduction is outlined in section \ref{sec.red}, followed 
by the event detection in section \ref{sec.det}. The analysis of these events are shown 
in section \ref{sec.ana}.
A discussion of our events, 
as well as results from previous M31 microlensing surveys are presented in 
section \ref{sec.dis}, with a summary and prospects in section \ref{sec.sum}.  

\section{Observations}
\label{sec.obs}

\begin{table*}[!ht]
  \setlength{\tabcolsep}{0.5mm}
  \begin{center}
  \begin{tabular}{cc|rrrr|rrrr|rrrr|c}
    \hline\hline
  Season    & Observatory & \multicolumn{4}{c}{R-band} & \multicolumn{4}{c}{I-band} & \multicolumn{4}{c}{R or I-band} & R or I-band \\
            &          &  F1 &  F2 &  F3 &  F4  &  F1 &  F2 &  F3 &  F4  &  F1 &  F2 &  F3 &  F4 & F1 or F2 or F3 or F4 \\
  \hline
  1997 1998 &    WS    &  36 &   7 &   1 &   4  &  33 &   7 &   0 &   3  &  37 &   7 &   1 &   4 &          38          \\
  1998 1999 &    WS    &  33 &   1 &   1 &   1  &  28 &   1 &   1 &   1  &  33 &   1 &   1 &   1 &          33          \\
  \hline
            &    WS    &  64 &   0 &   0 &   0  &  60 &   0 &   0 &   0  &  65 &   0 &   0 &   0 &          65          \\
            &    CA    &  89 &   0 &   0 &   0  &  84 &   0 &   0 &   0  &  95 &   0 &   0 &   0 &          95          \\
  1999 2000 & WS or CA & 128 &   0 &   0 &   0  & 124 &   0 &   0 &   0  & 134 &   0 &   0 &   0 &         134          \\
  \hline
            &    WS    &  75 &   0 &  16 &   0  &  68 &   0 &  15 &   0  &  75 &   0 &  16 &   0 &          75          \\
            &    CA    & 106 & 107 & 107 & 107  &  89 &  89 &  89 &  89  & 106 & 107 & 107 & 107 &         107          \\
  2000 2001 & WS or CA & 153 & 107 & 119 & 107  & 137 &  89 & 101 &  89  & 153 & 107 & 119 & 107 &         154          \\
  \hline
            &    WS    & 106 &   0 &  23 &   0  &  93 &   0 &  21 &   0  & 106 &   0 &  23 &   0 &         106          \\
            &    CA    & 134 & 136 & 136 & 136  & 119 & 119 & 119 & 119  & 137 & 137 & 137 & 137 &         137          \\
  2001 2002 & WS or CA & 200 & 136 & 148 & 136  & 176 & 119 & 129 & 119  & 201 & 137 & 148 & 137 &         201          \\
  \hline
            &    WS    &  27 &  11 &  16 &  11  &  24 &  10 &  17 &  12  &  27 &  11 &  18 &  12 &          45          \\
            &    CA    &   7 &   7 &   7 &   7  &   6 &   6 &   6 &   6  &   7 &   7 &   7 &   7 &           7          \\
  2002 2003 & WS or CA &  34 &  18 &  23 &  18  &  30 &  16 &  23 &  18  &  34 &  18 &  25 &  19 &          52          \\
  \hline
  2003 2004 &    WS    &  35 &  24 &  29 &  31  &  33 &  21 &  26 &  29  &  35 &  24 &  29 &  32 &          69          \\
  2004 2005 &    WS    &  25 &  23 &  26 &  25  &  19 &  16 &  19 &  19  &  26 &  23 &  26 &  25 &          47          \\
  2005 2006 &    WS    &  30 &  25 &  28 &  28  &  26 &  20 &  22 &  23  &  32 &  26 &  28 &  28 &          71          \\
  2006 2007 &    WS    & 107 & 106 & 103 & 103  &  48 &  45 &  46 &  47  & 107 & 108 & 104 & 103 &         124          \\
  2007 2008 &    WS    &  62 &  56 &  52 &  58  &  36 &  35 &  35 &  38  &  63 &  58 &  55 &  61 &          92          \\
  \hline
    total   &    WS    & 600 & 253 & 295 & 261  & 468 & 155 & 202 & 172  & 606 & 258 & 301 & 266 &         765          \\
    total   &    CA    & 336 & 250 & 250 & 250  & 298 & 214 & 214 & 214  & 345 & 251 & 251 & 251 &         346          \\
    total   & WS or CA & 843 & 503 & 530 & 511  & 690 & 369 & 402 & 386  & 855 & 509 & 536 & 517 &        1015          \\
  \hline\hline
  \end{tabular}
  \caption{The number of analyzed nights per year during the 11 WeCAPP seasons. 
    Note that from 1999 until 2002 we used both telescopes at Wendelstein (WS)
    and Calar Alto (CA). A season is defined
    to last from May 1$^{th}$ until 30$^{th}$ April of the next year.
    The total amount of observed nights are 1015 nights out of 11 years.
    A total of 4,432 stacked frames were analyzed in both filters and 4 fields. }
  \label{tab.time_histo}
  \end{center}
\end{table*}

\begin{figure}[!ht]
  \centering
  \includegraphics[width=0.8\textwidth]{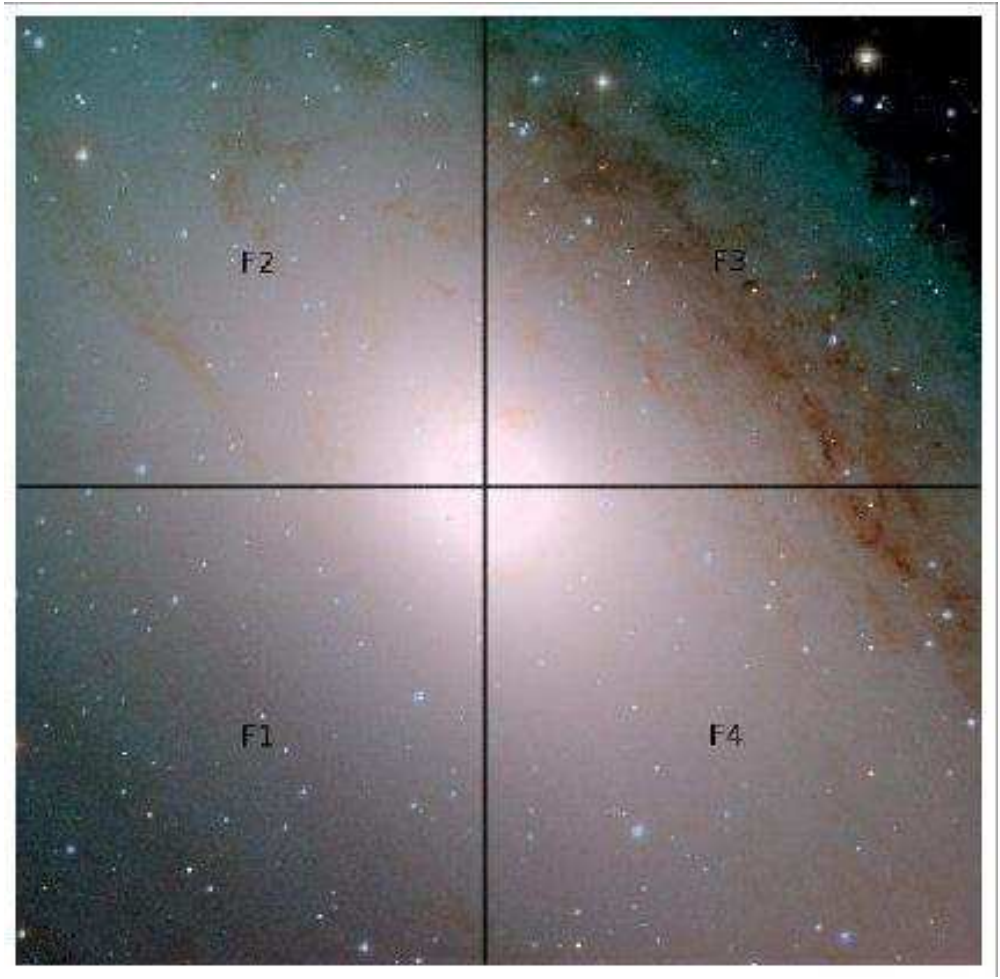}
  \caption{M31 composite image ($V$-, $R$-, and $I$-band) of the
    observed fields F1 to F4, taken at Calar Alto Observatory during
    the 2000/2001 season. The black lines mark the positions of
    fields F1 to F4. The Calar Alto camera covers all
    four fields simultaneously. Field F1 was observed during all 11 campaigns 
    from September 1997 until March 2008.}
  \label{fig.m31}
\end{figure}

WeCAPP continuously monitored M31 from August 1997 until March 
2008 using the Wendelstein 0.8m telescope \citep{2001A&A...379..362R}. The data were initially taken with a 
TEK CCD with 1,024$\times$1,024 pixels with a field-of-view of 8.3$\times$8.3
arcmin$^2$ pointing at the bulge of M31, optimally on a daily basis in both \textit{R}- 
and \textit{I}-filters. Following the suggestions of \cite{1996AJ....112.2872T} and 
\cite{1996ApJ...473..230H}, we pointed to the far side of the M31 disk (F1 in Fig. \ref{fig.m31}),
where the halo lensing probability is maximized. From June 1999 to December 2002 
we collected additional data using the 1.23 m (17'.2$\times$17'.2 FOV) telescope 
at Calar Alto
Observatory in Spain to increase the time sampling. This provided a
FOV which is four times the Wendelstein FOV, and enabled us to survey the major 
part of the M31 bulge. After 2002, we used the Wendelstein telescope
solely to mosaic the full Calar Alto field-of-view with four pointings, as 
indicated in Fig. \ref{fig.m31}.

\begin{figure*}[ht]
  \centering
  \includegraphics[width=0.48\textwidth]{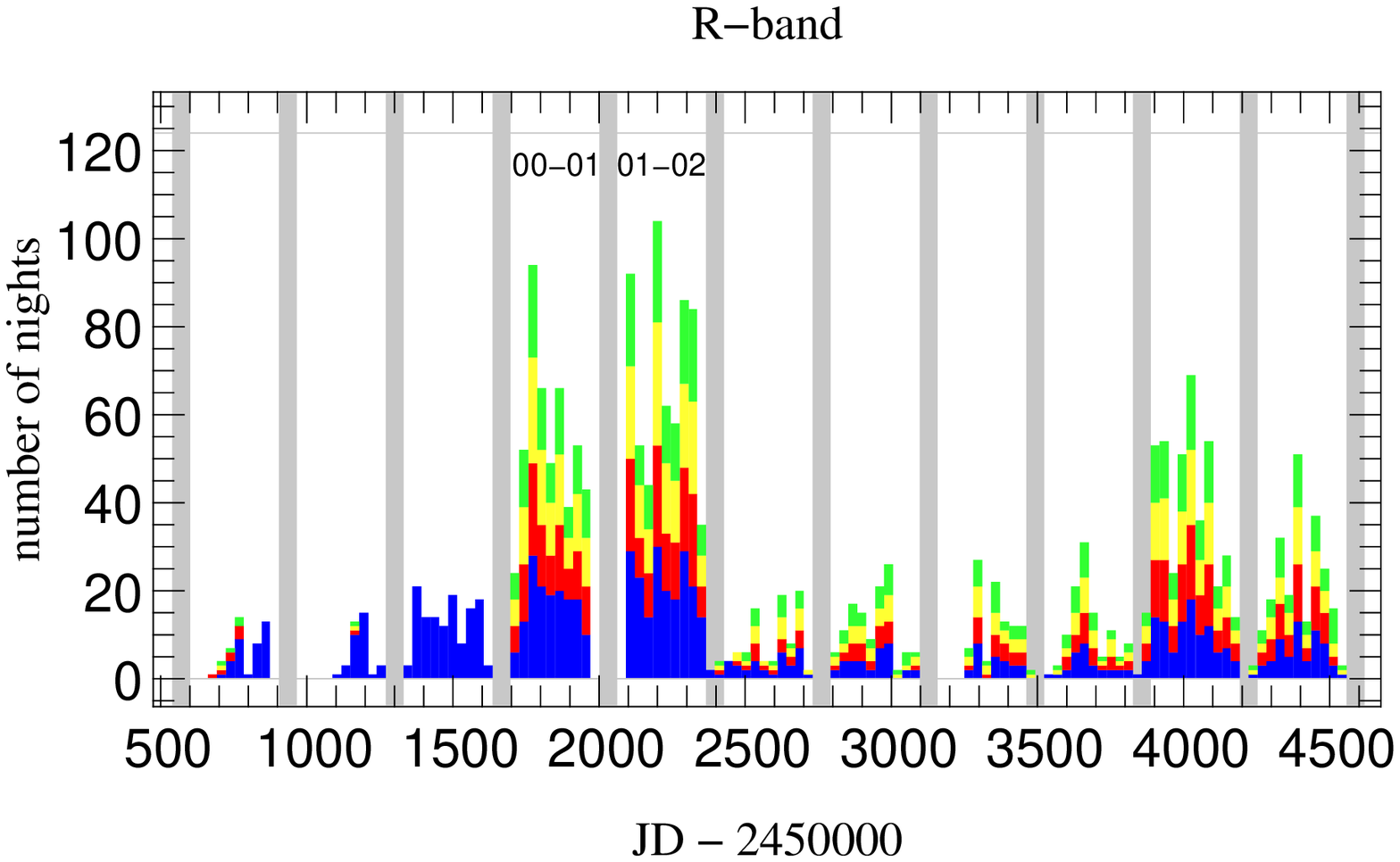}
  \includegraphics[width=0.48\textwidth]{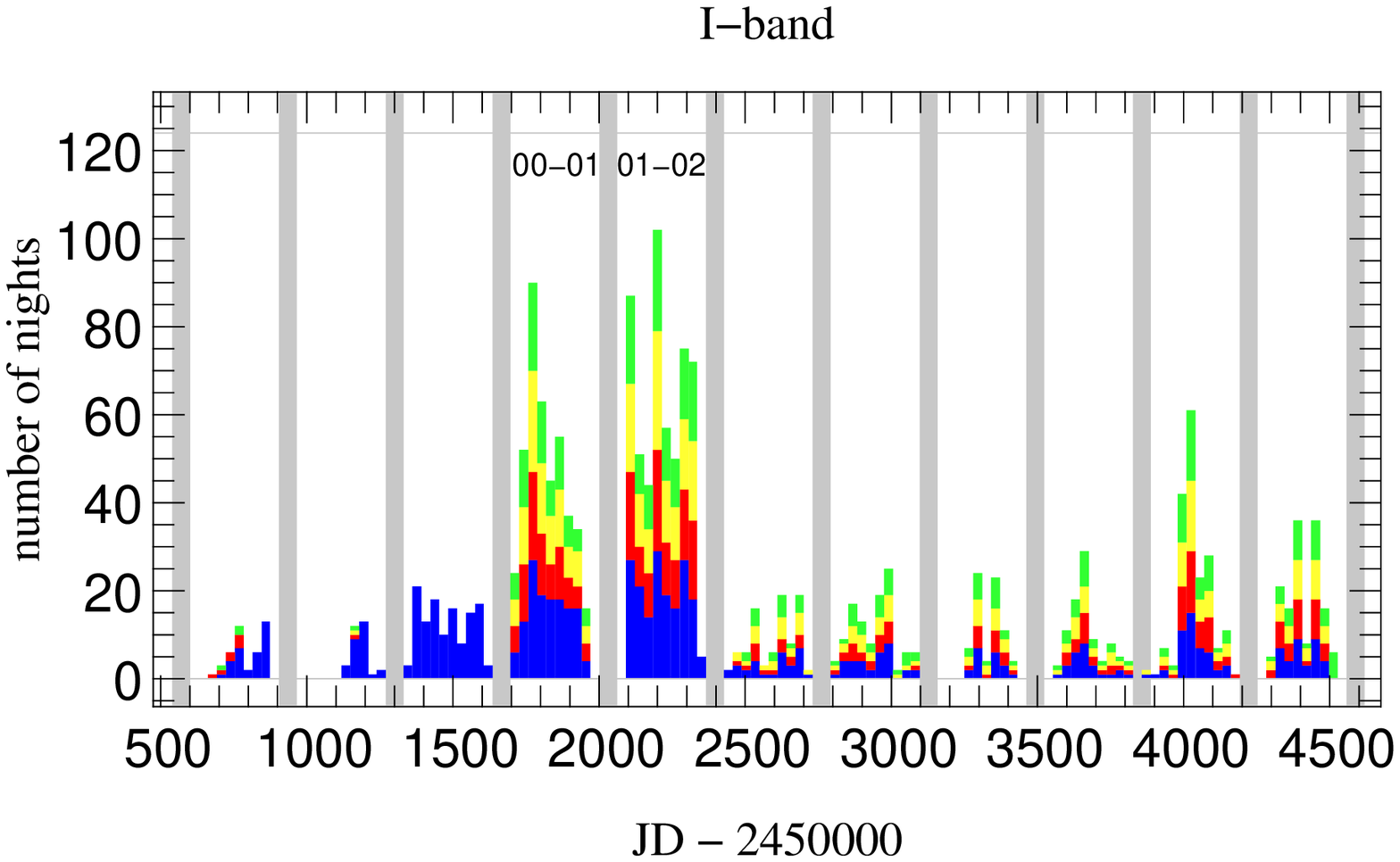}
  \caption{Histogram of the number of analyzed nights.  The 4 different fields
    F1, F2, F3, F4 are colored {\it blue} {\it red}, {\it yellow} and {\it green},
    respectively. Periods marked in {\it gray} show the 61 days (from
    $1^{st}$ April to $31^{st}$ May) during which M31 can hardly be
    observed (see also Table.~\ref{tab.time_histo}). Our most complete seasons (2000/2001 and 2001/2002) are marked with the texts ``00-01'' and ``01-02''.}
  \label{fig.hist_t0}
\end{figure*}

The amount of observations taken in the four pointings differs significantly 
during the 11 seasons. Fig. \ref{fig.hist_t0} shows a histogram of the 
number of observed 
nights by WeCAPP. The most complete seasons are 2000/2001 and 
2001/2002 with joint observations from both Wendelstein and Calar Alto (see also 
Table \ref{tab.time_histo}). 

\begin{figure}[!ht]
  \centering
  \includegraphics[width=0.8\textwidth]{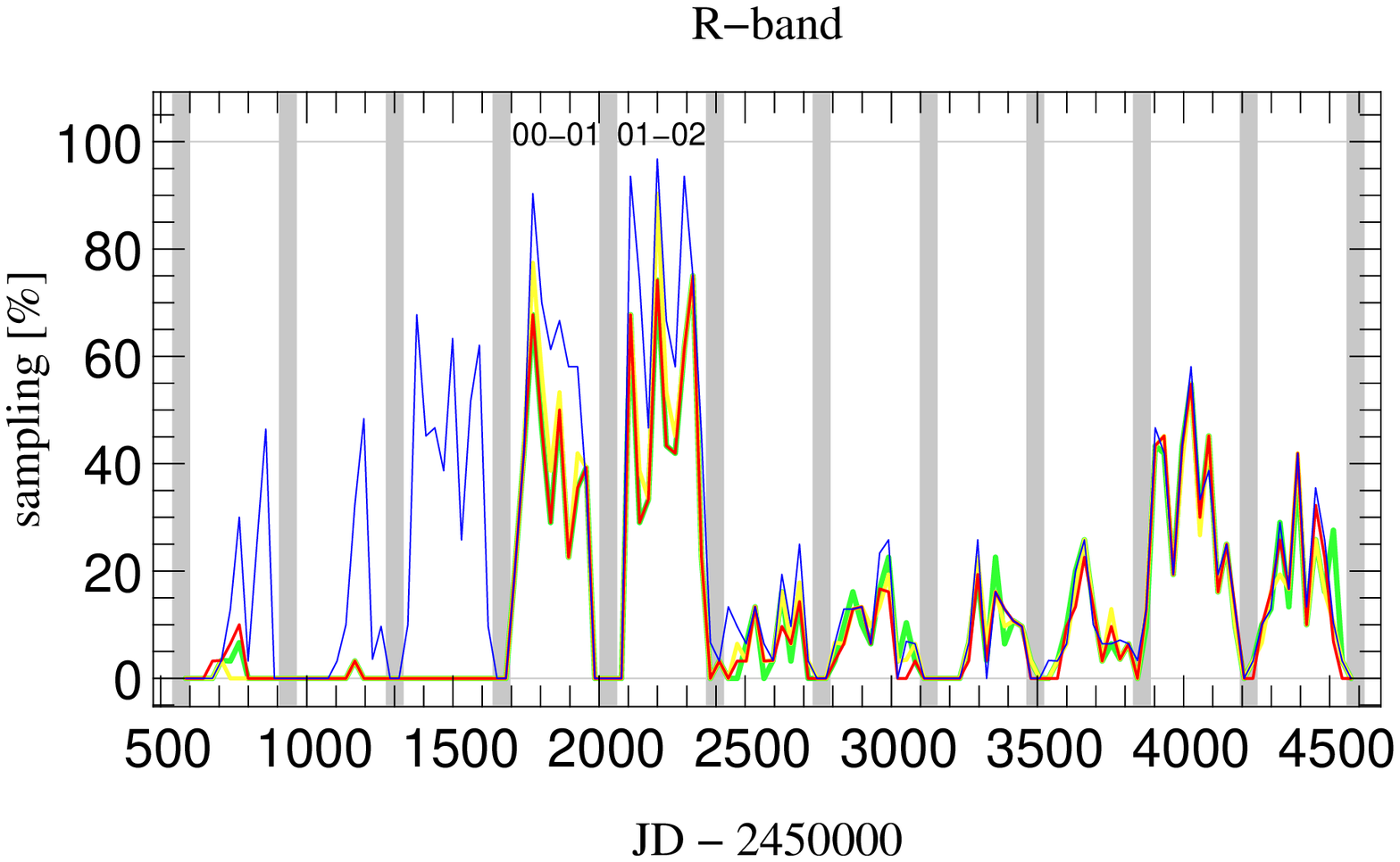}
\caption{Daily sampling of the 4 different fields
    F1, F2, F3, F4 are colored in {\it blue}, {\it red}, {\it yellow} and {\it green},
    respectively. Periods marked in {\it gray} show the 61 days (from
    $1^{st}$ April to $31^{st}$ May) during which M31 can hardly be
    observed. See also Table.~\ref{tab.time_histo}. The two seasons with the highest sampling
    are 2000/2001 and 2001/2002, i.e. those seasons where we could combine Wendelstein with 
    Calar Alto data.}
  \label{fig.Daily_sampling}
\end{figure}

To have an overview of the observing cadence, we show the daily sampling 
in Fig. \ref{fig.Daily_sampling}. 
Thanks to the joint observations at Wendelstein and Calar Alto, 
we achieved an average time coverage for F1 in R of 42\% during the 2000/2001 
season (peaking in August 2000 with 90\% on JD $\sim$ 2451770) and an average 
time coverage of 55\% during the 2001/2002 season, reaching more than 93\% in 
3 months (July and October 2001, and January 2002, around JD $\sim$ 2452110, 
2452200 and 2452290, respectively).
During the 11 seasons, we have obtained 
a total sampling efficiency\footnote{The sampling efficiency is a fraction with respect to the overall baseline length, i.e. including periods where no observations were scheduled.} of 14.9\% in $R$ and 11.5\% in $I$, with 
11.3\% for $R$ and $I$ combined for all fields (F1-F4), which means that 
in 11.3\% of the nights we have both $R$ and $I$ observations.

\begin{figure}[!ht]
  \centering
  \includegraphics[width=0.8\textwidth]{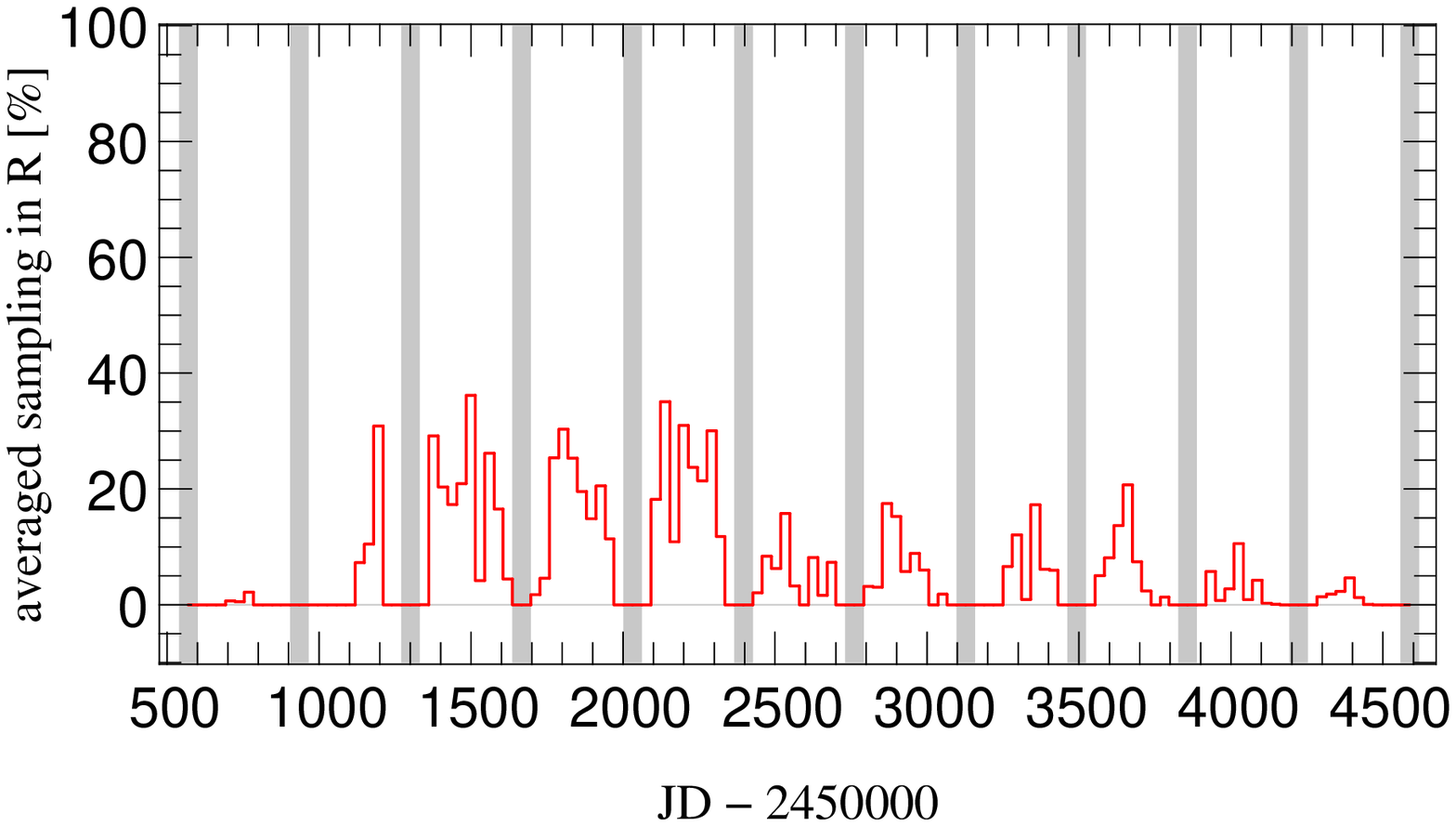}
  \caption{Sampling in $t$ averaged over $x$ and $y$ $<\M{S}(x,y,t)>_{xy}$}
  \label{fig.lownoise_sampling}
\end{figure}

\begin{figure}[!ht]
  \centering
  \includegraphics[width=0.8\textwidth]{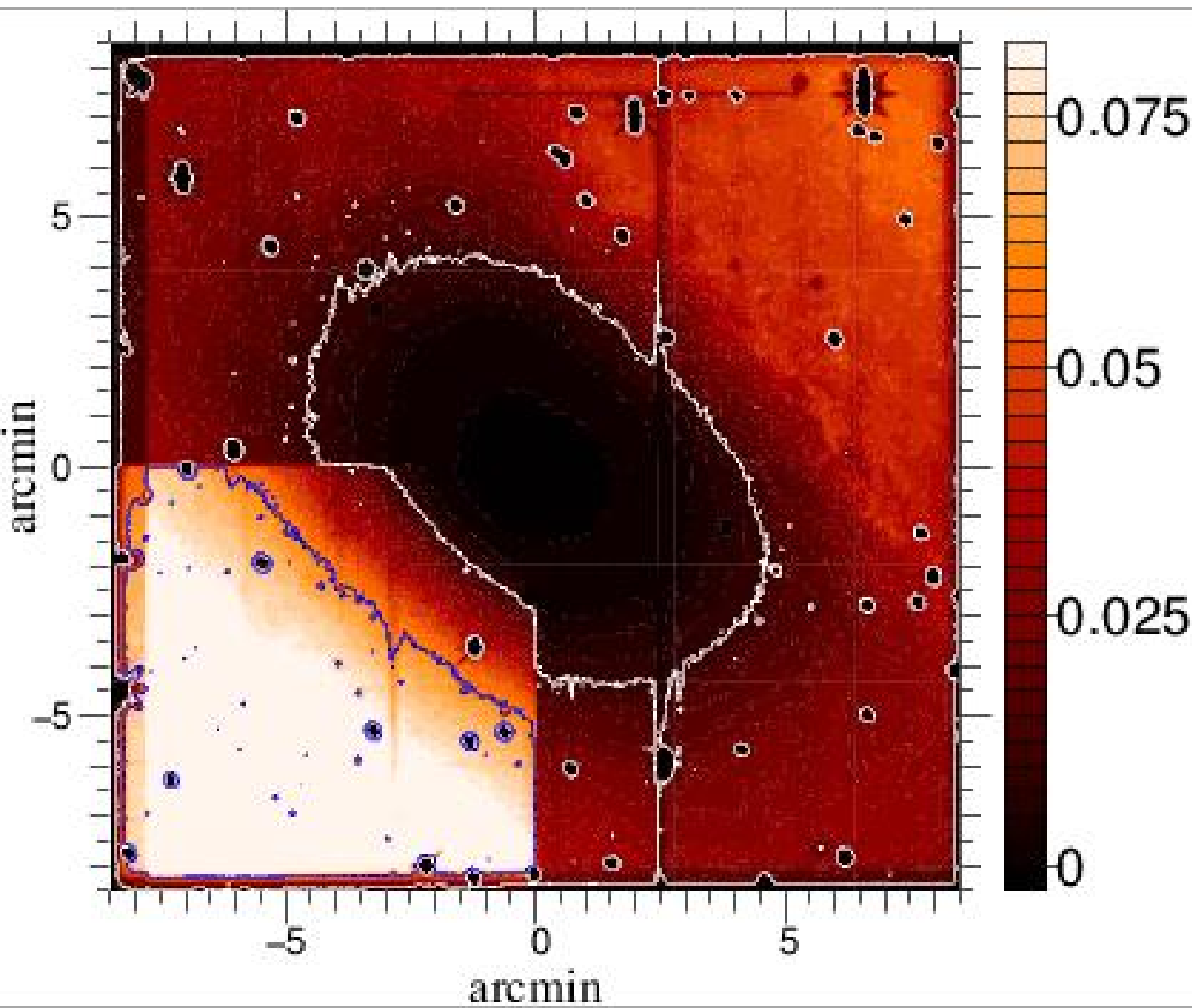}
\caption{Sampling in $(x,y)$ averaged over time $<\M{S}(x,y,t)>_t$. 
         Color levels range from 0\% to 10\%; the two contour levels
         show 1\% and 5\%, respectively. 58 arcmin$^2$ are well sampled
         by more than 5\% of the 11 year survey. 220 arcmin$^2$ are well
         sampled by more than 1\%.} 

  \label{fig.lownoise}
\end{figure}

However, not all images have the same quality. 
Rather than quantifying the fraction of nights we have observed through the
11 seasons we would like to have (as a function of location in M31) the fraction of nights where
the noise is below a certain threshold.
For this reason, we empirically chose a noise limit of 
the minimum S/N of 8.9 for our faintest event, i.e. 0.73 $\times$ 10$^{-5}$ Jy/8.9, 
which corresponds to the 8.9$\sigma$ detection criterion we present in section 4.
For pixels with noise levels above this value, the lensing signal would mix with 
the high noises and could not be detected. Hence these pixels cannot be used for the detection. 
The sampling thus depends on the $x$ and $y$ position of the pixel, as well as the 
observation time $t$, which we denote as $<S$($x,y,t$)$>$.
In Fig. \ref{fig.lownoise_sampling}, we show the area having a noise smaller
than our noise limit for every observed night. By averaging over time $t$, we will 
get the positional dependence as shown in Fig. \ref{fig.lownoise}.

\section{Data reduction}
\label{sec.red}

We process the data using our customized pipeline MUPIPE \citep{2002A&A...381.1095G}, 
where standard reduction processes -- such as bias subtraction, treatment of bad pixels, 
flat-fielding, cosmic ray removing -- are performed with per pixel error propagation. 
To identify unresolved variables, we employ the difference imaging analysis (DIA) 
proposed by \cite{1998ApJ...503..325A}, which enables us to detect variables with
amplitudes at the photon noise level and to measure their flux excesses relative to
high signal-to-noise reference images. 

After the difference imaging, we perform PSF photometry on each pixel in the following 
manner. We first extract the PSF profile from several isolated, bright and unsaturated 
reference stars. 
Then we fit this PSF to all pixels to generate light curves  of varying sources 
identified in the difference images.
The flux of the source is estimated by integrating the count rates over the 
area of the PSF.

The results from a subset of data of this project have been presented in 
\cite{2003ApJ...599L..17R, 2008ApJ...684.1093R} and partially contributed to 
\cite{2010ApJ...717..987C}. In addition to the original microlensing targets, 
the high cadence observations also yielded a sample of more than 20,000 variables 
in the bulge of M31 \citep{2006A&A...445..423F} and 91 candidate novae 
\citep{2007A&A...465..375P,2012A&A...537A..43L}.

\section{Event detection}
\label{sec.det}

\cite{1986ApJ...304....1P} provided an analytic formula to describe 
the amplification of a microlensing event:
\begin{equation}
A(t)=\frac{u^2+2}{u\sqrt{u^2+4}}, \quad u=\sqrt{\frac{(t-t_0)^2}{t_\M{E}^2}+u_0^2},
\end{equation}
where $t_0$ is the time of maximum amplification, $t_\M{E}$ is the Einstein ring
crossing time, $u_0$ is the impact parameter in units of the Einstein ring radius.
The light curve (or the measured flux as a function of time) of the microlensing 
event can thus be expressed as:
\begin{equation}
F(t) = F_0[A(t)-1]+B,
\label{eq.pac}
\end{equation} 
where $F_0$ is the un-lensed flux and $B$ is the blending within the PSF.
 
This conventional microlensing light curve formula is highly degenerate in $t_\M{E}$ and $u_0$ for
microlensing events towards M31, because one can not resolve the
source anymore. For M31 microlensing, the only observables are the flux
excess $\DF$ and the event time-scale $\tfwhm$. 
Gould (1996) has derived the pixel-lensing light curve 
formula in the context of high magnification. Riffeser et al. (2006)
further revised the microlensing light curve formula by Gould (1996), such
that microlensing light curves with moderate magnifications can be well-described
as well. The formula of pixel-lensing light curve from Riffeser et al. (2006) is 
expressed as:
\begin{equation}
  \DF(t) \approx F_\M{eff}
      \left[\frac{12 (t-t_0)^2}{\tfwhm^2} + 1\right]^{-1/2} + B
  \label{eq.gould}
\end{equation}
where $F_\M{eff}$ is the effective flux, which for high magnifications is
approximated by the flux excess $F_\M{eff} \equiv F_0/u_0 \approx \Delta_F$.
Since Paczynski (1986) was the first to present the simple analytical form of microlensing
light curve as equation (\ref{eq.pac}), we refer to equation (\ref{eq.pac}) as Paczynski-fit throughout
this paper.
Since Gould (1996) was the first to introduce the approximated form of microlensing
light curve as equation (\ref{eq.gould}), we refer to equation (\ref{eq.gould}) as Gould-fit throughout
this paper.

We use equation (\ref{eq.gould}) to identify microlensing events. 
The microlensing event detection is performed on the light
curve of each pixel (based on the aforementioned PSF photometry) with
several successive criteria. 
Compared to Riffeser et al. (2003), we introduce additional criteria to avoid human interaction
during the selection process. We now describe our criteria; an overview of the number of pixel light curves passing each
criterion is listed in Table \ref{tab.sel_new}. 

\begin{table*}[!ht]
  \setlength{\tabcolsep}{1.5mm}
  \begin{center}
  \begin{tabular}{rlrrr}
    \hline\hline
    & \multicolumn{1}{c}{criterion} & number & left LCs from I  & colors in Fig.~\ref{fig.sigmahist}\\
    \hline
    I    & Analyzed LCs with $>$ 50 data points                 & 3,872,240  & 100.0\% &    \\
    II   & Three consecutive 3$\sigma$ in $R$                   &   719,628  &  18.6\% &     \\ 
    III   & $\chi_R<1.5$ {\bf and} $\chi_I<2.1$                  &   152,753  &   3.9\% & yellow  \\ 
    IV    & $(S/N)>8.9$ in $R$ {\bf and} maximum with good PSF  &     2,247  &   0.5\% & black    \\ 
    V     &  $-1.5 \ge E_R^{peak} \le 1.5$                        &     1,379  &  10.4\% & green   \\ 
    VI    & $\tfwhm<1000$d                                      &       63   &  16.8\% & blue    \\ 
    VII   & $samp^{t<t_0}_R>$0.18 {\bf and} $samp^{t>t_0}_R>$0.08 {\bf and} & & & \\
          & $samp^{t<t_0}_I>$0.18 {\bf and} $samp^{t>t_0}_I>$0.08     &       12   &  9.3\% & red    \\ 
  \hline
    \vspace{-0.5cm}
  \end{tabular}
  \caption{Selection criteria for the 11 year data taken between MJD = 685.5 and 4535.3. 
    In the 3rd column we show how applying each criterion (II, III, IV, V, VI,
    VII) to the 3,872,240 light curves (criterion I) reduces the filtered light curves.}
  \label{tab.sel_new}
  \end{center}
\end{table*}

\begin{itemize}

\item Criterion I is applied to exclude pixel light curves which have 
too few data points to make them worth analyzing. Hence we exclude light curves
with less than 50 data points in either R- or I-filters. This leaves us with 
3,872,240 light curves.

\item Criterion II identifies varying sources warrant for further analysis. We
preselect variable pixel light curves with at least three consecutive 3$\sigma$ 
outliers\footnote{The difference between the individual flux measurement and the 
constant part/offset of the light curve is three times larger than the 
error of this individual flux measurement. The constant part/offset of 
the light curve is derived in two steps:
after removing all data points with very high errors (larger than 5 times 
the median of all errors) we determine the constant part/offset of the 
light curve by iteratively fitting a line and clipping away all
points which are more than 2 sigma/errors above."
} in the R-filter.

\item Criterion III is designed to find microlensing light curves with good $\chi^2$.
We select pixel light curves that are well described by microlensing light curves.
We use the microlensing light curve in equation (\ref{eq.gould}) with some modifications: 

\begin{equation}
  \DF(t) \approx F_\M{eff} 
      \left[\frac{12 (t-t_0)^2}{\left(t_\M{fit}^2+0.5\right)^2} + 1\right]^{-1/2} + c
  \label{eq.DF}
\end{equation}

where we set $t_\M{fit} \equiv \sqrt{\tfwhm-0.5}$, hence the full width
half-maximum event timescale is always greater than or equal to 0.5 day.
We use $t_\M{fit}$ instead of $\tfwhm$ to avoid un-physically small $\tfwhm$ 
due to limited time resolution (up to half day) of our observation. 
The term $c$ takes into account the shift of the baseline flux by a
constant in cases for which there is a variable source spatially close to the
microlensing event and data of these variable phases enter the
reference image.

We then filter out light curves with a $\chi_k \equiv
\sqrt{\chi_k^2}$ (where $\chi^2_\M{k}$ is the total $\chi^2$ divided by
the degrees of freedom $k$) larger than 1.5 in the R band and larger than
2.1 in the I band. This is less strict than the value we used
in (Riffeser et al. 2003). The allowed $\chi_\M{k}$ in I is slightly higher
than R, because the noise level is increased by unaccounted systematics 
such as the detector fringing and nearby variable stars in the I band.

\item Criterion IV evaluates whether a high flux excess relative to the baseline
is more likely caused by random noise or due to a true microlensing signal. 
We consider the S/N of such a high flux excess measurement, $\M{SN}_i \equiv \frac{\Delta_{\M{con},i}}{\sigma_i}$,
where the flux offset is $\Delta_{\M{con},i} \equiv y_i - c$, $y_i$ is the $i$th flux measurement at time $t_i$ and $\sigma_i$ is 
the error of the flux measurement.
We take into account the probability of such a high flux excess measurement
being close to the maximum of a microlensing 
light curve fit $p_{\M{fit},i} \propto \exp\left(-\frac{\Delta_{\M{fit},i}^2}{2 \sigma_i^2}\right)$,
where the flux offset is $\Delta_{\M{fit},i} \equiv y_i - \Delta_F(t_i)$, and $\Delta_F(t_i)$ is the model flux at time $t_i$ according to
Equation (\ref{eq.DF}). We also take into account the probability of such a high flux excess measurement 
being at the constant baseline $p_{\M{con},i} \propto \exp\left(-\frac{\Delta_{\M{con},i}^2}{2
\sigma_i^2}\right)$. We then combine these three factors and define the S/N probability (SNP):

\begin{equation}
  \M{SNP}_{i}= \M{SN}_i \times p_{\M{fit},i} \times \left( 1-p_{\M{con},i}
  \right)
\end{equation}

This ensures that light curves that have high $S/N$ outliers, that are outside the time
interval of the microlensing event, get a lower weight.

To avoid multiple detections of the same microlensing signal in neighboring pixels, 
we only use the pixel with good PSF detections\footnote{To find the good PSF detections we have carried out the following steps: 1. We subsample each pixel by a factor of 5; 2. We determine the flux at the position of each sub-pixel by fitting a PSF; 3. We compare the $\chi^2$ of the PSF with the adjacent 8 pixels. If the pixel has the lowest $\chi^2$ among the adjacent 8 pixels, we consider it as a good PSF detection. After the detection we refit the position of 
the microlensing events and determine their positions at sub-pixel level.} to evaluate SNP$_i$, i.e. where the PSF fit
has a minimum in $\chi^2$ with respect to neighboring pixels.

We empirically require one data point in the light curve to have a SNP$_i$ 
larger than 8.9 in the R band to efficiently reject faint variable sources.

\item Criterion V quantifies the temporal correlations of the 
model-data mismatch (best-fit microlensing light curve vs. measured light curve) 
and enables us to reject intrinsic variable sources more efficiently. 

We combine the probabilities for positive $p^{+}_i$ or negative flux
offsets $p^{-}_i$ from the best-fit microlensing light curve, and 
assign positive values for consecutive data points that are always on the same 
side (either above or below) of the best-fit model light curve. For 
consecutive data points alternating along the best-fit model light curve, 
we assign negative values. We then define the energy of a potential 
microlensing light curve as

\begin{equation}
E \equiv \frac{\pi}{\sqrt{n}} \sum_{i=1}^{n-1} 
             p^{+}_i p^{+}_{i+1} + p^{-}_i p^{-}_{i+1} -
             p^{+}_i p^{-}_{i+1} - p^{-}_i p^{+}_{i+1}
\end{equation}
where $n$ is the number of data points and
the probabilities are defined as $p^{+}_i \equiv 0.5 \left[ 1 +
\erf\left(\sqrt{\frac{\Delta_{\M{fit},i}^2}{2 e_i^2}}\right) \right]$
and $p^{-}_i \equiv 1 - p^{+}_i$.

For a random process the distribution of $E$ is a Gaussian with an
expectation value of zero and a standard deviation of one.

We derive the energy $E^{20}$ using the $n=20$ closest data points to $t_0$ in
the R-band.
We find empirically that a value between -1.5 and 1.5 efficiently
rejects periodic variables and allows to skip the previously used
by-eye detection in Riffeser et al. (2003).

\item Criterion VI filters out long-periodic variables.
Since our light curves span a baseline of 11 years, contaminations from
the long-periodic variables are less severe than any of the previous campaigns.
We inspected the candidate light curves and found that
false detection from systematics and moving objects are having timescales longer than 1000 days.
Hence we are able to empirically increase the upper  
$\tfwhm$-limit for micrlensing searches to 1000 days (compared to, e.g. $\tfwhm <$ 20 days in Riffeser et la. 2003). With this criterion we also filter
out objects with proper motion. A moving object that passes through the 
field with constant angular velocity will cause variabilities at some (stationary) pixels that can mimic microlensing signals.
These proper motion objects can of course be excluded by inspecting postage stamps, but we would like to 
have a completely automatic selection here.

\item Criterion VII rejects light curves which 
look like microlensing events from their overall light curves but
which are not well sampled close to the light curve maximum, hence 
could be variable sources. The contaminations are mostly from novae,
e.g. not well sampled in their rising parts.

We define the sampling quality for the falling and rising
parts of each light curve within $(t_0-15\E{days}, t_0)$ and
$(t_0,t_0+15\E{days})$. The contribution of a single data point is
then calculated by integrating the model light curve within
$(t_i-0.5\E{days},t_i+0.5\E{days})$.  As sampling criteria we require
a total sampling of the area under the light curve of at least 18\% on
the rising part of the light curve and of at least 8\% on the falling
side in the R and I band.

\end{itemize}

In the 4$^{th}$ column of Table~\ref{tab.sel_new} we show how applying
each criterion (II, III, IV, V, VI, VII) to the  
3,872,240 light curves (criterion I) reduces
the filtered light curves. For example, 152,753/3,872,240$\sim$3.9\% light curves 
that pass criterion III. Less efficient criteria show a high percentage of
remaining light curves while efficient criteria filter out more light curves. 
Therefore the criterion IV is our most efficient criterion. 
In the 3rd column we show how applying each criterion 
(II, III, IV, V, VI, VII) to the 3,872,240 light curves (criterion I) reduces 
the filtered light curves. 

\begin{figure}[ht]
  \centering
  \includegraphics[width=0.8\textwidth]{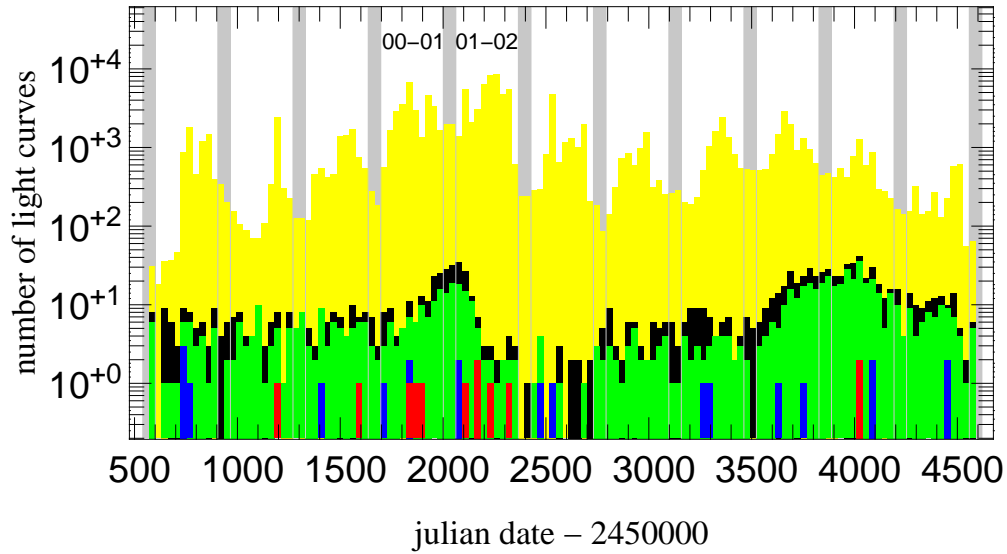}
  \caption{Histogram of the number of light curves passing
    different detection criteria. Our most complete seasons (2000/2001 and 2001/2002) are marked with ``00-01'' and ``01-02''. {\it Yellow}:
    $\chi$-limit (III). {\it Black:} $(S/N)$ constraint and good PSF (IV) 
    at light curve maximum. {\it Green:} Energy criterion for the 20 closest 
    data points to the maximum (V). {\it Blue:} timescale constraint (VI).
    {\it Red:} sampling criterion
    for the closest data points to the
    maximum (VII). In the end we are left with 12 microlensing events 
    detected (see Table~\ref{tab.sel_new} for details). Eight of them took place 
    during the two best observing seasons 2000/2001 and 2001/2002.}
  \label{fig.sigmahist}
\end{figure}

Fig. \ref{fig.sigmahist} shows the $t_0$ distribution of all
pixel light curves under analysis. The colored histograms show how the
number of light curves is reduced after each criterion.
The total numbers correspond to those given in Table \ref{tab.sel_new}.

Because of large dome seeing and an inappropriate
auto-guiding system, photometric errors are largest during the first season (1997/98). 
During the second season (1998/99) we were able to
decrease the FWHM of the PSF by a factor of two, and therefore the photometric
scatter also became smaller.

The vast majority of our microlensing events are 
detected between 2000/2001 and 2001/2002 seasons, where we have 
employed the Calar Alto telescope and the cadence is high. This
implies that deeper images with larger telescope along with densely sampled 
observations are crucial to detect microlensing events. Our observations
in other seasons are pivotal as well, as they serve the purpose to 
rule out contamination from variables.

\begin{figure*}[!ht]
  \centering
  \includegraphics[width=1.0\textwidth]{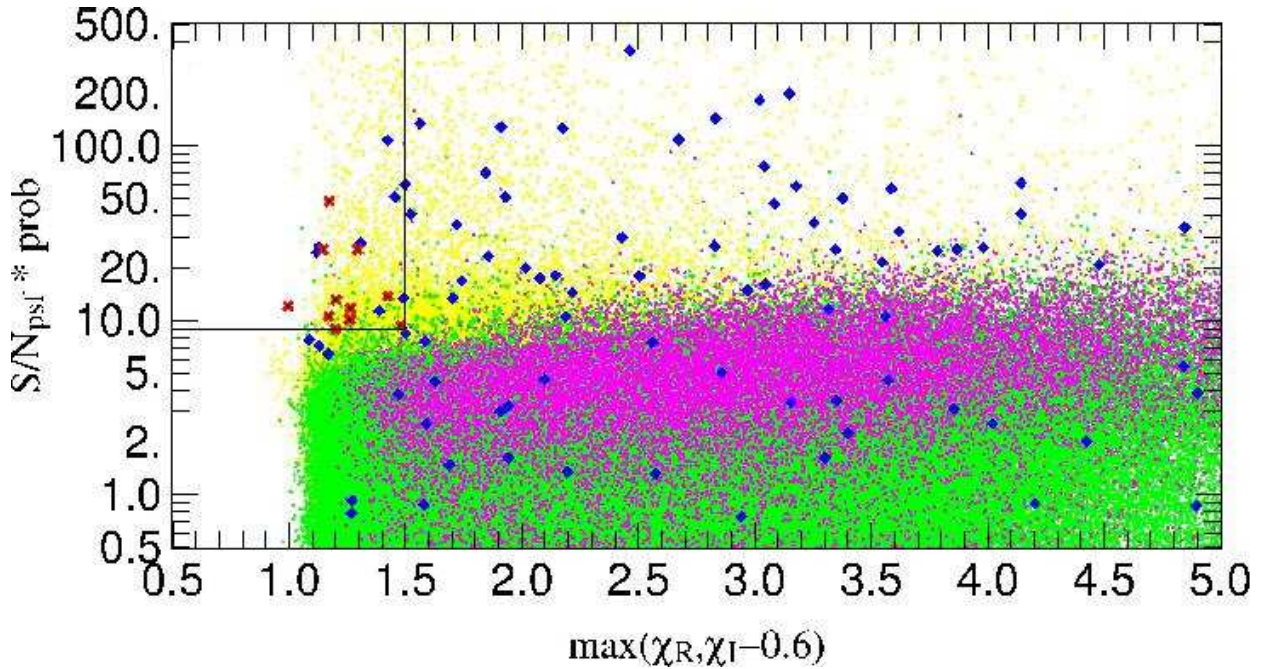}
  \caption{Criterion III and IV reduce the light curves
    from 719,628 (green) to 2247, as shown in the  
    box on the upper left side. The light curves which do no 
    pass criterion VI are marked in {\it yellow}. We
    show the 23,001 variables published in the WeCAPP variable star catalogue
    from \cite{2006A&A...445..423F} in {\it magenta}, as well as the 91 novae 
    from Lee et al. (2012) in {\it blue}. The final 12 microlensing events
    are shown in {\it red}. We note that the decline parts of nova light curves 
    resemble microlensing events and often make short timescale novae the major contaminations 
    to microlensing detections. In this regard, we show that criterion III and 
    IV can efficiently discriminate microlensing events from novae, hence 
    remove their contaminations.}
  \label{fig.detect_tfwhm_chi}
\end{figure*}

Fig. \ref{fig.detect_tfwhm_chi} shows the properties of the 719,628 
light curves passing criterion II. Since the major contaminations to
microlensing detections are variables and novae, we over-plot the 23,001 
variable sources published in \cite{2006A&A...445..423F} in 
\textit{magenta} and the 91 novae in Lee et al. 2012 in \textit{blue}. 
In Fig. \ref{fig.detect_tfwhm_chi} it is clear that criterion III 
and IV are efficient to filter out these two major contamination 
sources.

We note that not all of the criteria are independent.
To disentangle the issue of criteria overlapping with each other, we perform
a test with a subset of the criteria. The results are shown in Table \ref{tab.crit_overlap}.

\begin{table*}[!ht]
  \centering
  \begin{tabular}{l|rr}
    Criteria  & Number & Percentage \\
\hline\hline
 (1) I~II~& 719628 & 100.00 \\
\hline
 (2) I~II~III~~~~     & 152753 & 21.23=(2)/(1) \\
 (3) I~II~~~~~~IV~     &  18803 &  2.61=(3)/(1) \\
 (4) I~II~~~~~~~~~~V    & 402746 & 55.97=(4)/(1) \\
 (5) I~II~III~~~~~~V    & 106200 & 14.76=(5)/(1) \\
\hline
 (6) I~II~~~~~~IV~V~VI~VII &  1338 &  0.90=(10)/(6) \\
 (7) I~II~III~~~~~V~VI~VII & 41214 &  0.03=(10)/(7) \\
 (8) I~II~III~IV~~~~VI~VII &    15 & 80.00=(10)/(8) \\
 (9) I~II~~~~~~IV~~~~VI~VII &  5382 &  0.22=(10)/(9) \\
\hline
(10) I II III IV V VI VII & 12 & \\
  \end{tabular}
  \caption{Number of events that pass a subset of the criteria.}
  \label{tab.crit_overlap}
\end{table*}

Crit II and IV overlap already by definition. All detections passing Crit IV are a 
subset of the detections passing II, because we fit only light curves which have 3 
times 3-sigma outliers (Crit II). As can be seen in row (3) only 2.6\% of 
the Crit II light curves pass the Crit IV, therefore Crit IV is much stricter 
than Crit II. A comaprision between row (7) and (10) shows that Crit IV is able 
to reduce the preselected light curves from all other criteria by a factor 3400.
  
The comparision between Crit III and V shows that Crit V is weaker than III, 
because it reduces the light curves only by 56\% compared to 21\% from Crit III. 
Two-thirds of the light curves passing Crit II overlap with the light curves 
passing Crit V (14.8\% of 21.2\%), this shows a quite strong overlap. 
Also row (8) shows that Crit V has only a small impact to the detection 
as it reduces the number of detection only from 15 to 12.
  
In short, we are aware of the overlapping among criteria. However, in the cases of 
Crit II and IV, we need Crit II prior to Crit IV to preselect the light curves of 
interest, so that we can analyze the full data-set in a reasonable amount of time. 
Nevertheless overlapping criteria should not be an issue if the efficiency study 
is running through the same detection criteria.

Note that in Criterion VI we deliberately set the upper limit of $\tfwhm$ 
to be 1000 days. The fact that our final set comprises only events with $\tfwhm$ smaller than 15 days, 
can be hint that no rather long microlensing events in M31 exists. If we would have set the limit to
20 days, we could not address this speculation.

\begin{figure*}[!ht]
\centering
\includegraphics[width=1.0\textwidth]{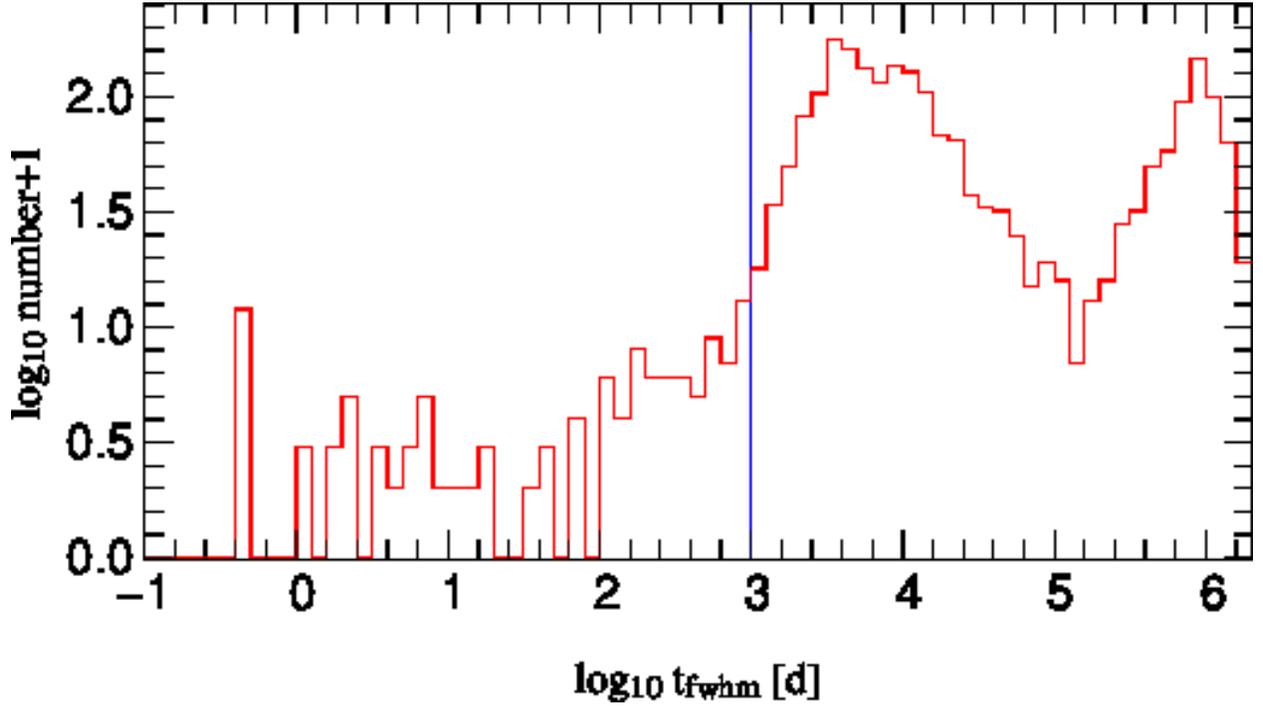}
\caption{$\tfwhm$ distrubtions of the 2247 detections passing criterion IV. The blue vertical line
indicates the 1000 days upper limit in our selection criterion.}
\label{fig.tfwhm_histo_SN_chi_good}
\end{figure*}

This $\tfwhm$ limit is empirically picked by looking into the $\tfwhm$ distributions of
the 2247 detections fulfilling passing criterion IV, as shown in Fig. \ref{fig.tfwhm_histo_SN_chi_good}.

We note that the long timescale fits may arise from slowly moving objects. To test for this
we inspect the positions of all detections passing criterion IV with $\tfwhm$ $>$ 1000 days.
We find that they are homogeneously distributed (they appear in some spikes of bright stars) 
except a small region close to a bright high proper motion object. This object influences the
difference imaging kernel and therefore mimicking a proper motion of all other sources.

\begin{figure*}[!ht]
  \centering
  \includegraphics[width=1.0\textwidth]{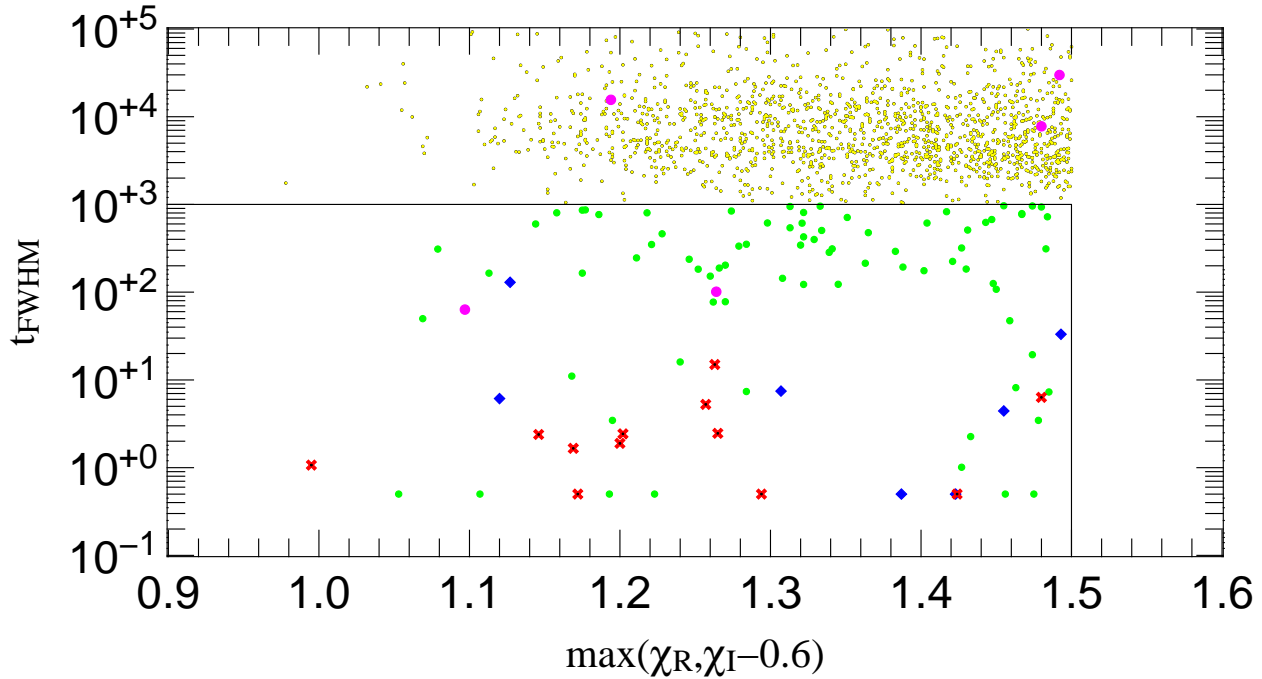}
  \caption{The events passing criterion IV, as well as our selection criteria of $\tfwhm$ and $\chi^2$. The yellow points are all detections passing criterion IV and with $\tfwhm \ge$ 1000d. The green points are all detections passing criterion IV and with $\tfwhm <$ 1000d. The magenta points are variables from Fliri et al. (2006). The blue points are novae from Lee et al. (2012). The red points with black dot are the 12 microlensing events presented in this paper.}
  \label{fig.detect_tfwhm_chi_good}
\end{figure*}

To demonstrate that 1000-d is a reasonable limit for $\tfwhm$, we plot the $\tfwhm$ and $\chi^2$
distributions of the events in Fig. \ref{fig.detect_tfwhm_chi_good}. The yellow points are 
all detections passing criterion IV and with $\tfwhm \ge$ 1000d. The green points are all detections 
passing criterion IV and with $\tfwhm <$ 1000d. The magenta points are variables from Fliri et al. (2006). 
The blue points are novae from Lee et al. (2012). 
The red points with black dot are the 12 microlensing events presented in this paper.

\section{WeCAPP M31 lensing events}
\label{sec.ana}

In this section we present the 12 microlensing events found in the 
WeCAPP data. 
Table \ref{tab.pacfit} gives 
an overview on the properties of these 12 events. In this table we focus on two 
observables $\tfwhm$ and $\DF$. 
We fit the light curves with equation \ref{eq.pac}.

As shown in Table \ref{tab.pacfit}, the Paczynski-fit provides 
additional information on the Einstein timescale $t_E$ and the impact parameter
$u_0$. 

We note that the parameters $\tfwhm$ and $\DF$ are degenerate, which often leads to 
overestimated errors in these two parameters for very short time-scale events ($\tfwhm$ below 2 day). 
Irrespective of this degeneracy, we can still determine these two parameters with reasonable uncertainty. 
To demonstrate the degeneracy, and our confidence in determining these two parameters, 
we show the 1, 2, and 3-$\sigma$ contour plot of $\tfwhm$ vs. $\DF$ in the appendix.

\begin{sidewaystable*}[!ht]
  \footnotesize
  \centering
  \setlength{\tabcolsep}{0.1mm}
      \begin{tabular}{|l|c|c|c|c|c|c|c|c|c|c|c|c|c}
    \hline\hline
          & RA(J2000)   & Dec(J2000)  & $\Delta_{\mathrm{M31}}$ & $t_0$ & $\tfwhm$ & $m_R$  & $\DF_{R}$ & $m_I$  & $\DF_{I}$ & (R-I) & log($t_E$) & log($u_0$) & $\chi^2_{dof}$ \\  
        &  & & [arcmin] & MJD [day]  & [day] & [mag] & [$10^{-5}$ Jy] & [mag] & [$10^{-5}$ Jy] & [mag] & [day] & [$\theta_E$] & \\
    \hline
WeCAPP-01$\ddag$ & 00:42:30.03 & 41:13:01.5  & 4.08 & $1850.84_{-0.02}^{+0.02}$ & $1.62_{-0.10}^{+0.10}$ & 18.68 & $10.29_{-0.53}^{+0.54}$ & 17.81 & $18.22_{-0.98}^{+
0.99}$ & $0.88_{-0.03}^{+0.03}$ & $7.85_{-0.01}^{+0.01}$ & $-8.18_{-0.02}^{+0.02}$ & 1.21       \\
WeCAPP-02 & 00:42:33.01 & 41:19:58.5  & 4.36 & $1895.41_{-0.62}^{+0.71}$ & $5.85_{-2.55}^{+3.04}$ & 20.81 & $ 1.45_{-0.27}^{+0.37}$ & 19.64 & $ 3.38_{-0.66}^{+
0.88}$ & $1.17_{-0.09}^{+0.09}$ & $1.61_{-0.35}^{+0.29}$ & $-1.36_{-0.49}^{+0.55}$ & 1.20       \\  
WeCAPP-03 & 00:42:57.03 & 41:12:27.9  & 4.40 & $1585.35_{-0.18}^{+0.14}$ & $2.44_{-0.26}^{+0.29}$ & 20.47 & $ 1.99_{-0.04}^{+0.10}$ & 19.11 & $ 5.51_{-0.11}^{+
0.26}$ & $1.36_{-0.04}^{+0.04}$ & $2.64_{-0.62}^{+0.62}$ & $-2.79_{-0.62}^{+0.62}$ & 1.23       \\  
WeCAPP-04 & 00:42:54.07 & 41:14:37.0  & 2.48 & $2178.86_{-0.06}^{+0.06}$ & $3.40_{-0.27}^{+0.27}$ & 20.18 & $ 2.59_{-0.10}^{+0.10}$ & 19.58 & $ 3.56_{-0.16}^{+
0.16}$ & $0.60_{-0.05}^{+0.05}$ & $0.99_{-0.11}^{+0.11}$ & $-0.93_{-0.15}^{+0.15}$ & 1.40       \\  
WeCAPP-05 & 00:43:02.03 & 41:18:29.2  & 4.09 & $2177.85_{-0.40}^{+0.39}$ & $6.34_{-1.10}^{+1.06}$ & 20.94 & $ 1.29_{-0.09}^{+0.12}$ & 20.38 & $ 1.71_{-0.16}^{+
0.20}$ & $0.56_{-0.11}^{+0.11}$ & $1.47_{-0.32}^{+0.33}$ & $-1.18_{-0.42}^{+0.39}$ & 1.38       \\  
WeCAPP-06 & 00:42:49.01 & 41:14:55.3  & 1.52 & $2317.23_{-0.03}^{+0.03}$ & $0.43_{-0.13}^{+0.16}$ & 18.15 & $16.79_{-5.56}^{+4.47}$ & 17.36 & $27.50_{-9.77}^{+
7.84}$ & $0.79_{-0.04}^{+0.03}$ & $8.27_{-0.02}^{+0.02}$ & $-9.18_{-0.11}^{+0.14}$ & 1.67       \\  
WeCAPP-07 & 00:42:55.03 & 41:18:50.9  & 3.38 & $1847.44_{-0.15}^{+0.14}$ & $1.24_{-0.34}^{+0.26}$ & 20.47 & $ 1.99_{-0.30}^{+0.37}$ & 19.74 & $ 3.08_{-0.43}^{+
0.59}$ & $0.73_{-0.13}^{+0.13}$ & $8.68_{-0.07}^{+0.06}$ & $-9.13_{-0.06}^{+0.05}$ & 1.02       \\  
WeCAPP-08 & 00:42:50.03 & 41:18:40.6  & 2.75 & $2111.55_{-0.06}^{+0.10}$ & $0.58_{-0.14}^{+0.16}$ & 19.11 & $ 6.93_{-0.94}^{+0.87}$ & 18.36 & $10.97_{-1.88}^{+
1.59}$ & $0.75_{-0.07}^{+0.06}$ & $1.01_{-0.31}^{+0.27}$ & $-1.77_{-0.38}^{+0.39}$ & 1.17       \\  
WeCAPP-09 & 00:42:44.07 & 41:12:54.9  & 3.22 & $2231.67_{-0.41}^{+0.40}$ &$14.04_{-1.38}^{+1.38}$ & 21.17 & $ 1.05_{-0.05}^{+0.05}$ & 20.48 & $ 1.55_{-0.10}^{+
0.11}$ & $0.68_{-0.07}^{+0.07}$ & $1.53_{-0.15}^{+0.14}$ & $-0.86_{-0.20}^{+0.20}$ & 1.42       \\
WeCAPP-10 & 00:42:12.01 & 41:09:21.5  & 9.07 & $4026.88_{-0.13}^{+0.13}$ & $2.75_{-0.48}^{+0.42}$ & 21.01 & $ 1.20_{-0.09}^{+0.12}$ & 20.10 & $ 2.21_{-0.22}^{+
0.26}$ & $0.92_{-0.11}^{+0.11}$ & $0.81_{-0.28}^{+0.33}$ & $-0.83_{-0.46}^{+0.39}$ & 1.38       \\
WeCAPP-11 & 00:42:46.02 & 41:15:11.9  & 1.01 & $1198.81_{-0.18}^{+0.18}$ & $6.41_{-0.79}^{+0.70}$ & 21.09 & $ 1.13_{-0.07}^{+0.08}$ & 19.89 & $ 2.69_{-0.16}^{+
0.19}$ & $1.20_{-0.08}^{+0.08}$ & $1.21_{-0.36}^{+0.47}$ & $-0.88_{-0.58}^{+0.43}$ & 1.62       \\
WeCAPP-12 & 00:43:07.08 & 41:17:40.7  & 4.64 & $4018.87_{-0.21}^{+0.24}$ & $4.92_{-1.20}^{+1.08}$ & 20.64 & $ 1.69_{-0.25}^{+0.30}$ & 19.80 & $ 2.90_{-0.43}^{+
0.53}$ & $0.84_{-0.12}^{+0.12}$ & $1.02_{-0.23}^{+0.25}$ & $-0.79_{-0.41}^{+0.36}$ & 1.17       \\ 
    \hline\hline
  \end{tabular}
  \caption{Paczynski parameters of the WeCAPP microlensing events. 
           In the fourth column we also show the distance from the M31 center.
           $^\ddag$To determine the parameters of the light curve, we include data from POINT-AGAPE,
            as in Riffeser et al. (2003).}
  \label{tab.pacfit}
\end{sidewaystable*}

\begin{figure*}[!ht]
  \centering
    \includegraphics[width=8cm]{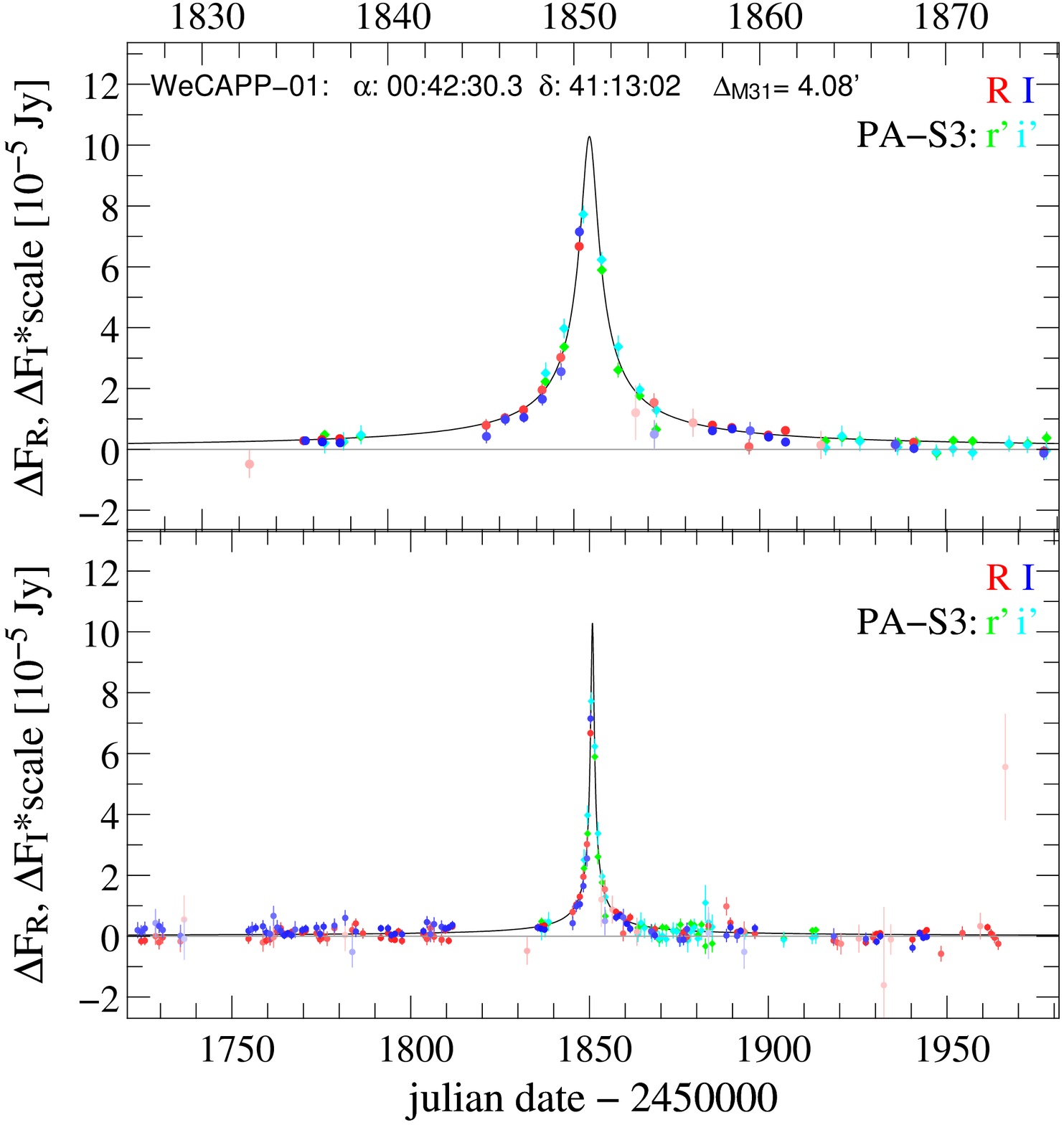}
    \includegraphics[width=8cm]{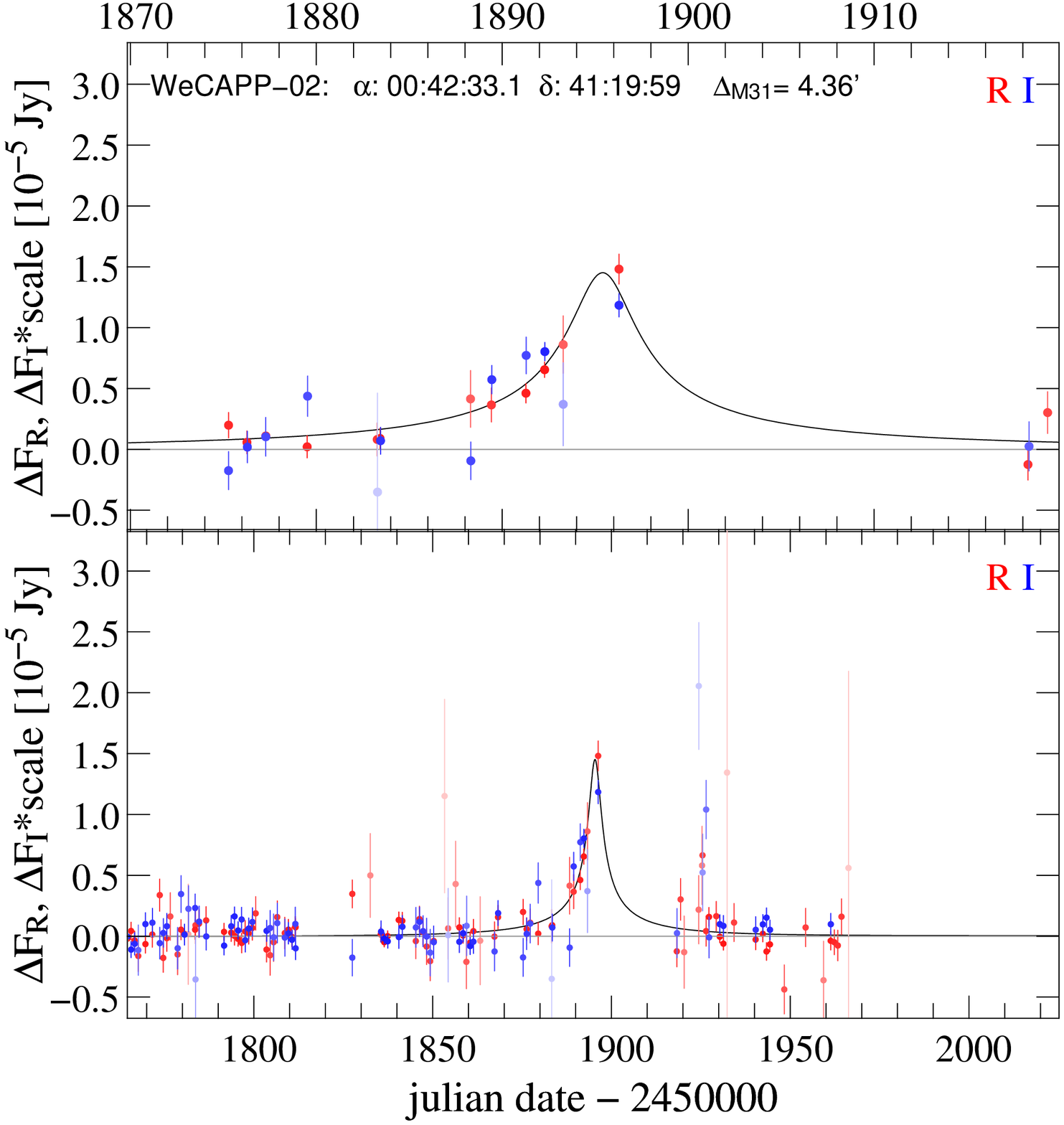}
    \includegraphics[width=8cm]{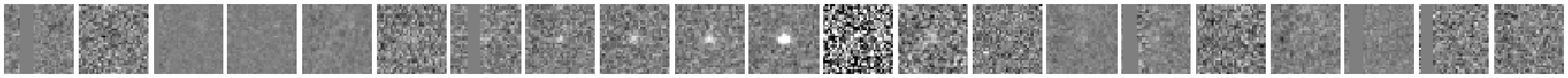}
    \includegraphics[width=8cm]{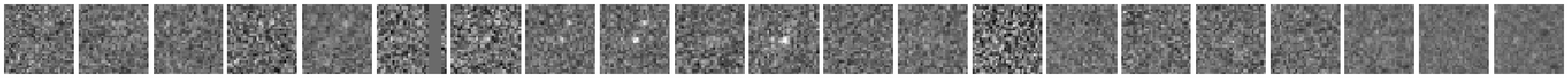}
    \includegraphics[width=8cm]{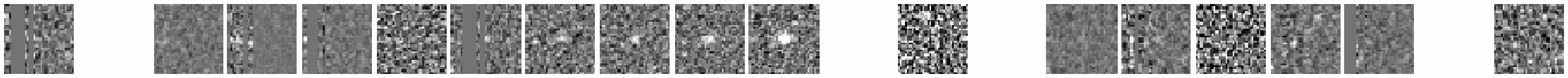}
    \includegraphics[width=8cm]{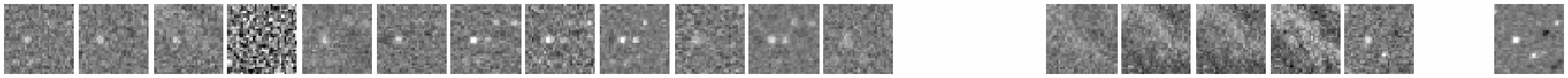}
    \includegraphics[width=8cm]{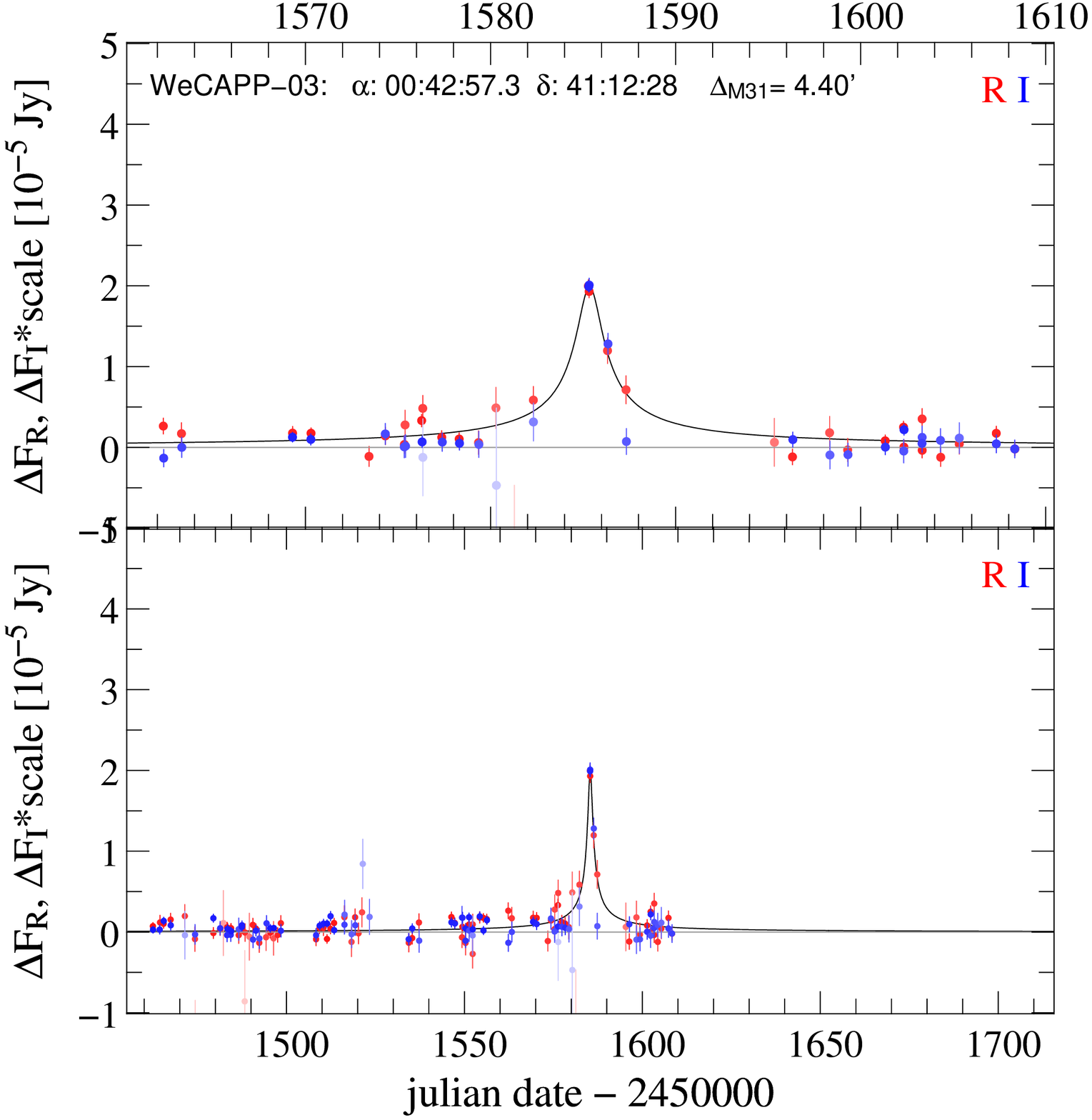}
    \includegraphics[width=8cm]{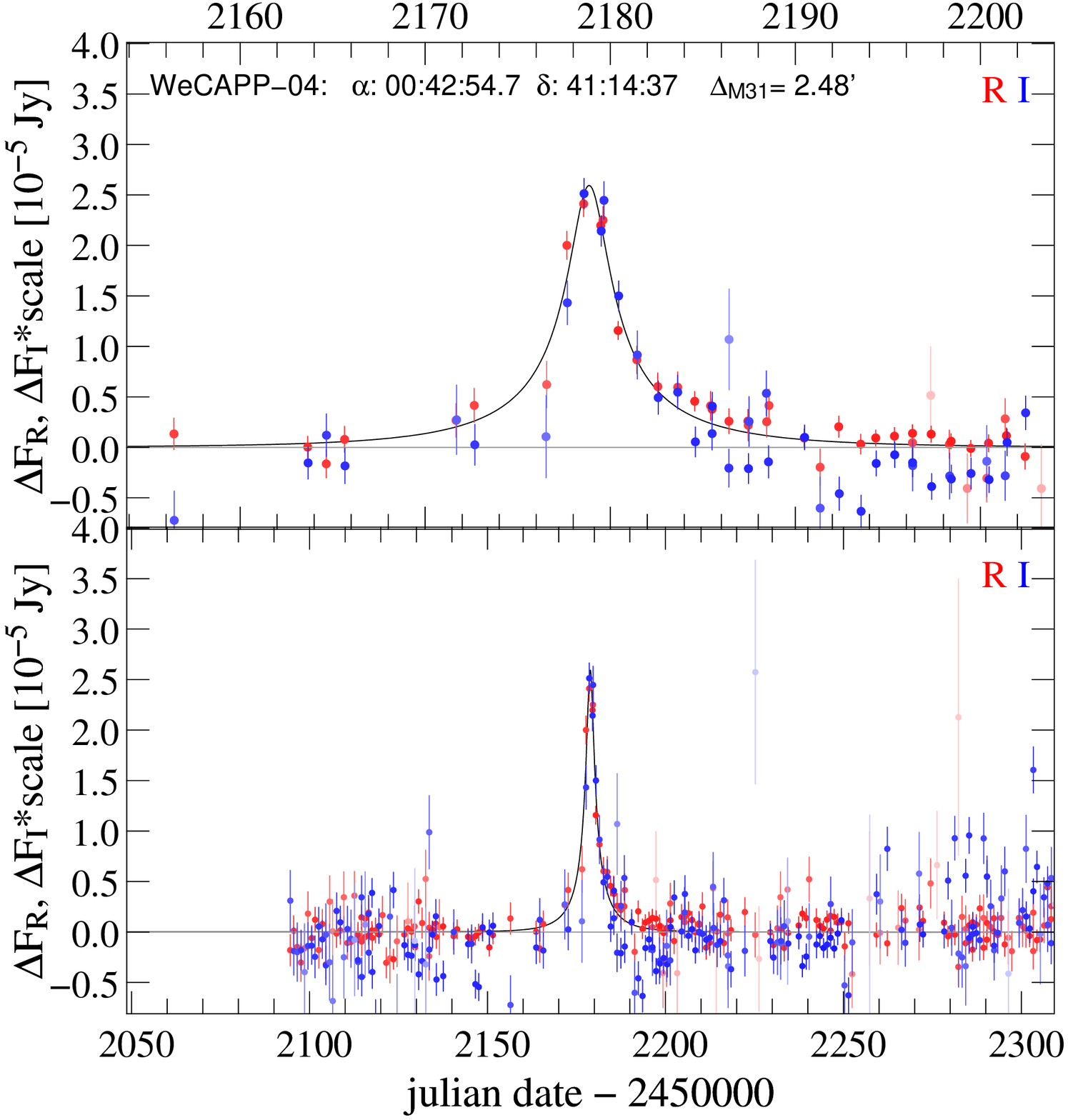}
    \includegraphics[width=8cm]{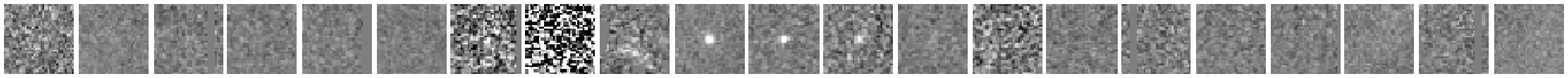}
    \includegraphics[width=8cm]{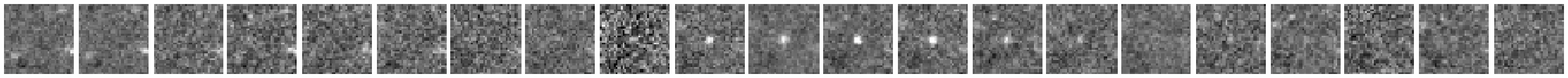}
    \includegraphics[width=8cm]{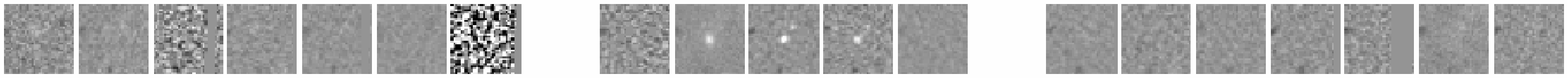}
    \includegraphics[width=8cm]{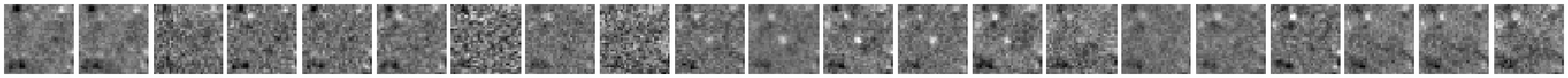}
  \caption{WeCAPP microlensing event light curves: WeCAPP 1-4 with
  corresponding cut-outs of the difference frames in R and I band.
  The data points are color-coded in grey-scale according to their errors; measurements 
  with larger errors in R (I) are shown in light-red (light-blue), while measurements with smaller errors are shown in dark-red (dark-blue).}
  \label{fig.event_1-4}
\end{figure*}

\begin{figure*}[!ht]
  \centering
    \includegraphics[width=8cm]{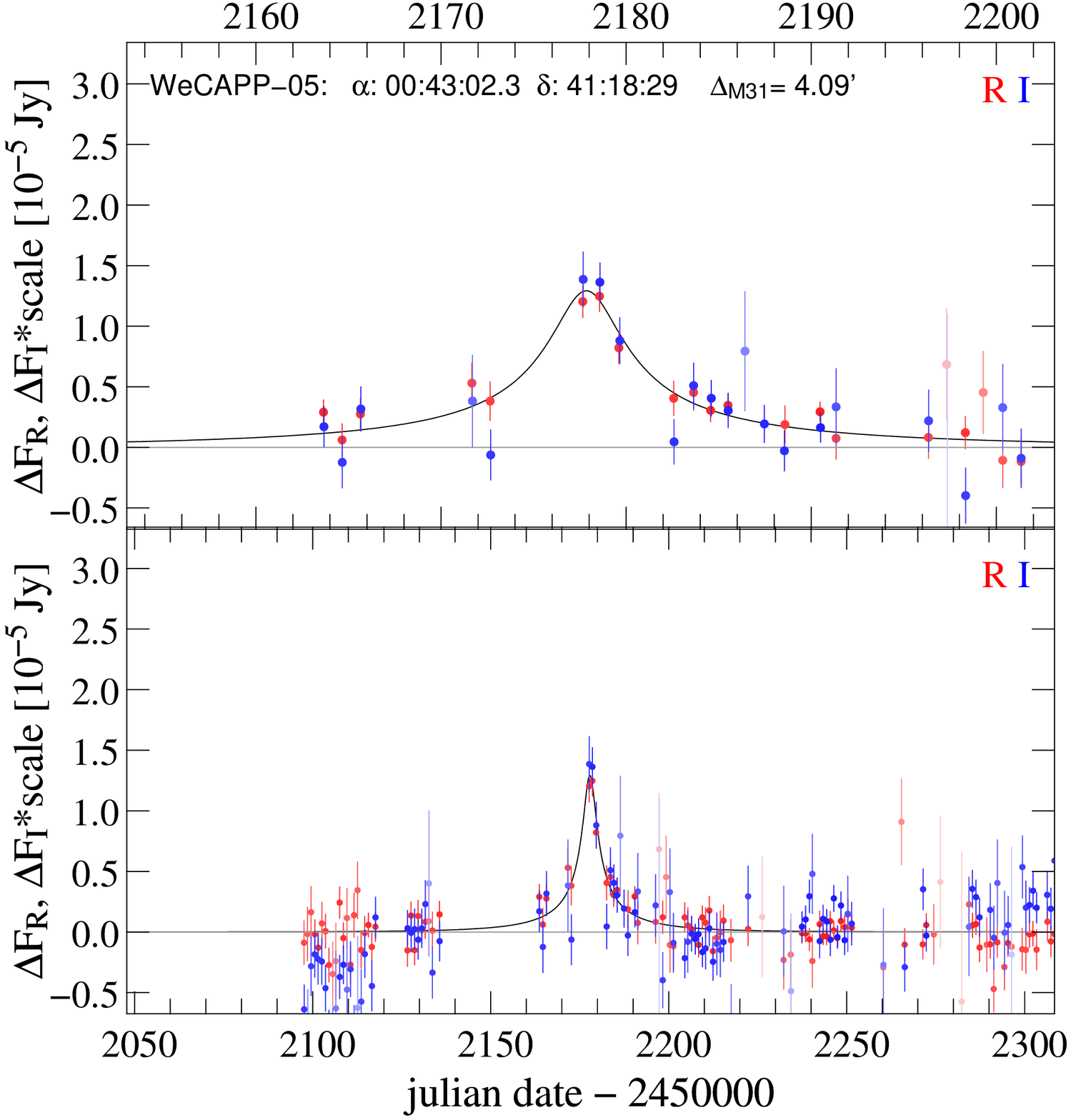}
    \includegraphics[width=8cm]{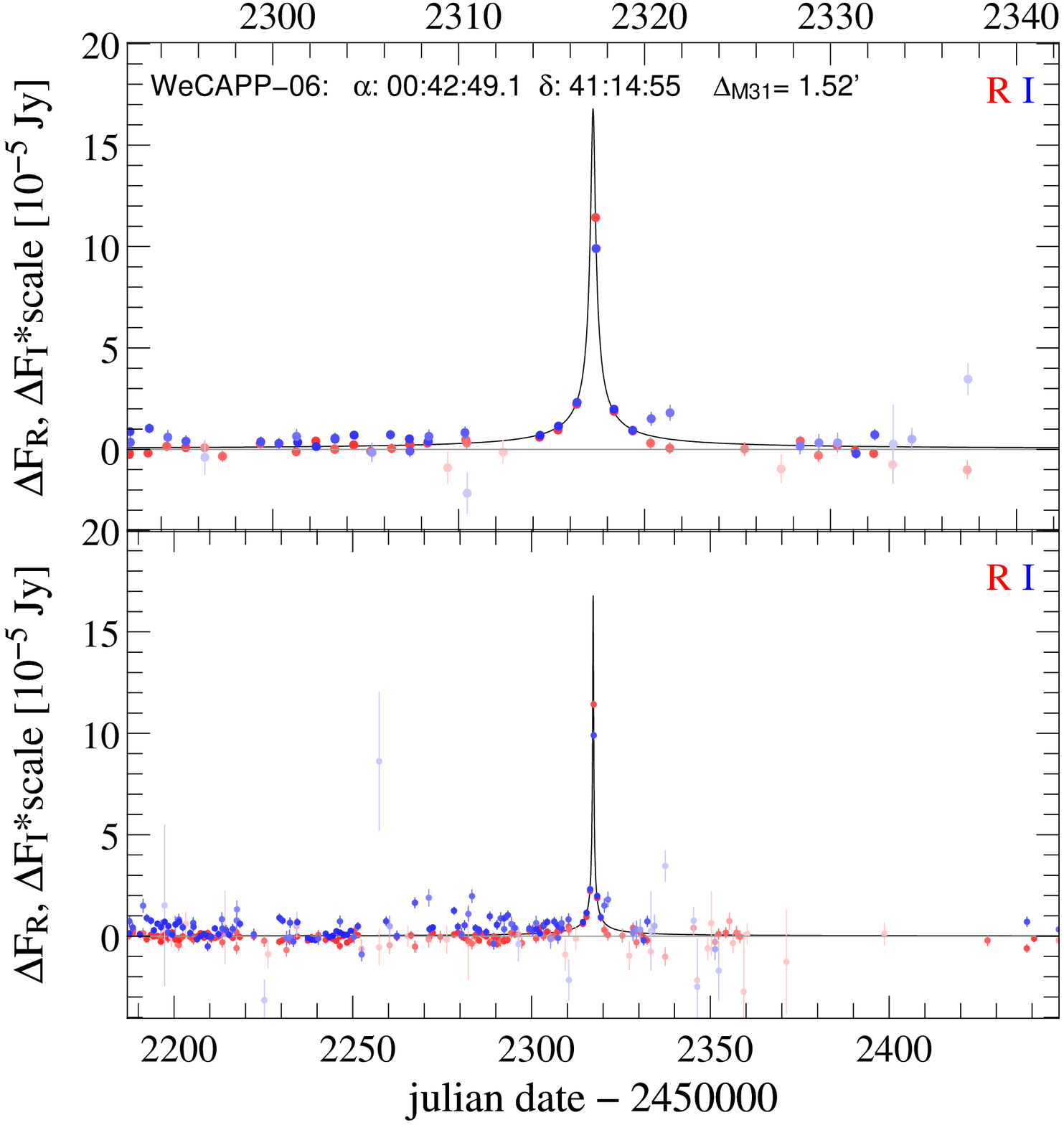}
    \includegraphics[width=8cm]{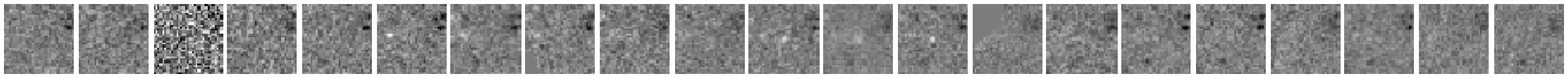}
    \includegraphics[width=8cm]{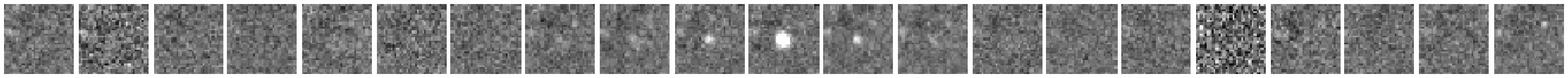}
    \includegraphics[width=8cm]{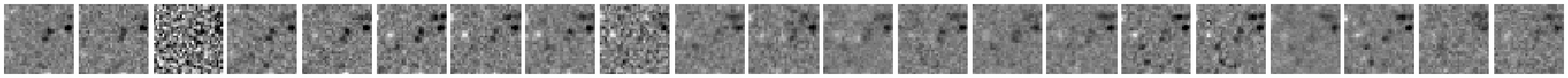}
    \includegraphics[width=8cm]{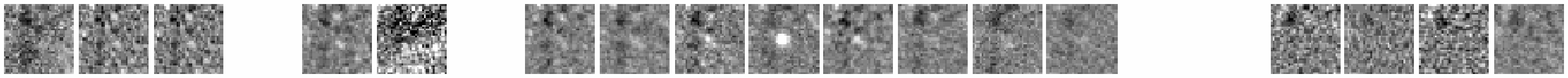}
    \includegraphics[width=8cm]{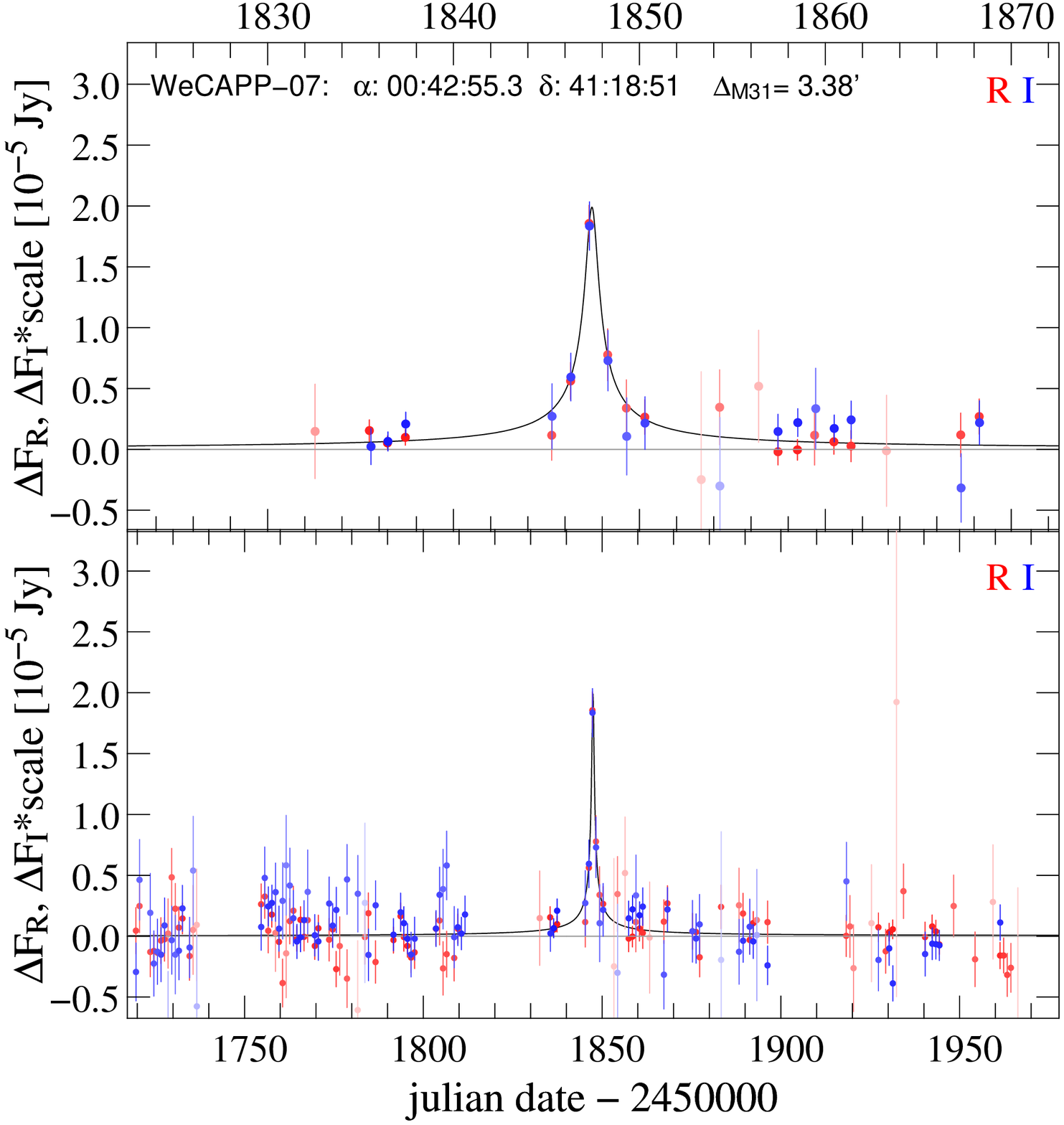}
    \includegraphics[width=8cm]{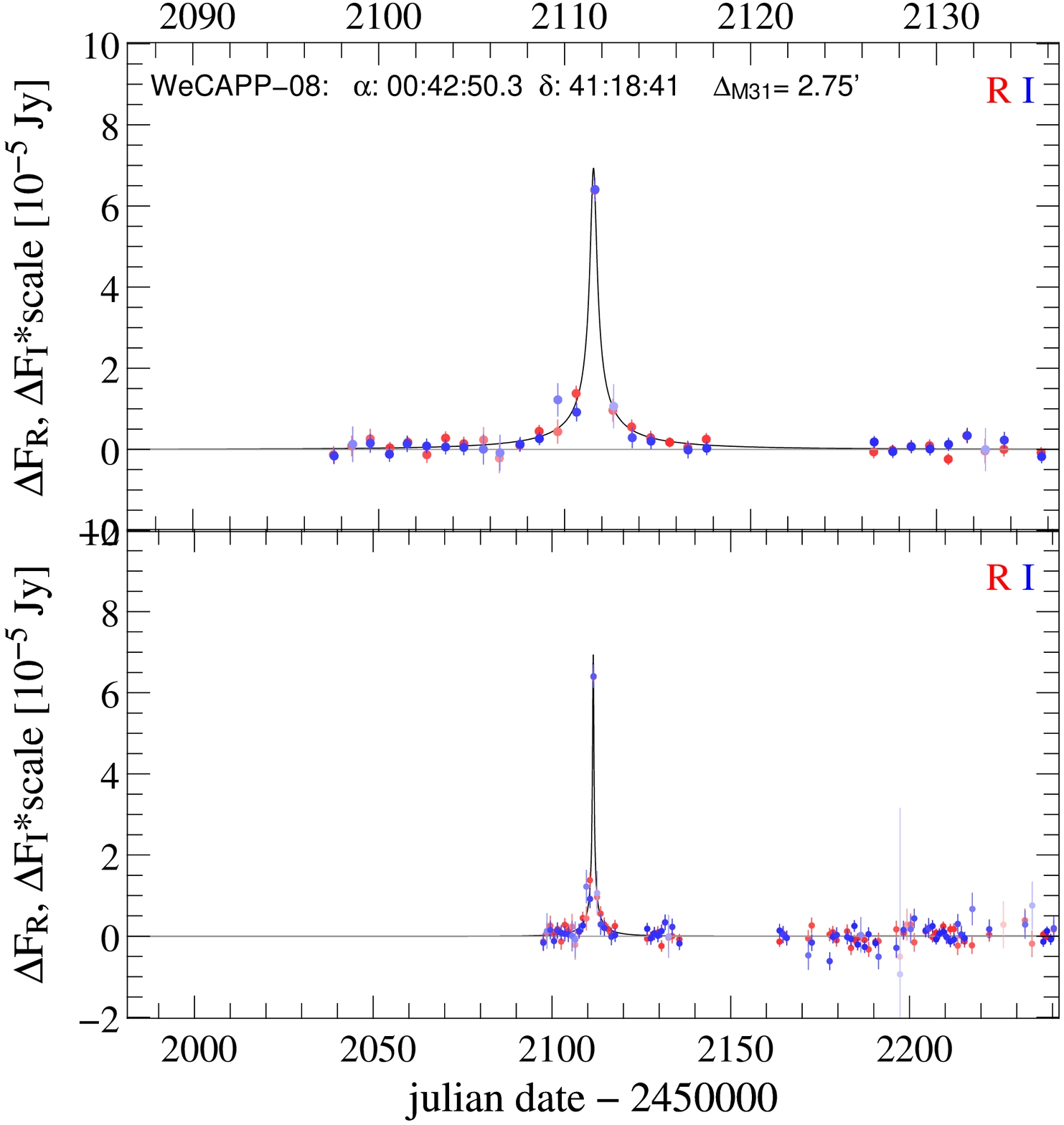}
    \includegraphics[width=8cm]{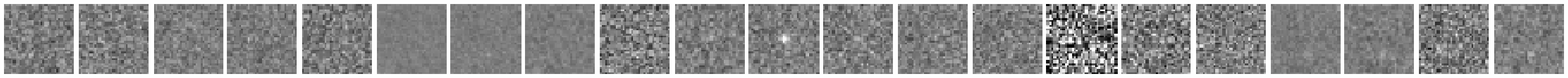}
    \includegraphics[width=8cm]{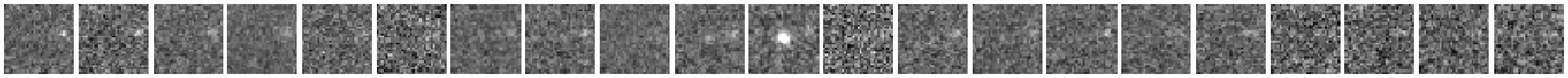}
    \includegraphics[width=8cm]{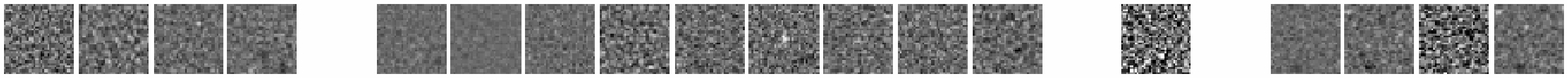}
    \includegraphics[width=8cm]{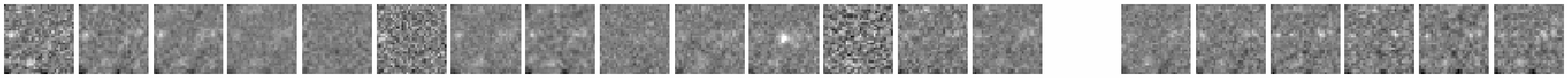}
  \caption{WeCAPP microlensing event light curves: WeCAPP 5-8 with
  corresponding cut-outs of the difference frames in R and I band.}
  \label{fig.event_5-8}
\end{figure*}

\begin{figure*}[!ht]
  \centering
    \includegraphics[width=8cm]{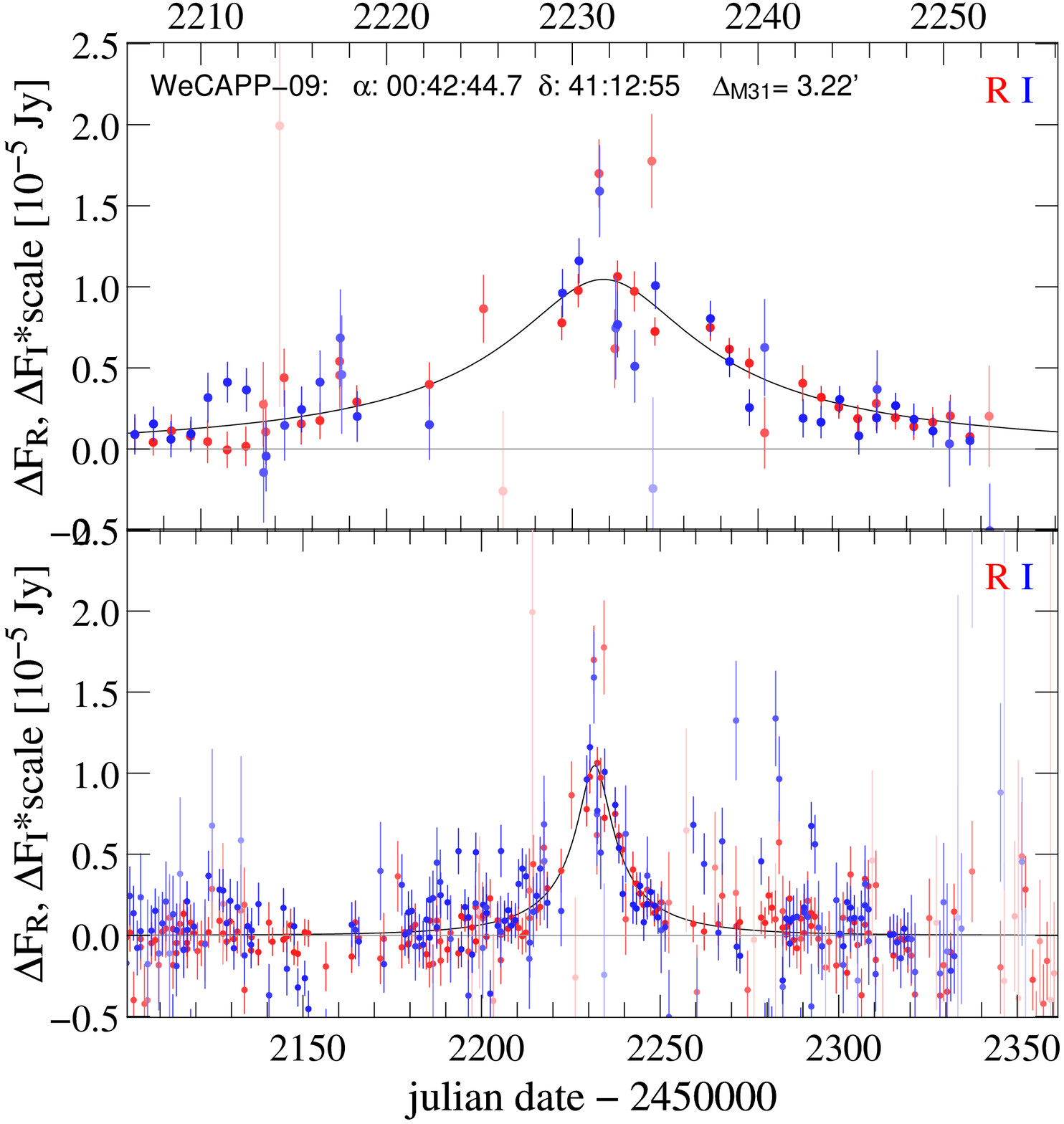}
    \includegraphics[width=8cm]{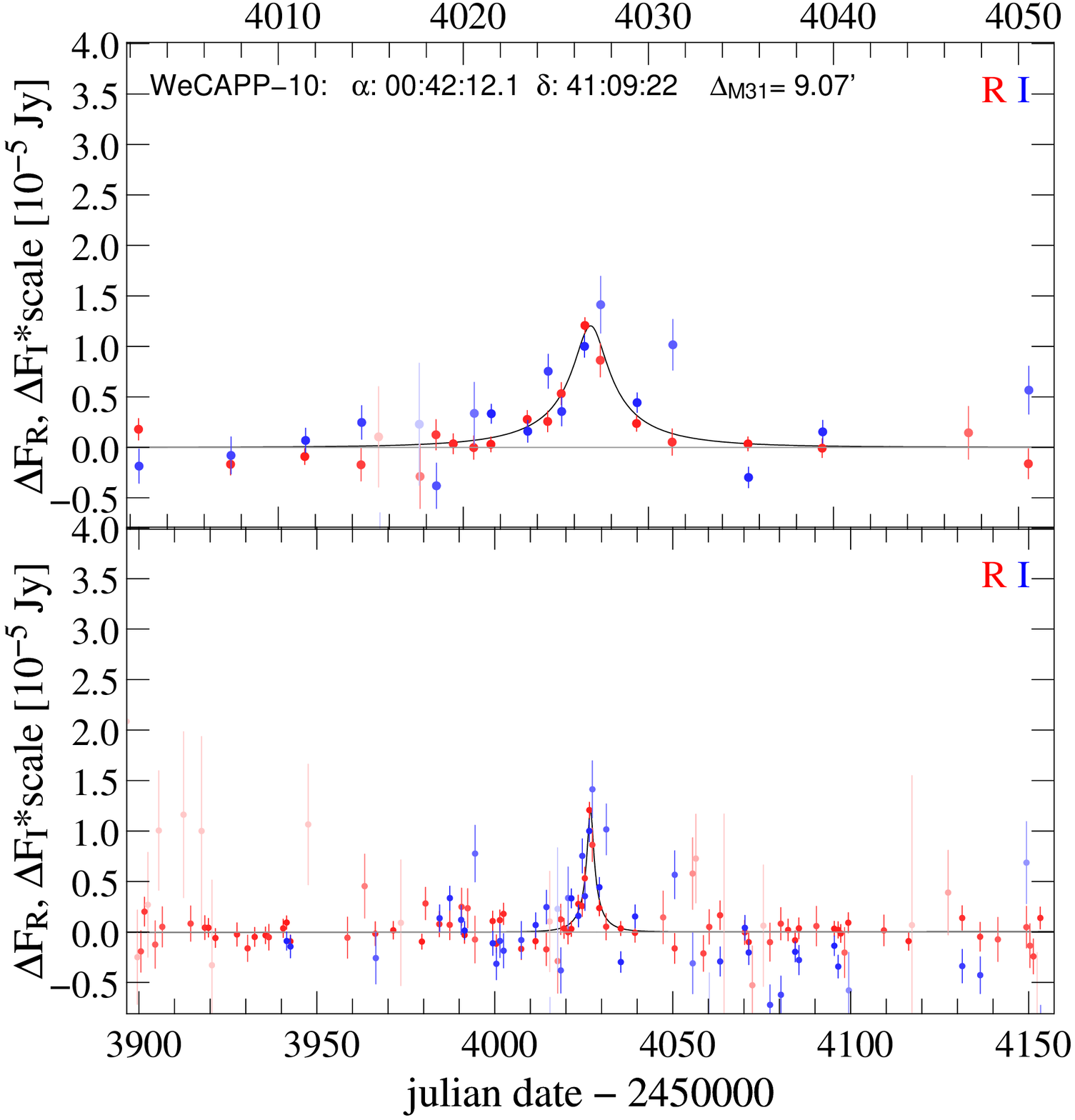}
    \includegraphics[width=8cm]{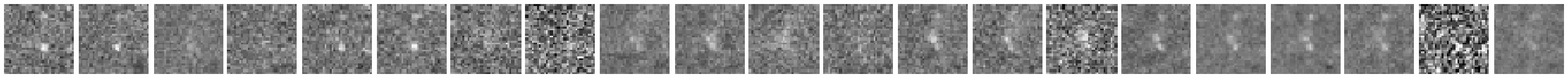}
    \includegraphics[width=8cm]{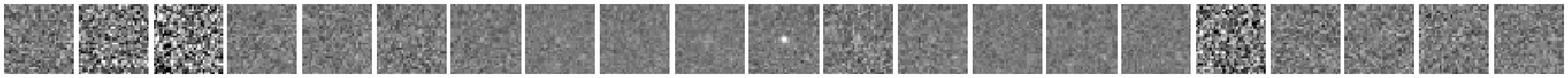}
    \includegraphics[width=8cm]{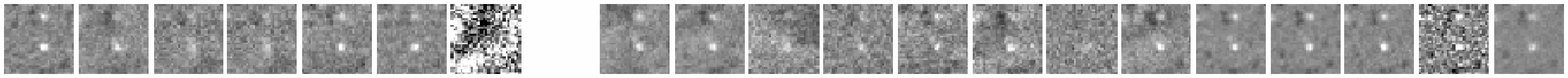}
    \includegraphics[width=8cm]{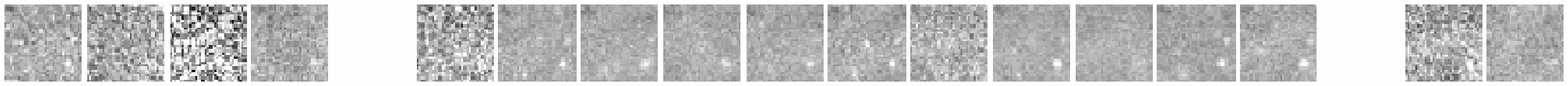}
    \includegraphics[width=8cm]{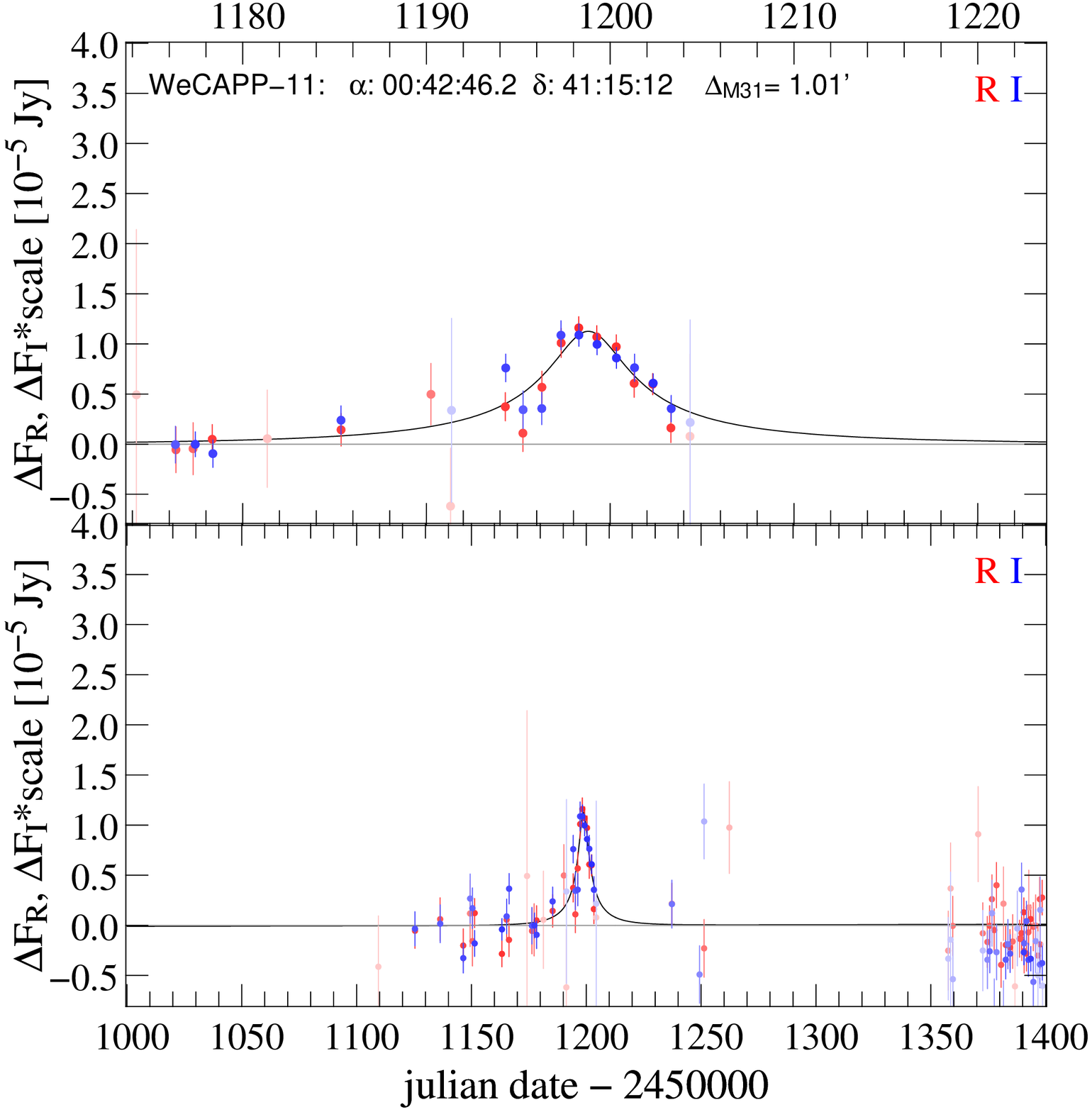}
    \includegraphics[width=8cm]{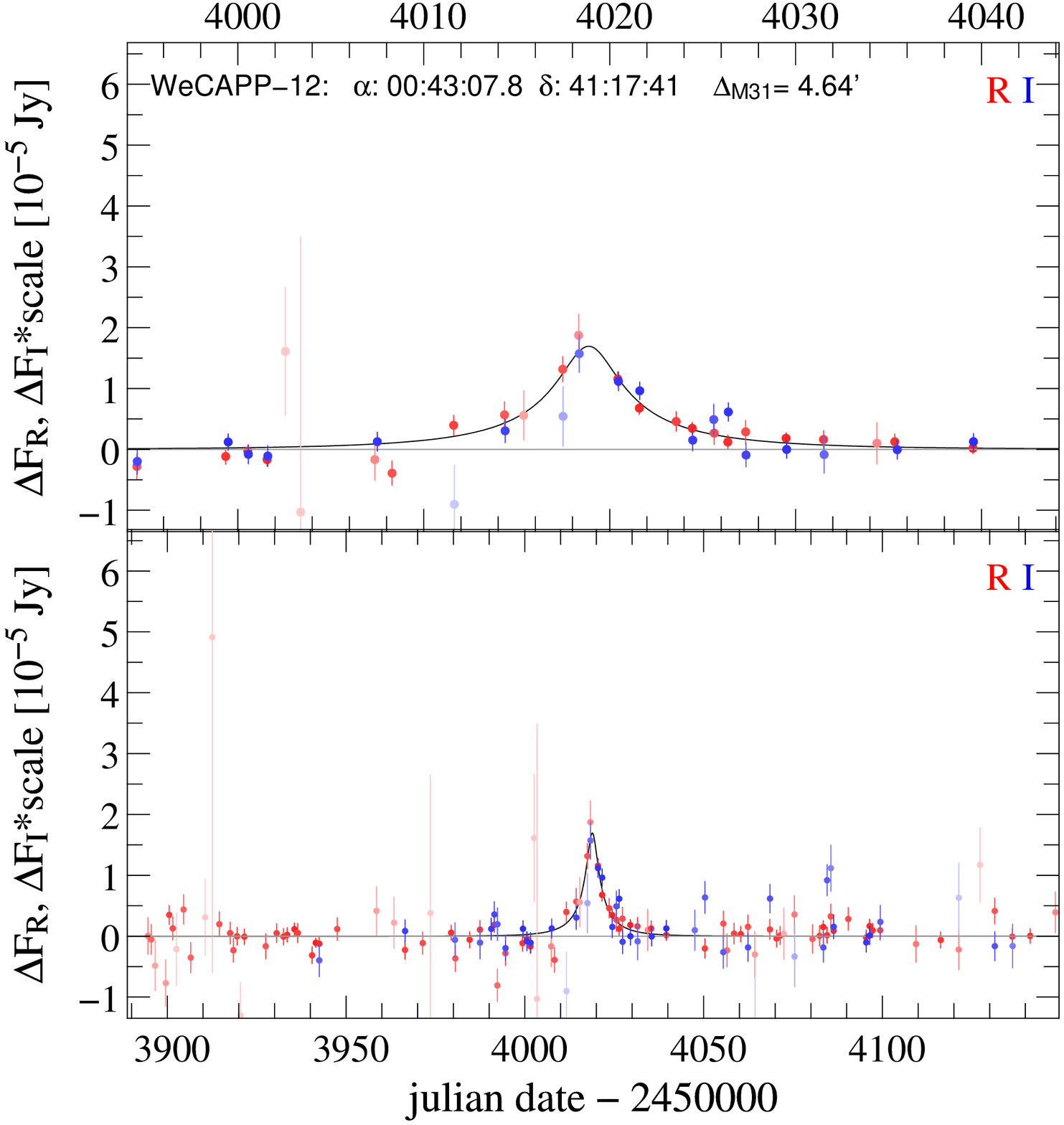}
    \includegraphics[width=8cm]{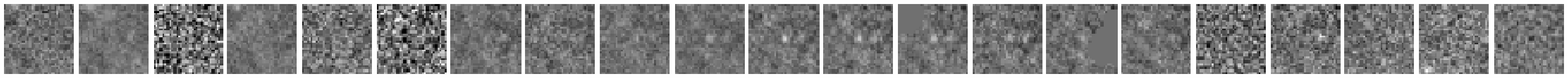}
    \includegraphics[width=8cm]{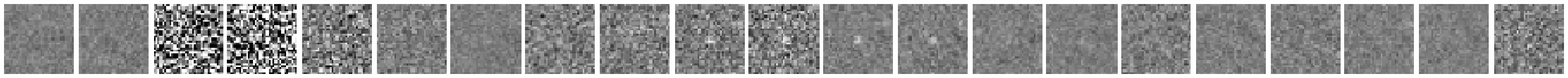}
    \includegraphics[width=8cm]{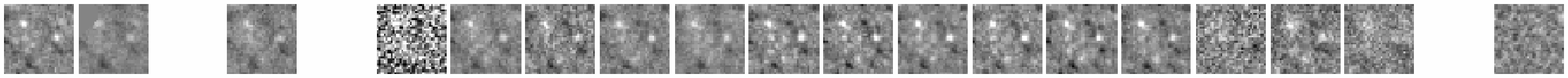}
    \includegraphics[width=8cm]{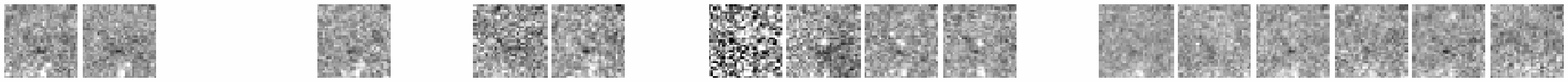}
  \caption{WeCAPP microlensing event light curves: WeCAPP 9-12 with
  corresponding cut-outs of the difference frames in R and I band.}
  \label{fig.event_9-12}
\end{figure*}

The light curves of the 12 WeCAPP
events are shown in Figs \ref{fig.event_1-4}-\ref{fig.event_9-12}. 
To illustrate the microlensing nature we also present 
postage stamps from the difference images close to the light curve 
maximum. These images help to rule out artifacts overseen by the pipeline, like hot pixels or cosmic 
rays as origin for the events. Indeed, as the inspection of the postage stamps show, 
none of our events is due to such an overseen artifact. In some
cases (W2, W5, W7) the postage stamps also show positions of nearby variables.
Because of improved photometric methods \citep[see][]{2006ApJS..163..225R}, the light curves for 
W1 and W2 slightly differ from the published ones in \cite{2003ApJ...599L..17R}, 
but agree within the error bars.

\section{Discussion} 
\label{sec.dis} 

\begin{table*}[!ht]
  \centering
  \begin{tabular}{lccccl}
    project           & events  & multi. & label in  & & recent \\ 
                      &         & detect. & Fig~\ref{fig.events_tfwhm_distribution},\ref{fig.events_positions} & & citation \\
    \hline\hline                         
    VATT/Columbia     & 6       &           &  VC     & {\it violet } & {\cite{1996ApJ...473L..87C}} \\ 
    AGAPE             & 1       &           &  Z      & {\it gray   } & {\cite{1999A&A...344L..49A}} \\ 
    POINT-AGAPE       & 1       &           &  N1     & {\it blue   } & {\cite{2001ApJ...553L.137A}} \\ 
    POINT-AGAPE       & 1       &           &  S4     & {\it blue   } & {\cite{2002ApJ...576L.121P}} \\ 
    POINT-AGAPE       & 2       & 2         &  N, S   & {\it blue   } & {\cite{2003A&A...405...15P}} \\ 
    WeCAPP            & 1       & 1         &  W      & {\it red    } & {\cite{2003ApJ...599L..17R}} \\ 
    POINT-AGAPE (MDM) & 3       &           &  C      & {\it brown  } & {\cite{2003A&A...405..851C}} \\
    VATT/Columbia     & 4       &           &  VC     & {\it black  } & {\cite{2004ApJ...612..877U}} \\
    MEGA              & 8       & 2         &  ML     & {\it green  } & {\cite{2004A&A...417..461D}} \\
    POINT-AGAPE       &         & 1         &  N2     & {\it blue   } & {\cite{2004ApJ...601..845A}} \\
    POINT-AGAPE       & 3       & 4         &  N, S   & {\it blue   } & {\cite{2005A&A...443..911C}} \\ 
    POINT-AGAPE       & 4       & 2         &  L1, L2 & {\it cyan   } & {\cite{2005MNRAS.357...17B}} \\ 
    Nainital          & 1       &           &  NMS    & {\it magenta} & {\cite{2005A&A...433..787J}} \\
    MEGA              &         & 5         &  ML     & {\it green  } & {\cite{2005ApJ...633L.105C}} \\
    MEGA              & 3       & 11        &  ML     & {\it green  } & {\cite{2006A&A...446..855D}} \\ 
    WeCAPP            &         & 1         &  W      & {\it red    } & {\cite{2008ApJ...684.1093R}} \\
    PLAN              & 2       &           &  OAB    & {\it yellow } & {\cite{2009ApJ...695..442C}} \\ 
    PLAN              &         & 1         &  OAB    & {\it yellow } & {\cite{2010ApJ...717..987C}} \\
    PAndromeda        & 6       &           &  PAnd   & {\it orange } & {\cite{2012AJ....143...89L}} \\            
    PLAN              &         & 3         &  OAB    & {\it yellow } & {\cite{2014ApJ...783...86C}} \\
    WeCAPP            & 10      & 2         &  W      & {\it red    } & this work                    \\ 
    \hline\hline                
    total             & 56      &           &         &         &                              \\
  \end{tabular}
  \caption{Overview of the 56 microlensing events in M31 presented 
    in different papers. In the 3rd column for the total number 
    we counted re-published detection only once. 
    For comparison see Figure~\ref{fig.events_positions}.}
  \label{tab.events_overview}
\end{table*}

In this section we summarize M31 microlensing events from previous surveys we 
are aware of and we compare our 12 events with previous studies. 

\begin{figure*}
  \centering
  \includegraphics[width=0.5\textwidth]{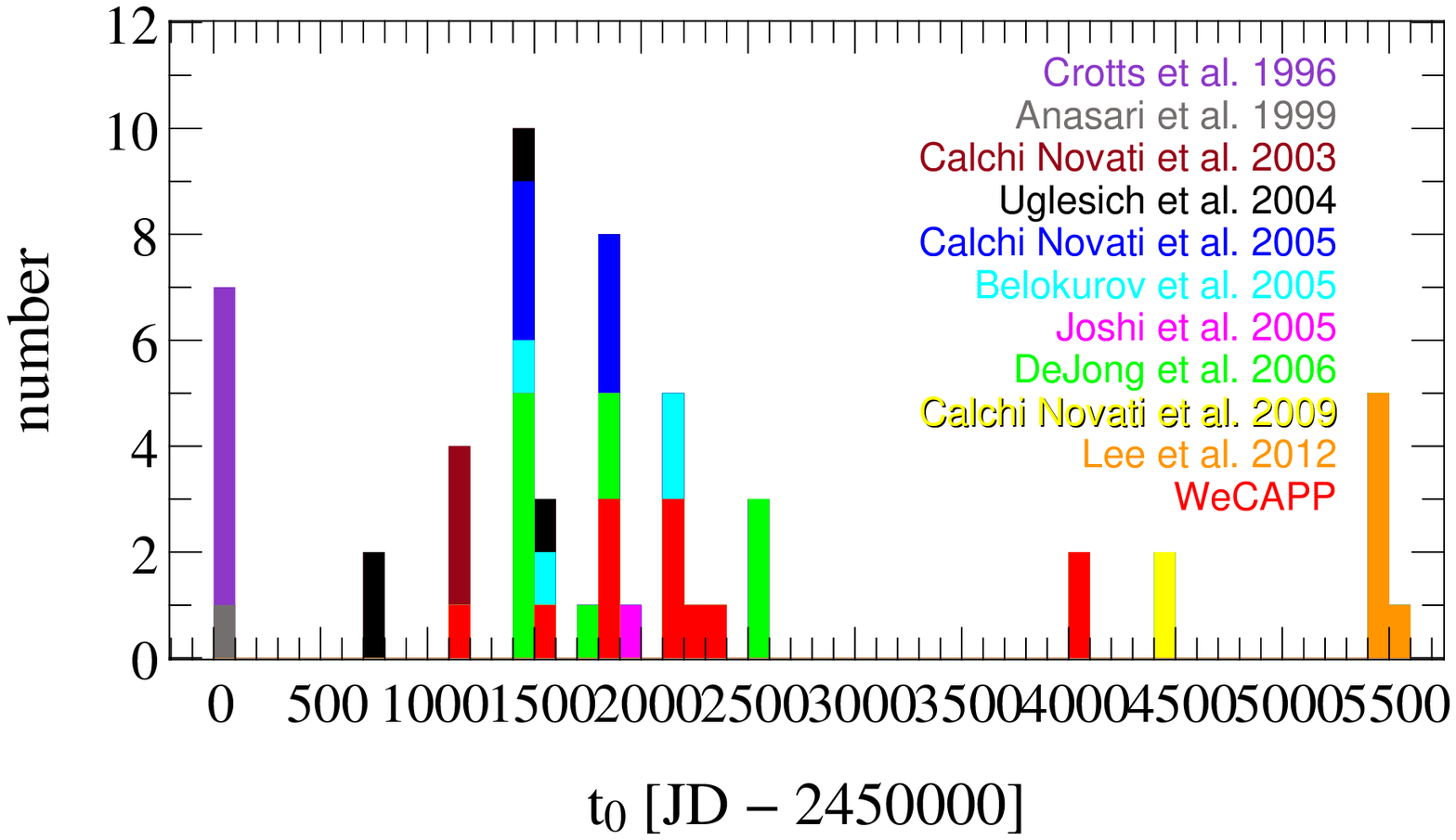}
  \includegraphics[width=0.5\textwidth]{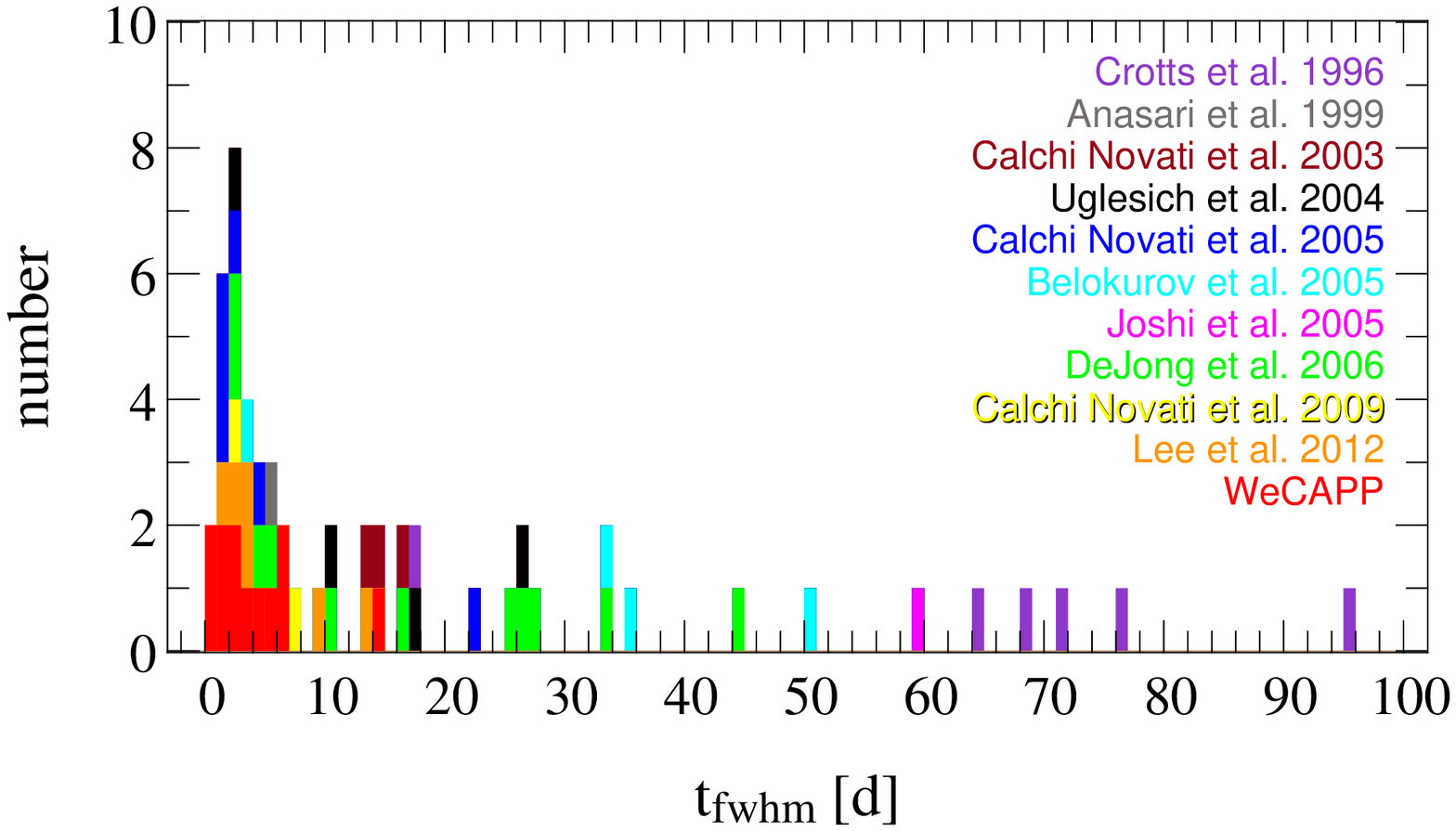}
  \includegraphics[width=0.5\textwidth]{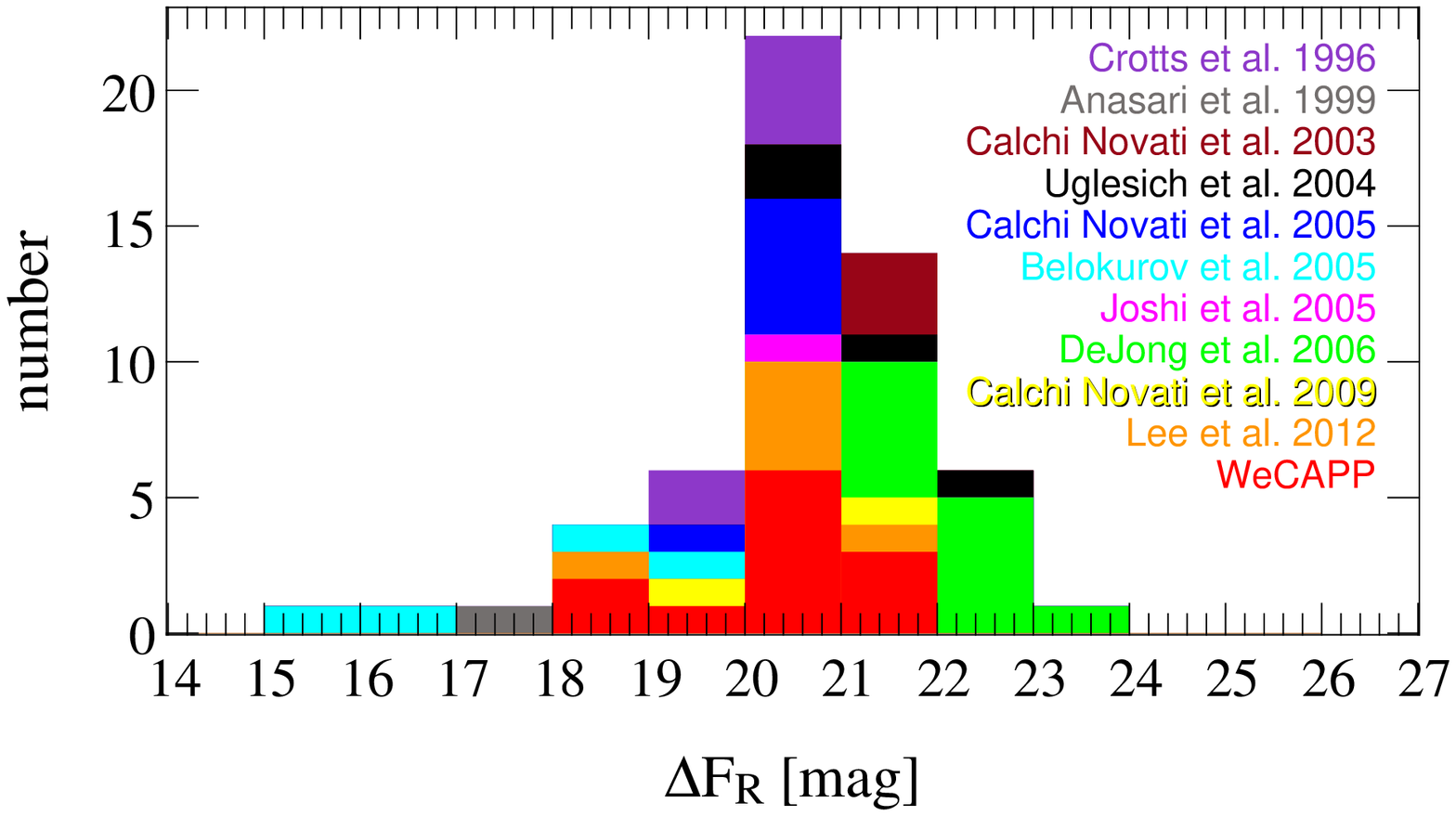}
  \caption{$t_0$, $\tfwhm$ and $\DF_R$ distribution of all microlensing events in M31
    reported up to now. Short events seem to be frequent. Around 20 days 
    there is a gap which separates short from long
    lensing events.  The WeCAPP events are plotted in {\it
      red}, VATT/Columbia from Crotts \& Tomaney (1996) in {\it violet}, POINT-AGAPE in {\it blue}, 
    Belokurov-AGAPE in {\it cyan},
    MEGA in {\it green}, AGAPE Z1 in {\it gray}, NMS in {\it magenta},
    PLAN in {\it yellow}, SLOTT-AGAPE in {\it brown}, VATT/Columbia from Uglesich et al. (2004) in {\it black}, PAndromeda in {\it orange}.  
    Note that the above distributions can not be directly 
    taken to study the nature of events (halo vs. self-lensing) 
    and to constrain the halo MACHO fraction, but instead have to 
    be corrected for the surveys' detection efficiencies.}
  \label{fig.events_tfwhm_distribution}
\end{figure*}

\begin{figure*}
  \centering
  \includegraphics[scale=0.9]{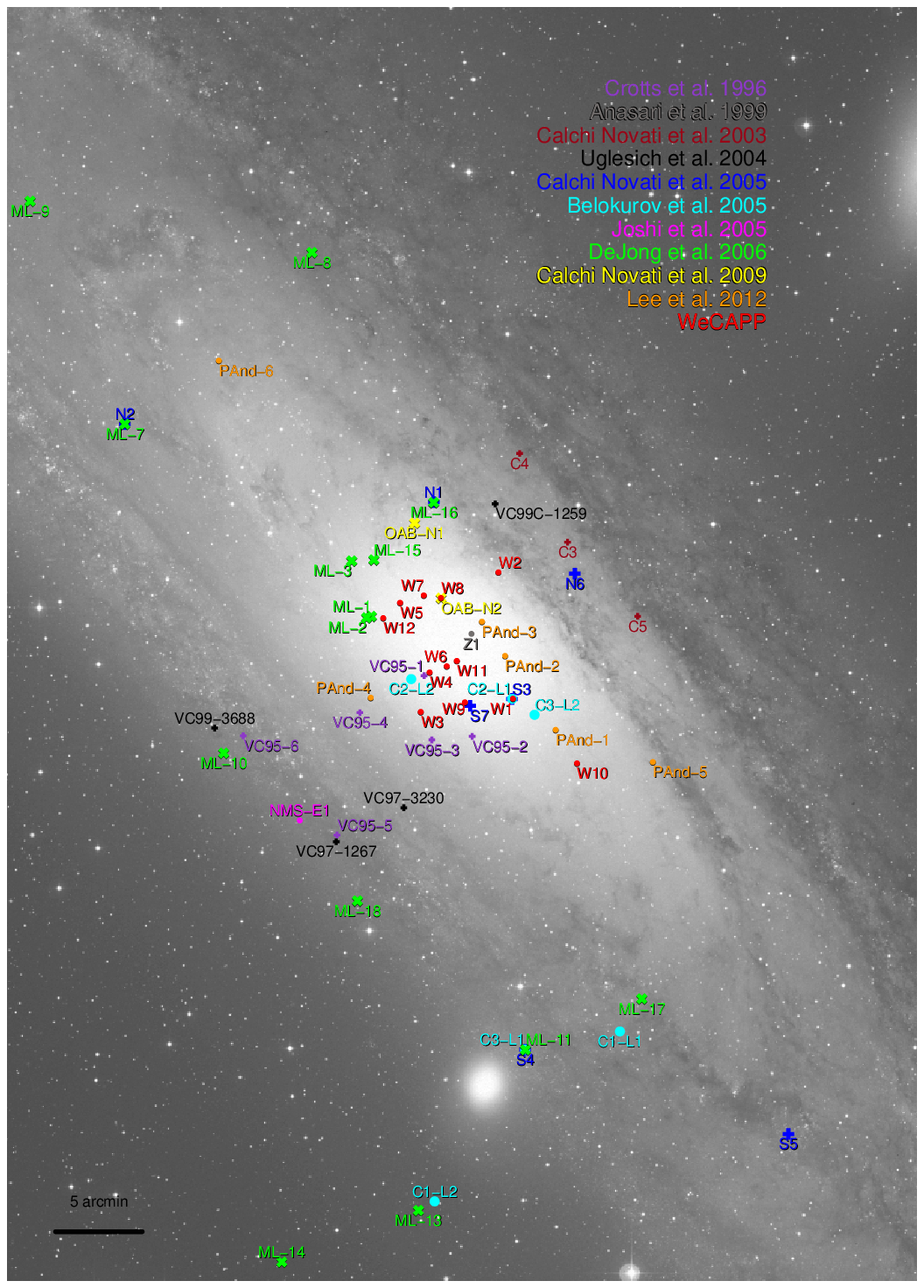}
  \caption{56 microlensing events in M31.  The image shows a wide
    field ($50 \times 70$ arcmin) from the $2^{nd}$ Palomar Sky Survey
    (ESO Online Digitized Sky Survey - DSS-2-blue).  The WeCAPP events
    are plotted in {\it red}, POINT-AGAPE events in {\it blue},
    Belokurov events in {\it cyan}, MEGA events in {\it green}, AGAPE Z1 
    event in {\it black}, NMS event in {\it magenta}, 
    PLAN events in {\it yellow}, PAndromeda events in {\it orange}, Slott-AGAPE in {\it brown}, VATT from Crotts \& Tomaney (1996) in {\it violet}, and VATT from Uglesich et al. (2004) in {\it black}. 
    We note that although our event W8 appears to be on top 
    of the OAB-N2 event in this figure, their $t_0$ differ by more than 6.5 years (W8 has a $t_0$ of 
    JD=2452112 and OAB-N2 has a $t_0$ of JD=2454467). In addition, from our difference images 
    their positions differ by 0.5 arcsec.} 
\label{fig.events_positions}
\end{figure*}

Table \ref{tab.events_overview} lists the project name, number of detected events,
their label and color as used in Figures \ref{fig.events_tfwhm_distribution} and 
\ref{fig.events_positions}, and the references for the events. 
In the 2$^{nd}$ column we report the number of events that were solely detected
by the corresponding survey. 
In the 3$^{rd}$ column we list the number of events that were detected by more 
than one groups. For example, Paulin-Henriksson et al. (2003) published 4 
events from the POINT-AGAPE project, where 2 of them were only detected by POINT-AGAPE, and 2 of them were also detected by other projects.
Altogether 56 different M31 microlensing events have been published to date (including those presented in this work).
Their $t_0$, $\tfwhm$, $\DF$ distributions are shown in Fig. \ref{fig.events_tfwhm_distribution}.
In Fig. \ref{fig.events_positions} we show the positions of microlensing events 
in a 50$\times$70 arcmin$^2$ field from the 2$^{nd}$ Palomar Sky Survey.

The first fact to note in the middle panel of Fig. \ref{fig.events_tfwhm_distribution} 
is that WeCAPP (red), POINT-AGAPE (blue), and PLAN (yellow) surveys
detected very short timescale events and hardly any events with $\tfwhm > $ 15 days. 
In contrast, the first events ever announced
by VATT/Columbia \citep[][marked in violet color in Fig. \ref{fig.events_tfwhm_distribution}]{1996ApJ...473L..87C} 
have very long timescales (5 of 6 events have $\tfwhm >$ 65 days). 
In the same paper \cite{1996ApJ...473L..87C} suspected that part of their events may be caused by long period 
variables. The MEGA (green) and Belokurov L2 (cyan) events show an almost flat timescale distribution. 
Five of the MEGA events are (in de-projection) located in the 
disk, almost at the same distance with respect to the M31 center. This makes them more suspicious,
since they could be intrinsic varying disk stars that exist in a particular evolutionary time scale,
that is overrepresented at this particular radius within the disk.

Predicting event rates and their characteristics
from the theoretical studies, Riffeser et al. (2008) from their M31 microlensing model show 
the expected event rate of long timescale
events ($\tfwhm > $ 30 days) is an order of magnitude smaller than the event rate of 
short timescale events ($\tfwhm < $ 10 days), see e.g. their Fig. 12. The fact that the 
vast majority of the events presented in this paper are short timescale events is therefore in good agreement with
expectations. In contrast, VATT/Columbin and MEGA surveys show an
almost flat timescale distribution. 
Unless surveys analyzed by MEGA and Belokurov et al. (2005)
have detection efficiencies that strongly suppress events with 
timescales $<$ 10 days, they should have detected more short timescale
microlensing.
An alternative and more plausible explanation is that a large fraction 
of these long events are indeed long period variables, but were not ruled 
out from the microlensing candidate light curves due to the relative short 
time-span of the survey. 

The VATT/Columbia team uses their observations in the 1994-1995 season to 
rule out objects of masses in the range of 0.003-0.08$\Msun$ as the primary 
constituents of the mass of M31 \citep{1996ApJ...473L..87C}. 
\cite{2004ApJ...612..877U} find 
4 probable microlensing events with data collected between 1997 and 1999.  
They conclude that 29$^{+30}_{-13}$\% of the halo masses are composed by 
MACHOs assuming a nearly singular isothermal sphere model. They also
provide a poorly constrained lensing component mass (0.02-1.5 $\Msun$ at 1$\sigma$ limits).

\cite{2005A&A...443..911C} present an analysis of the POINT-AGAPE data. 
They indicate that their microlensing event rate is much larger than 
self-lensing expectation alone. This leads to a lower limit of 
20\% of the halo mass in the form of MACHOs with mass between 0.5-1 $\Msun$ at 
the 95\% confidence level. The lower limit drops to 8\% for MACHOs
$\sim$ 0.01 $\Msun$.

On the contrary, the MEGA team \citep{2006A&A...446..855D} conclude 
that their 14 events are consistent with self-lensing prediction. This 
rules out a MACHO halo fraction larger than 30\% at the 95\% confidence 
level. However, recent studies by \cite{2006A&A...445..375I,2007A&A...462..895I}
show that, when compared with the timescales, maximum fluxes, and spatial 
distributions from Monte-Carlo simulations, the MEGA events cannot be fully 
explained by self-lensing alone.

The evidence for MACHOs can be inferred from individual events provided
good sampling of the light curves. For example, the WeCAPP-GL1/POINT-AGAPE-S3
 event was independently discovered by the POINT-AGAPE \citep{2003A&A...405...15P} 
and the WeCAPP \citep{2003ApJ...599L..17R} collaboration. Joint analysis of the 
light curves from these two collaborations has led to the conclusion that this
event is very unlikely to be a self-lensing event \citep{2008ApJ...684.1093R}.
This is because with a realistic model of the three-dimensional light distribution,
stellar population and extinction of M31, an self-lensing event with the parameters of 
WeCAPP-GL1/POINT-AGAPE-S3 is only expected to occur every 49 years. In contrast, a halo lensing
event (with 20\% of the M31 halo consists of 1 M$_\odot$ MACHOs) would occur every 10 years.
In addition, combining the data from PLAN and WeCAPP, \cite{2010ApJ...717..987C}
have revealed the finite-source effect in the event OAB-N2. 
\cite{2010ApJ...717..987C} then used the finite-source effect to determine 
the size of the Einstein ring radius, and combined the Einstein ring radius
with the time required for the lens to travel the Einstein ring radius ($t_\M{E}$)
to derive the lens proper motion.
Hence they were able to put a 
strong lower limit on the lens proper motion. 
Since the lens proper motion in the halo lensing scenario is different (and 
much larger) than the self-lensing scenario \citep[see e.g. Fig. 4 of][]{2010ApJ...717..987C}, 
their result thus favors the MACHO lensing
scenario over self-lensing for this event.

To conclude, some statistical studies and individual microlensing events
point to a non-negligible MACHO population, though the fraction in the halo 
mass remains uncertain. To pinpoint the MACHO mass fraction, a better 
understanding of the luminous population, i.e. self-lensing rate, 
is needed.

Determining the MACHO mass fraction requires precise detection efficiency
studies which is beyond the scope of this publication. We
have started such efficiency studies (Koppenhoefer et al., in preparation). 
By simulating
artificial events into the WeCAPP images we will be able to account for
true noise and systematics introduced by the processing and the detection
procedure. The results will enable us to derive a robust estimate of the
MACHO mass fraction using an improved M31 model for disk, bulge and halo.

\section{Summary and Outlook}
\label{sec.sum}

WeCAPP has monitored the bulge of M31 for 11 years, 
which is the longest time-span among M31 microlensing surveys.
We have established an automated selection pipeline and present the final 
12 microlensing events from WeCAPP. 
The brightest event of it \citep[see e.g.][]{2008ApJ...684.1093R} is hard to reconcile with self-lensing alone, hence hints 
at the existence of MACHOs in the halo of M31.
A similar case has been found by Calchi Novati et al. (2010).

To gain insights to the MACHO fraction, in-depth understanding of self-lensing is 
required, which is only possible with microlensing experiments that monitor both the 
bulge and the disk of M31 simultaneously. With the advent of Pan-STARRS 1, we have conducted 
a dedicated project monitoring M31 with a field-of-view of $\sim$ 7 degree$^2$. 
Preliminary results of six microlensing events 
in the bulge have been published in \cite{2012AJ....143...89L}, and a 
full analysis will follow soon (Seitz et al. in prep.).

\section{Acknowledgments}

We are grateful to the referee for the very useful comments.
We would like to thank Juergen Fliri, Christoph Ries, Otto Baernbantner, Claus Goessl, 
Jan Snigula and Sarah Buehler for their contributions in observation. 
This work was supported by the DFG cluster of excellence 'Origin and Structure of the Universe' (www.universe-cluster.de).

\section{Appendix}

\begin{figure*}[!th]
  \centering
  \includegraphics[width=3.5cm]{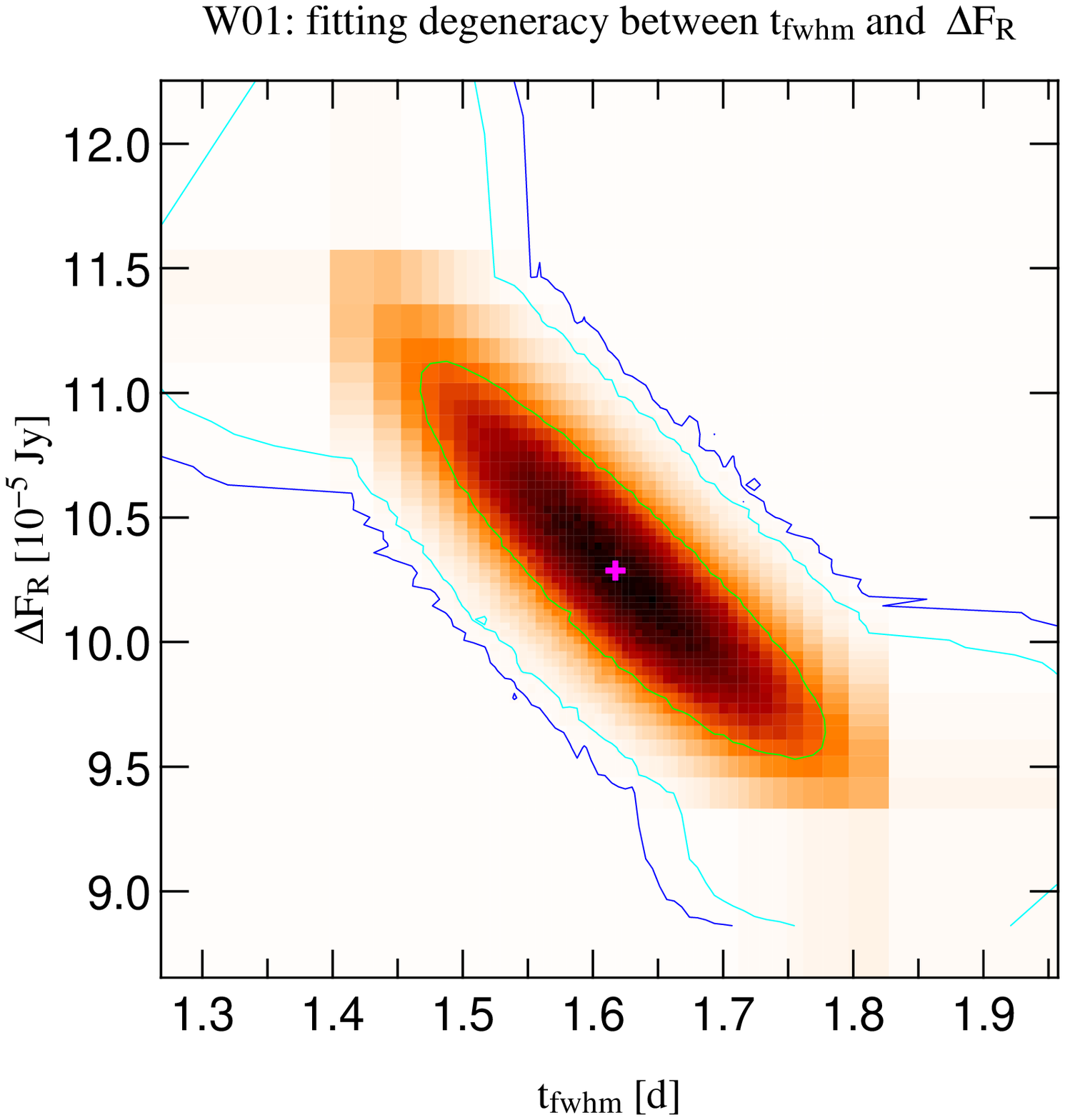}
  \includegraphics[width=3.5cm]{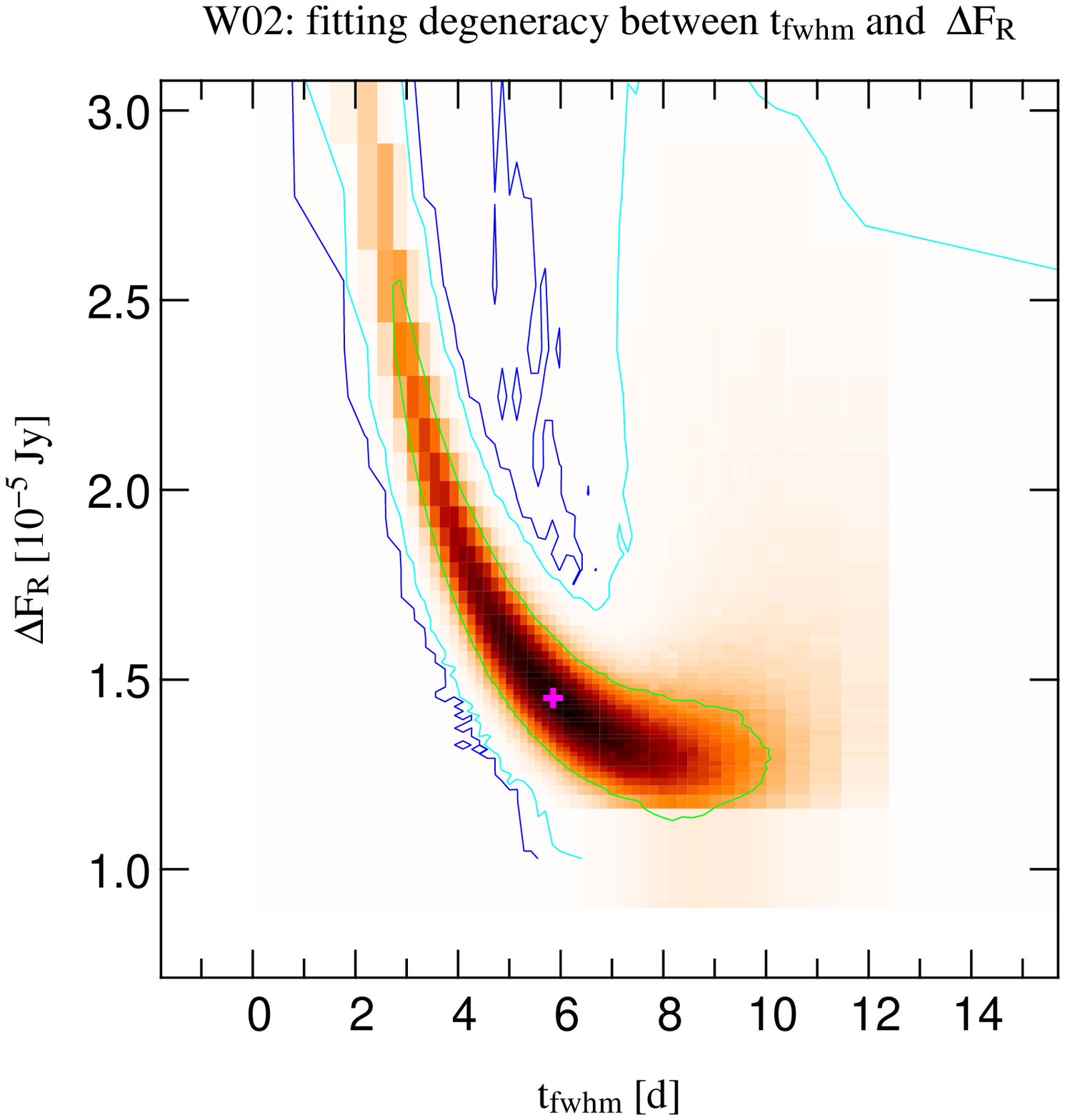}
  \includegraphics[width=3.5cm]{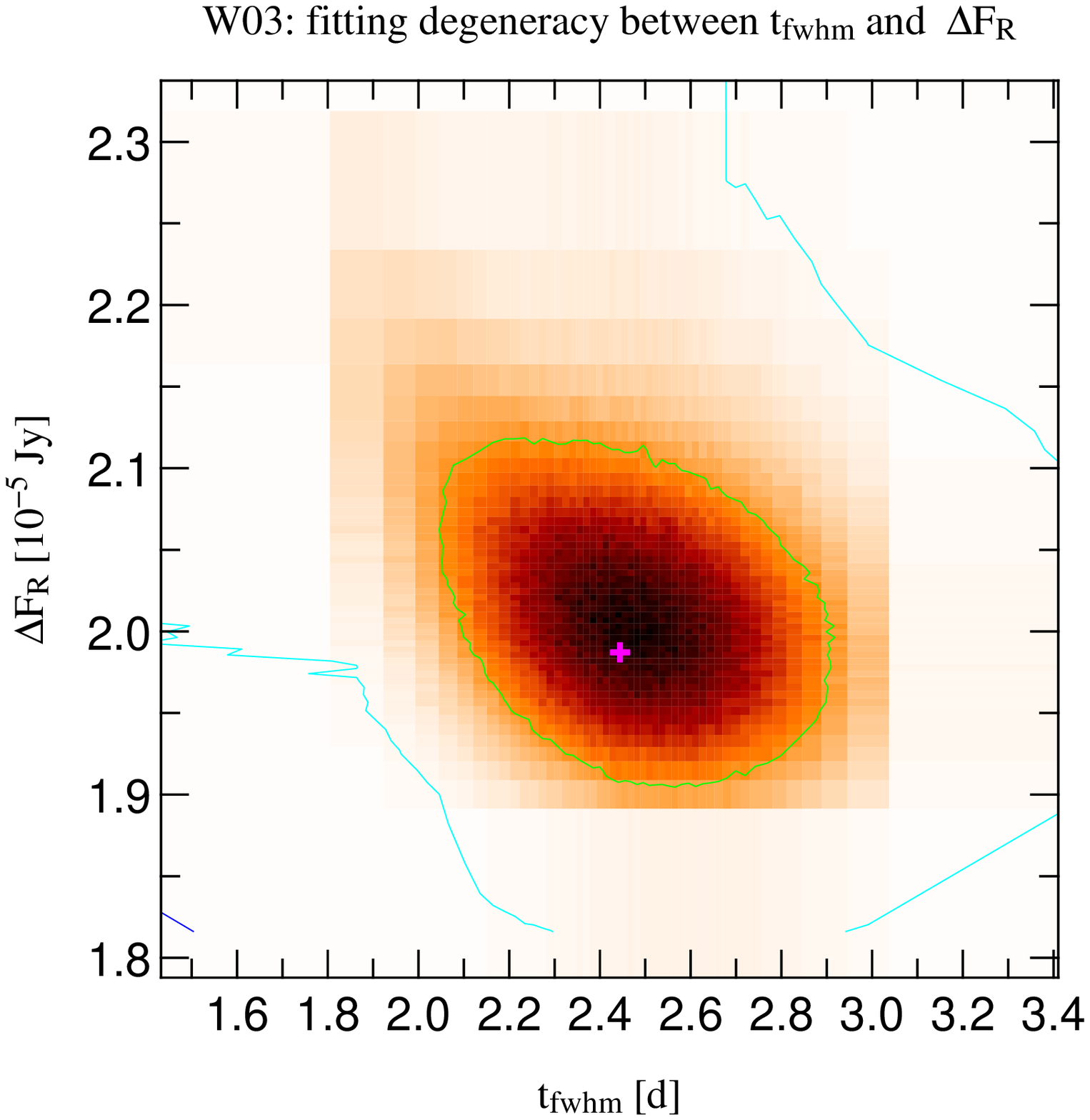}
  \includegraphics[width=3.5cm]{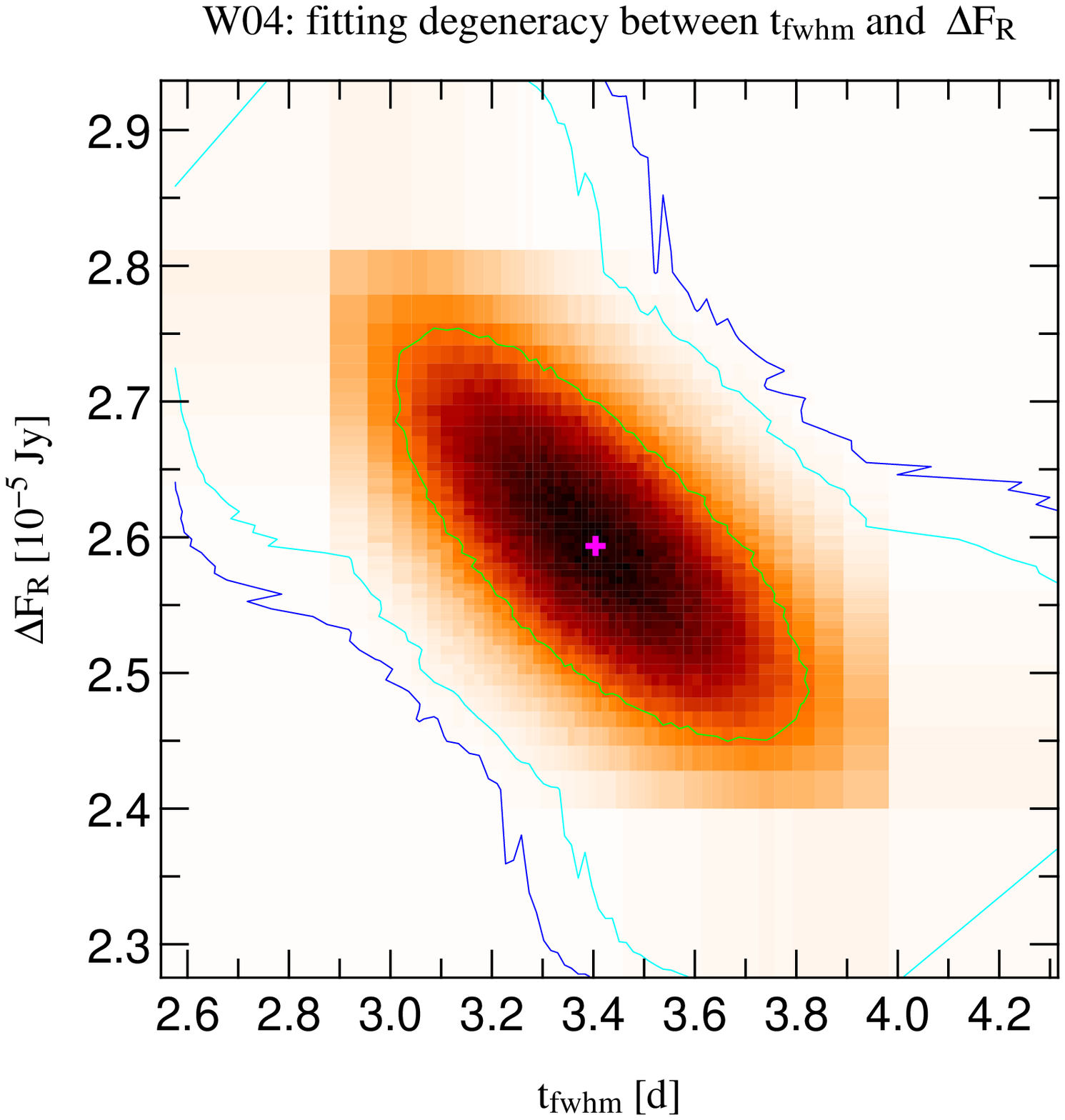}
  \includegraphics[width=3.5cm]{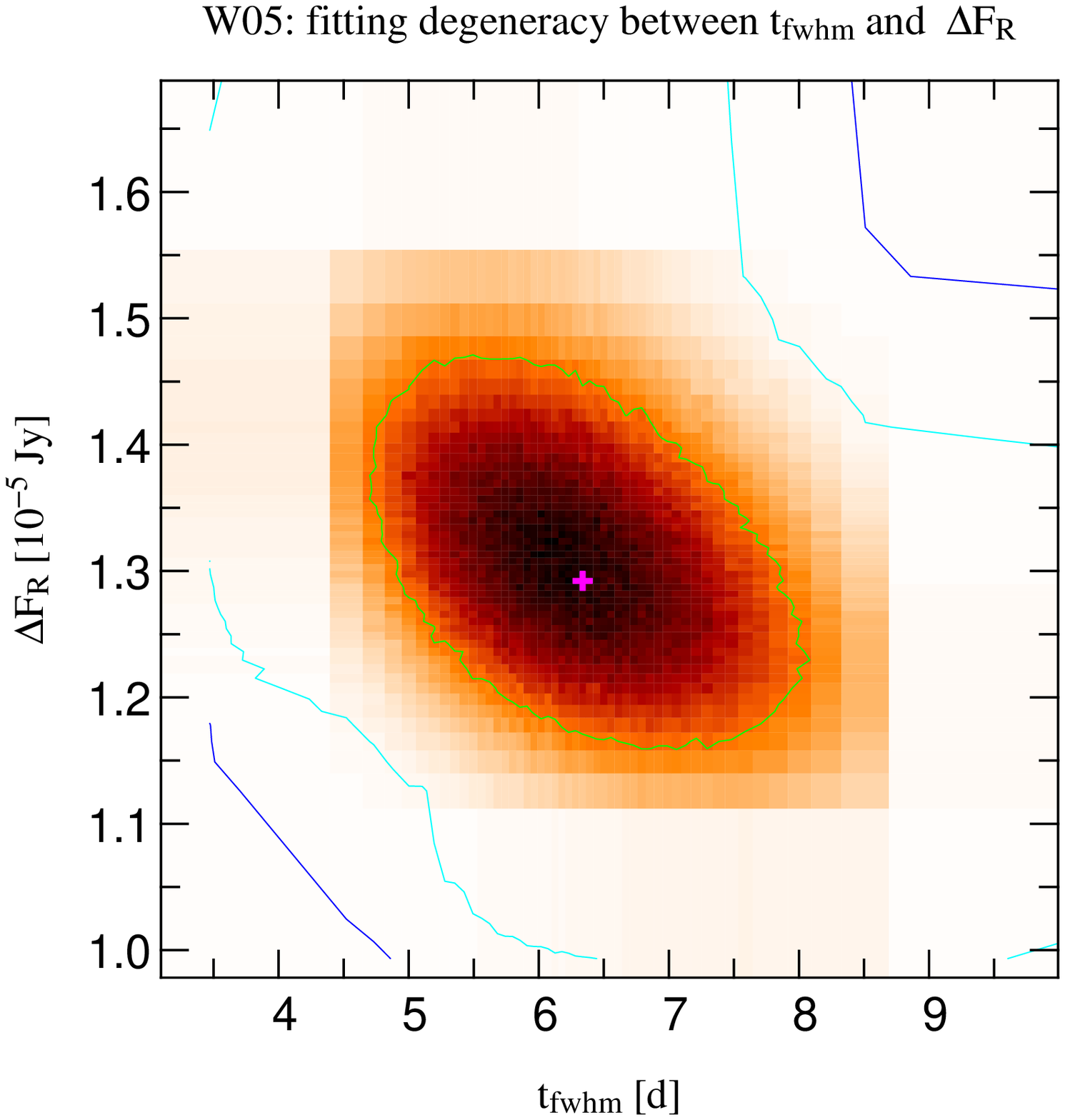}
  \includegraphics[width=3.5cm]{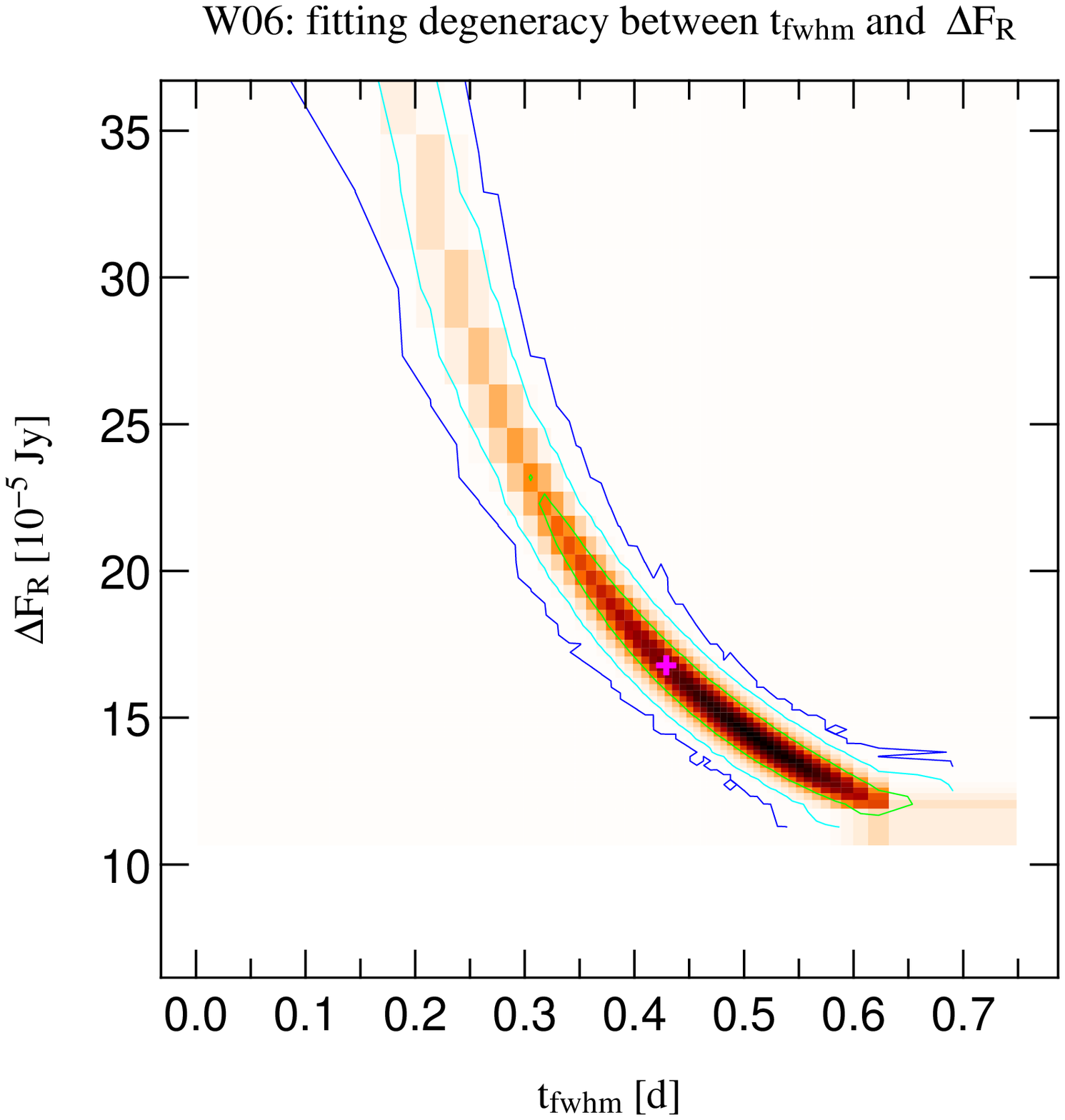}
  \includegraphics[width=3.5cm]{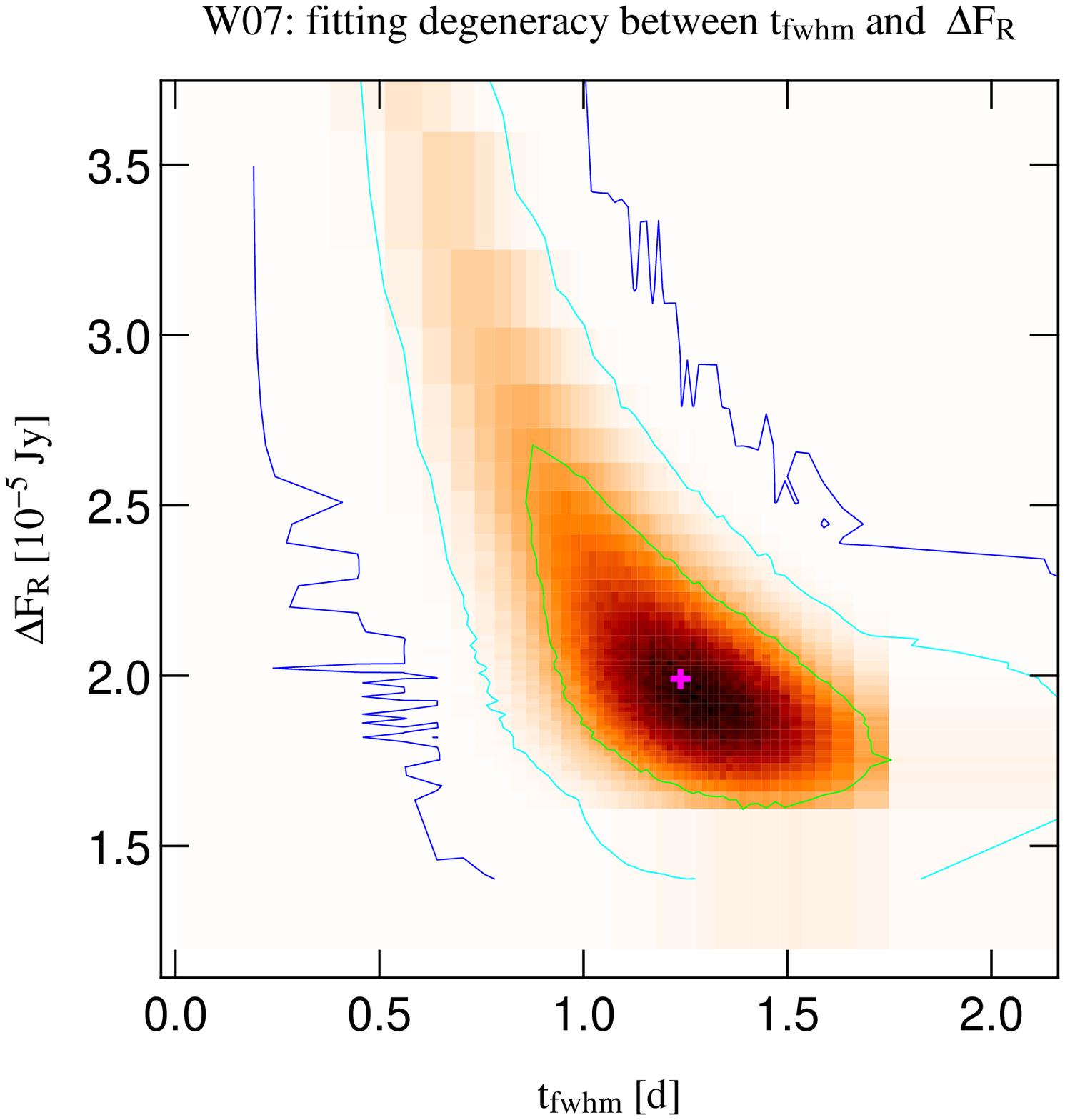}
  \includegraphics[width=3.5cm]{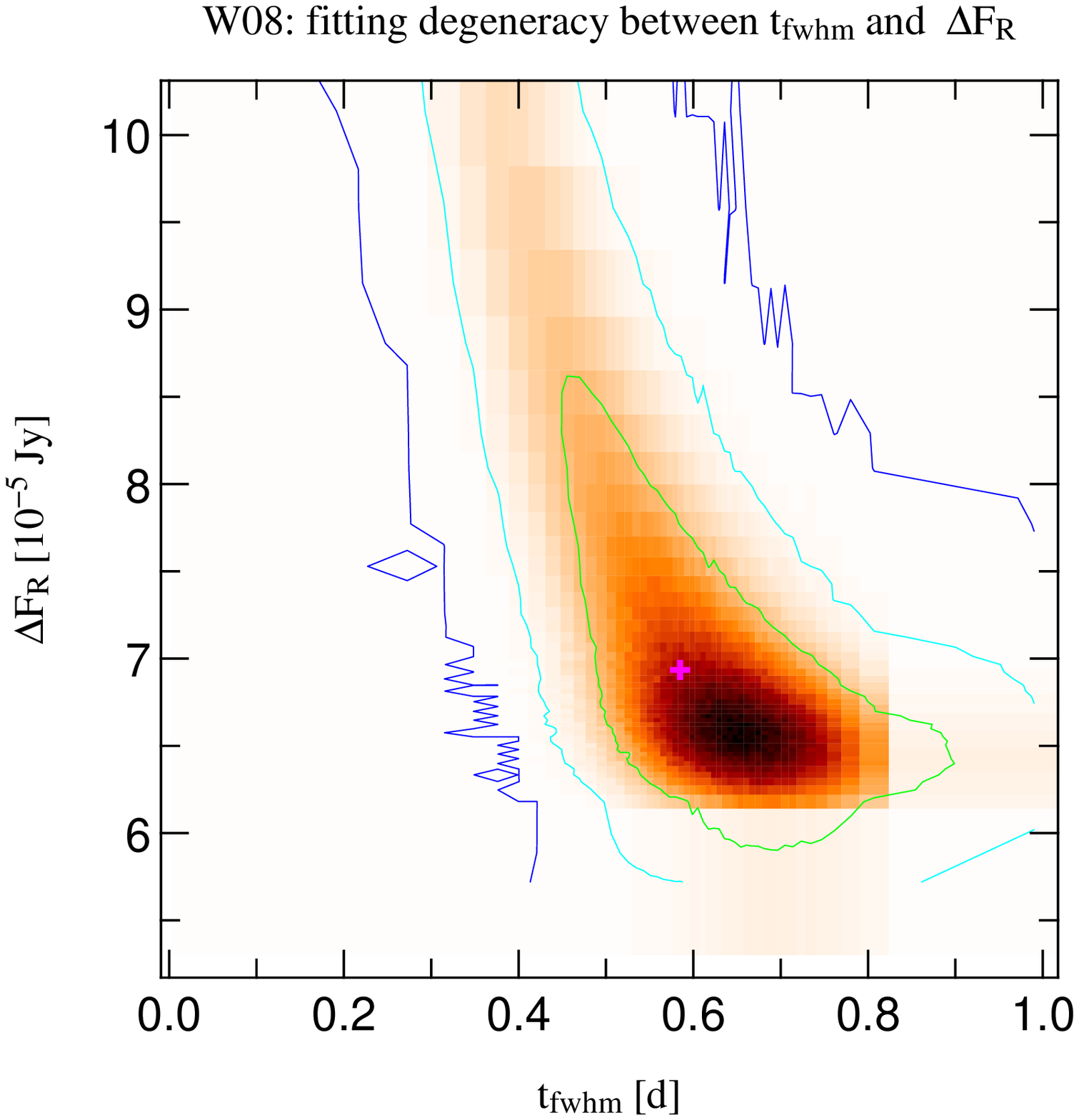}
  \includegraphics[width=3.5cm]{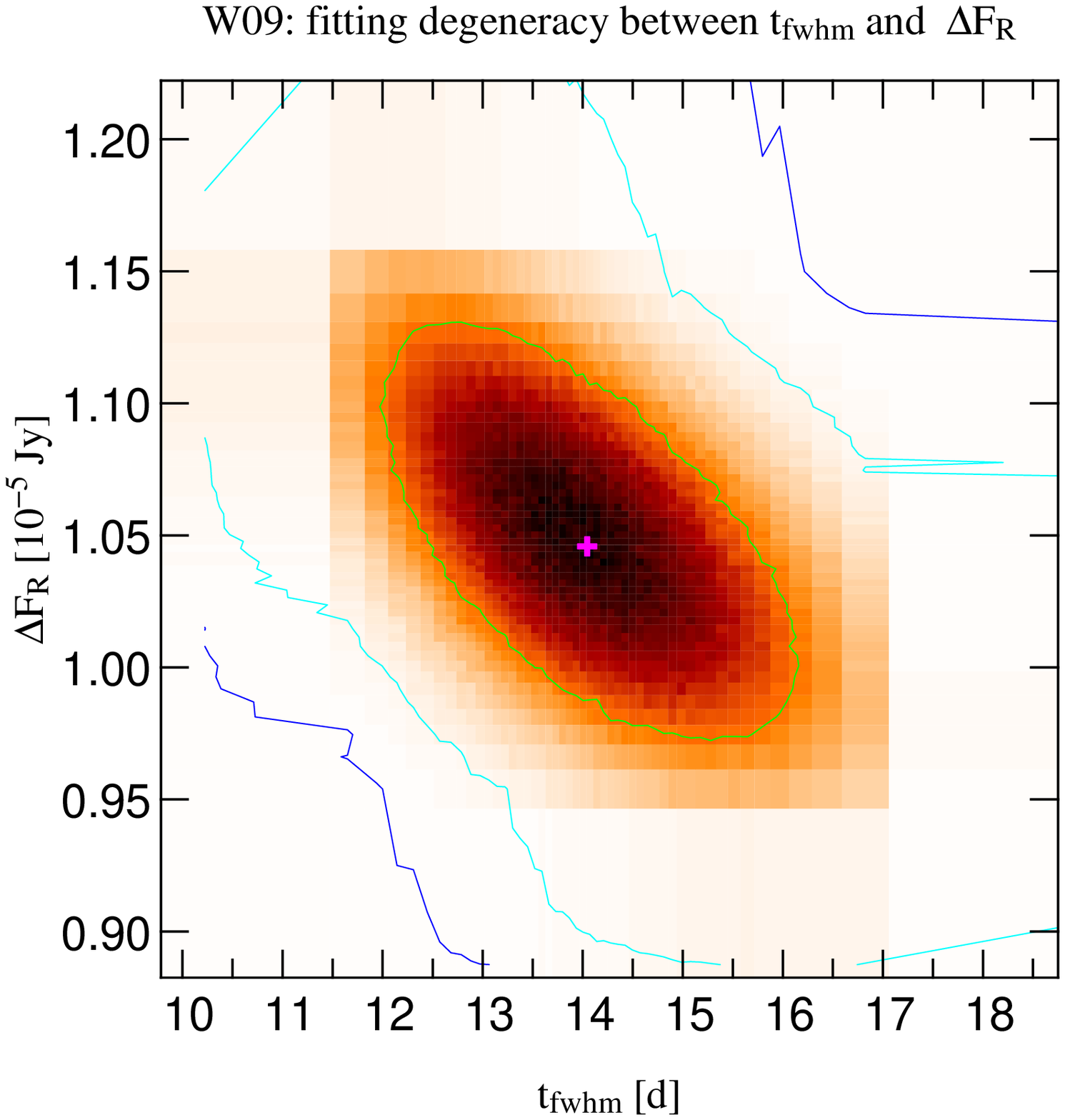}
  \includegraphics[width=3.5cm]{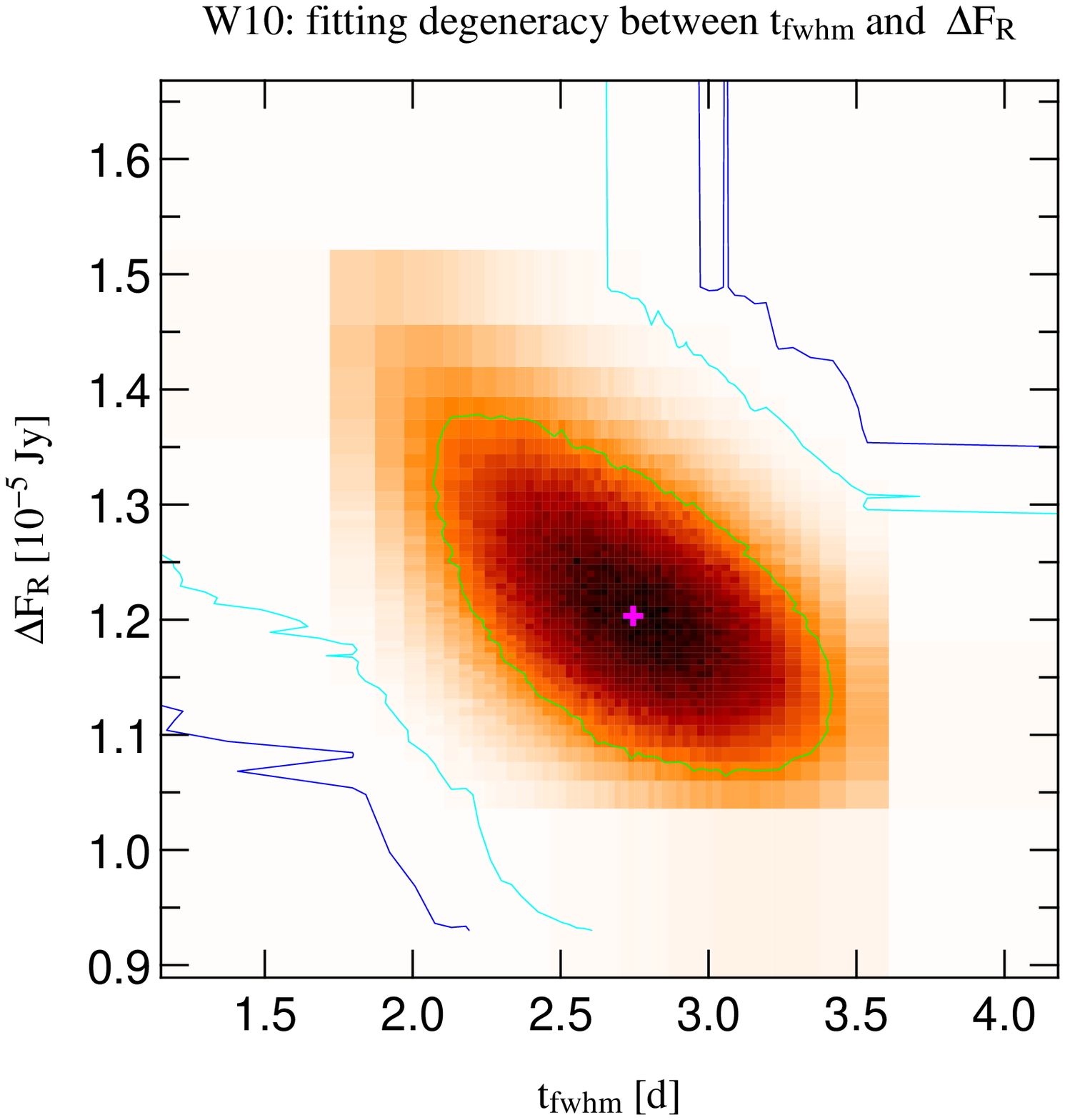}
  \includegraphics[width=3.5cm]{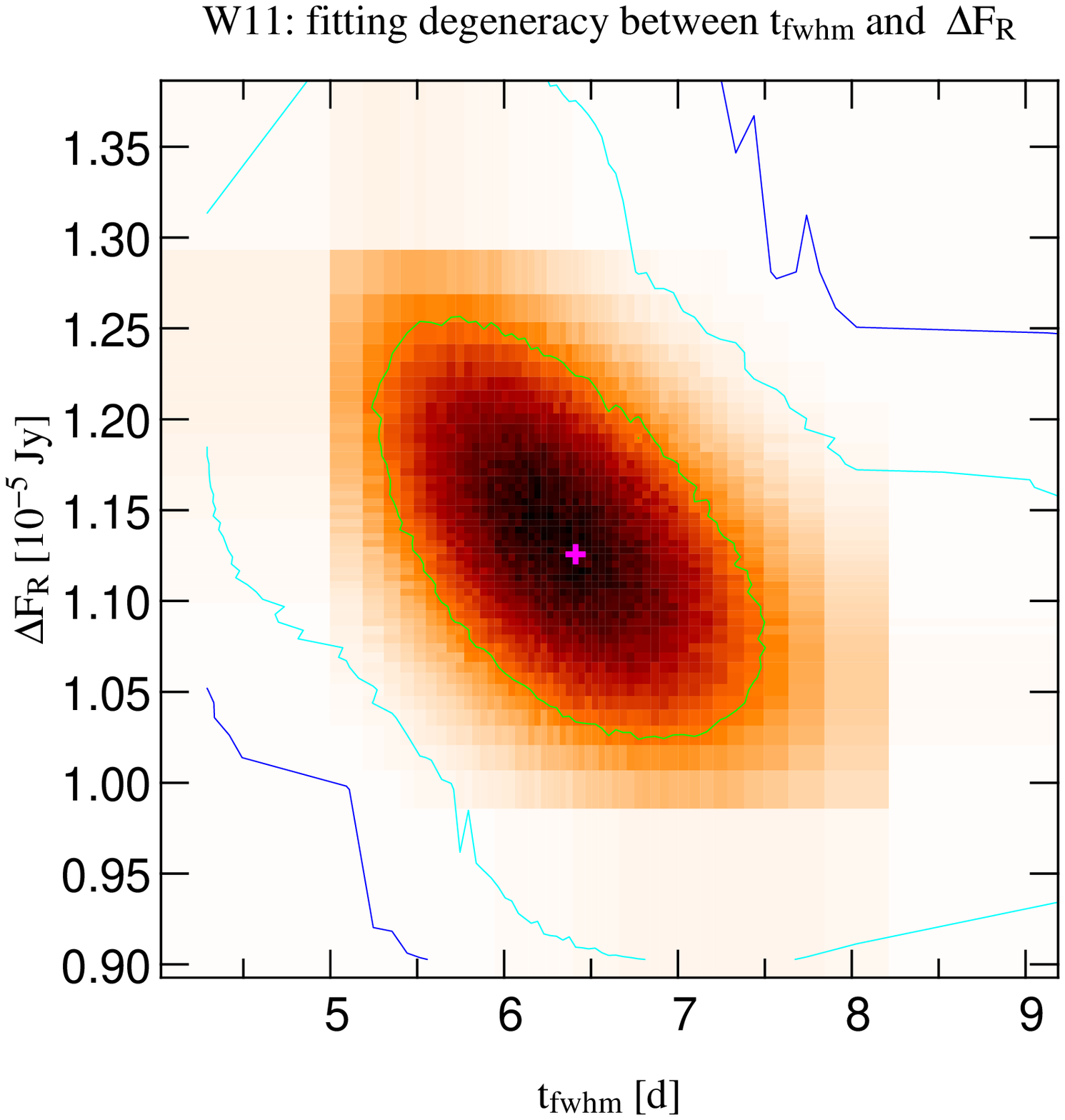}
  \includegraphics[width=3.5cm]{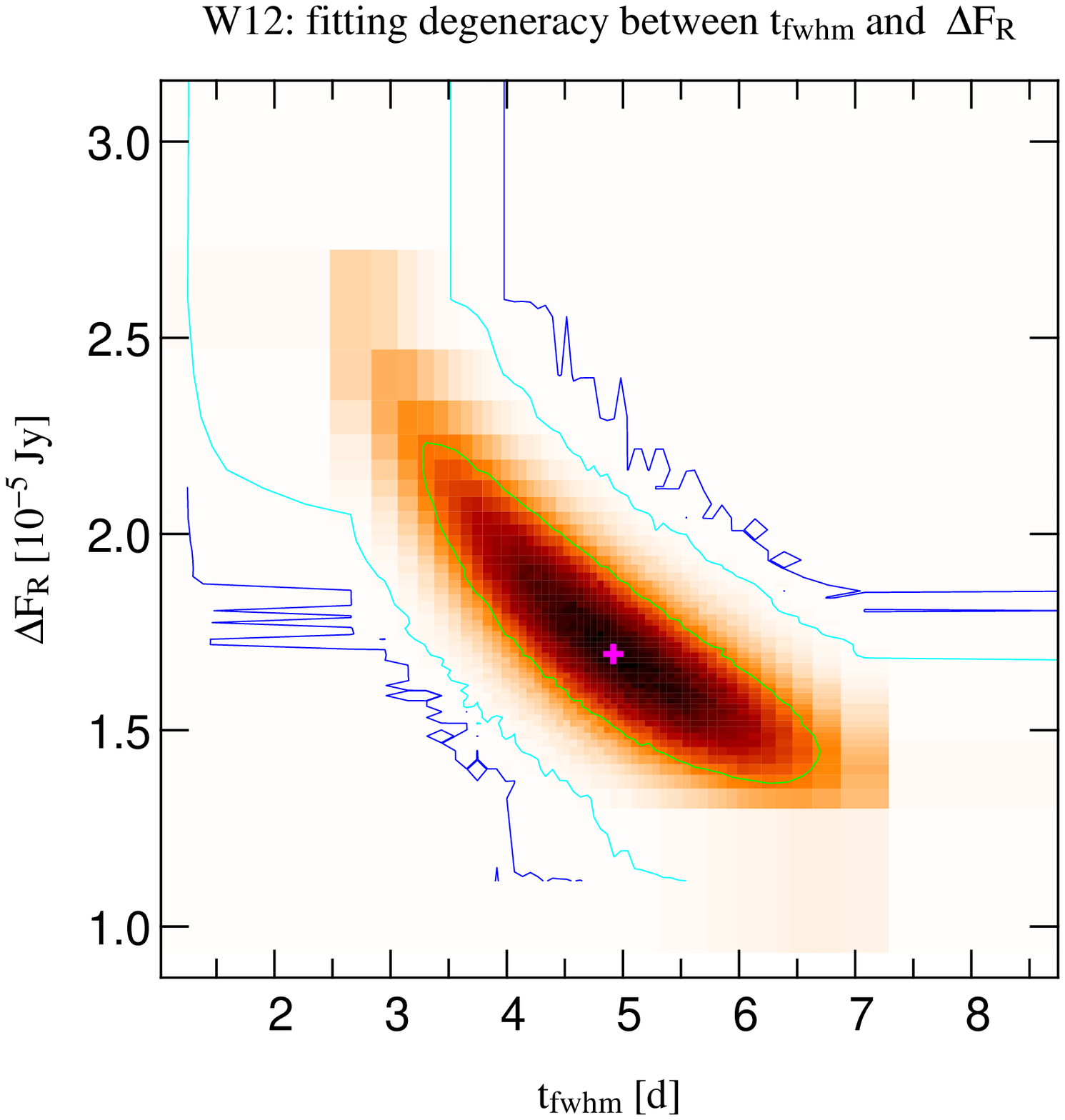}
  \caption{$\tfwhm$ vs. $\DF_R$ distribution of the 12 WeCAPP microlensing events. We note that the $\tfwhm$ and $\DF_R$-scale in the figures are different for each event. The 1, 2, and 3-$\sigma$ contours are marked in green, cyan and blue, respectively. The best-fit parameter is marked in magenta.}
  \label{fig.contour}
\end{figure*}

We show the degenercy between $\tfwhm$ and $\DF$ for our 12 WeCAPP microlensing events in Fig. \ref{fig.contour}. These two parameters are highly degenerate for very short time-scale events ($\tfwhm$ below 2 day). For events with long time-scale, i.e. W3, W4, W5, W9, W10 and W11, their $\tfwhm$ and $\DF$ are well constrained.

\begin{figure*}[!th]
  \centering
  \includegraphics[width=3.5cm]{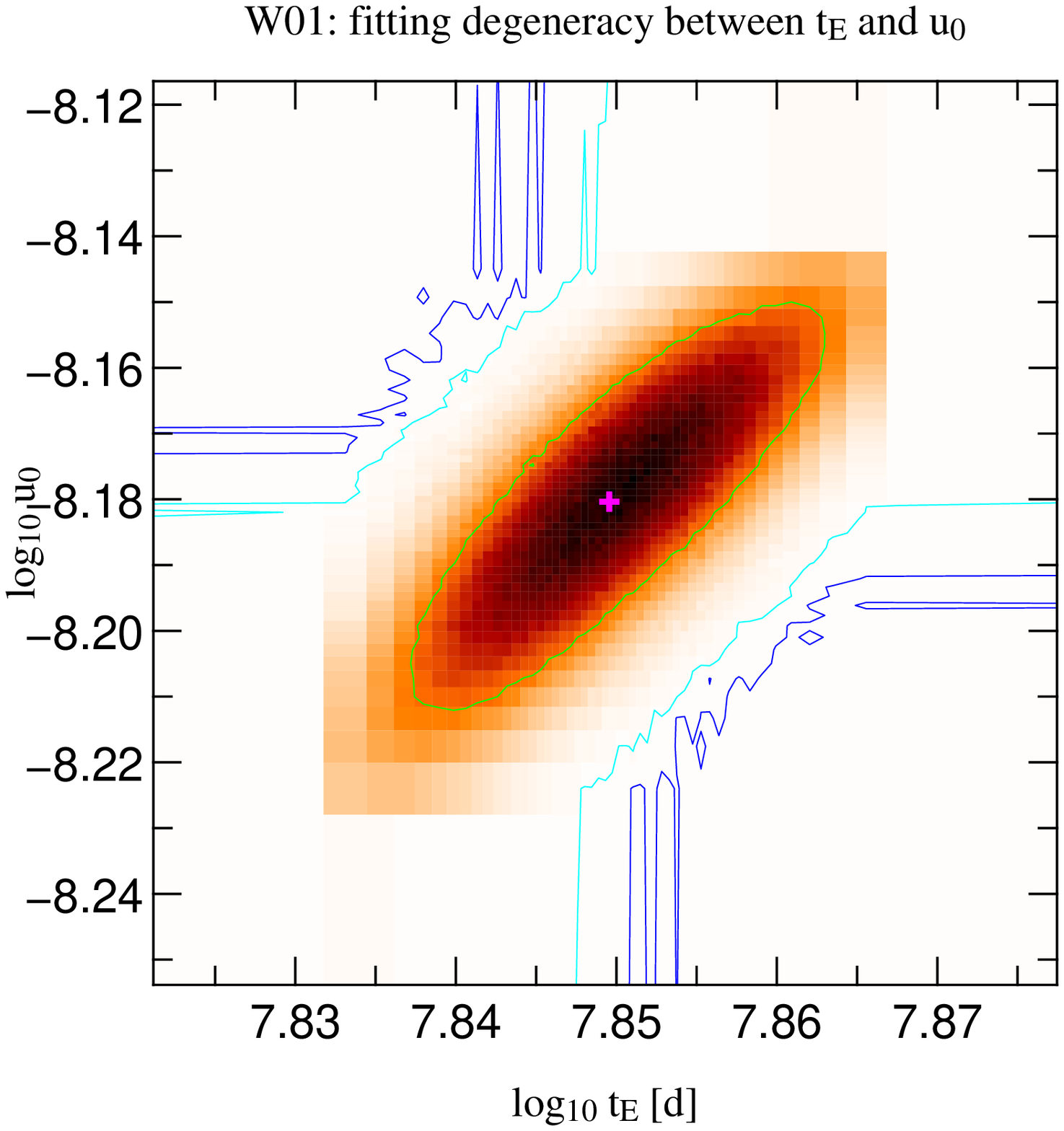}
  \includegraphics[width=3.5cm]{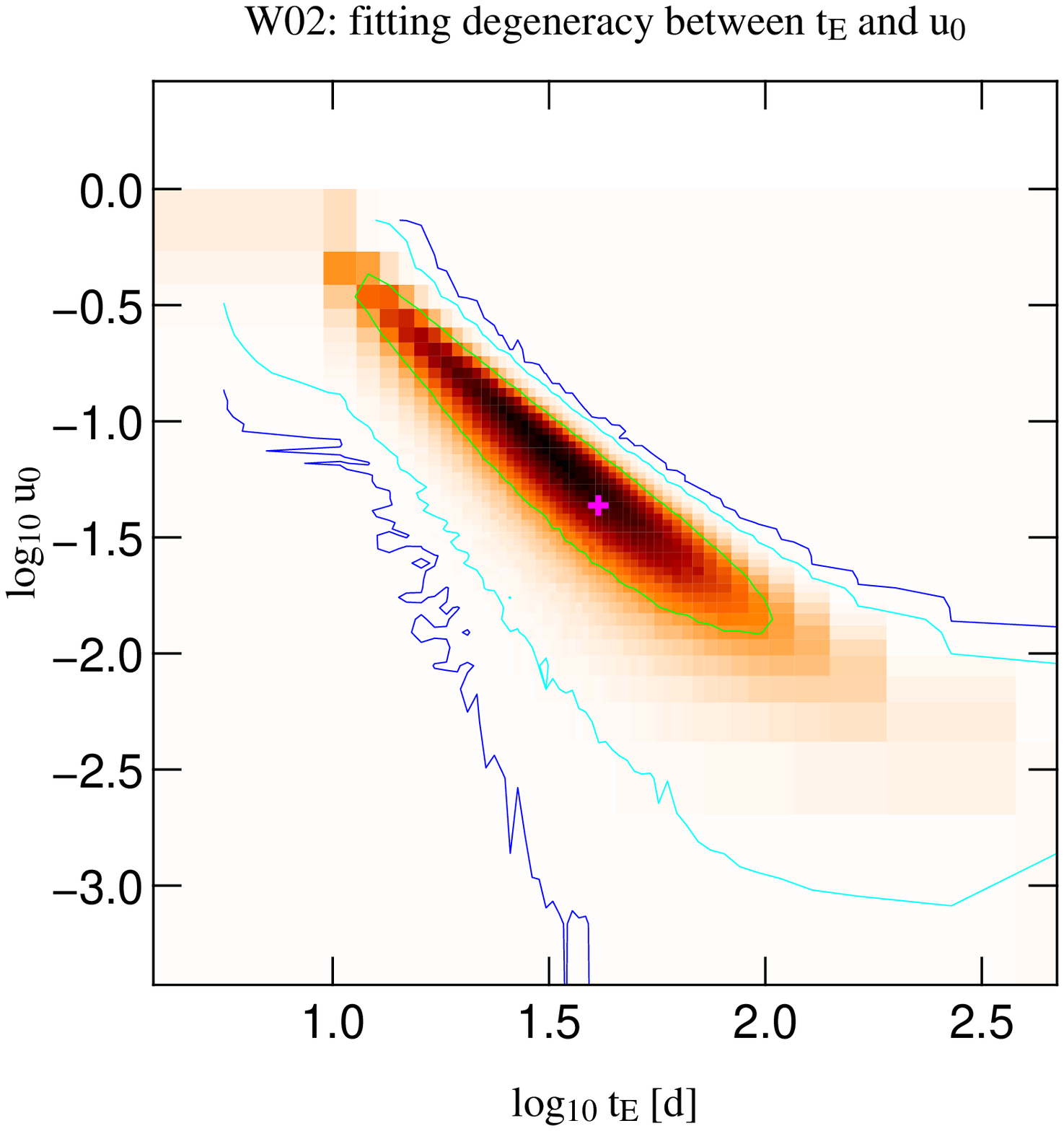}
  \includegraphics[width=3.5cm]{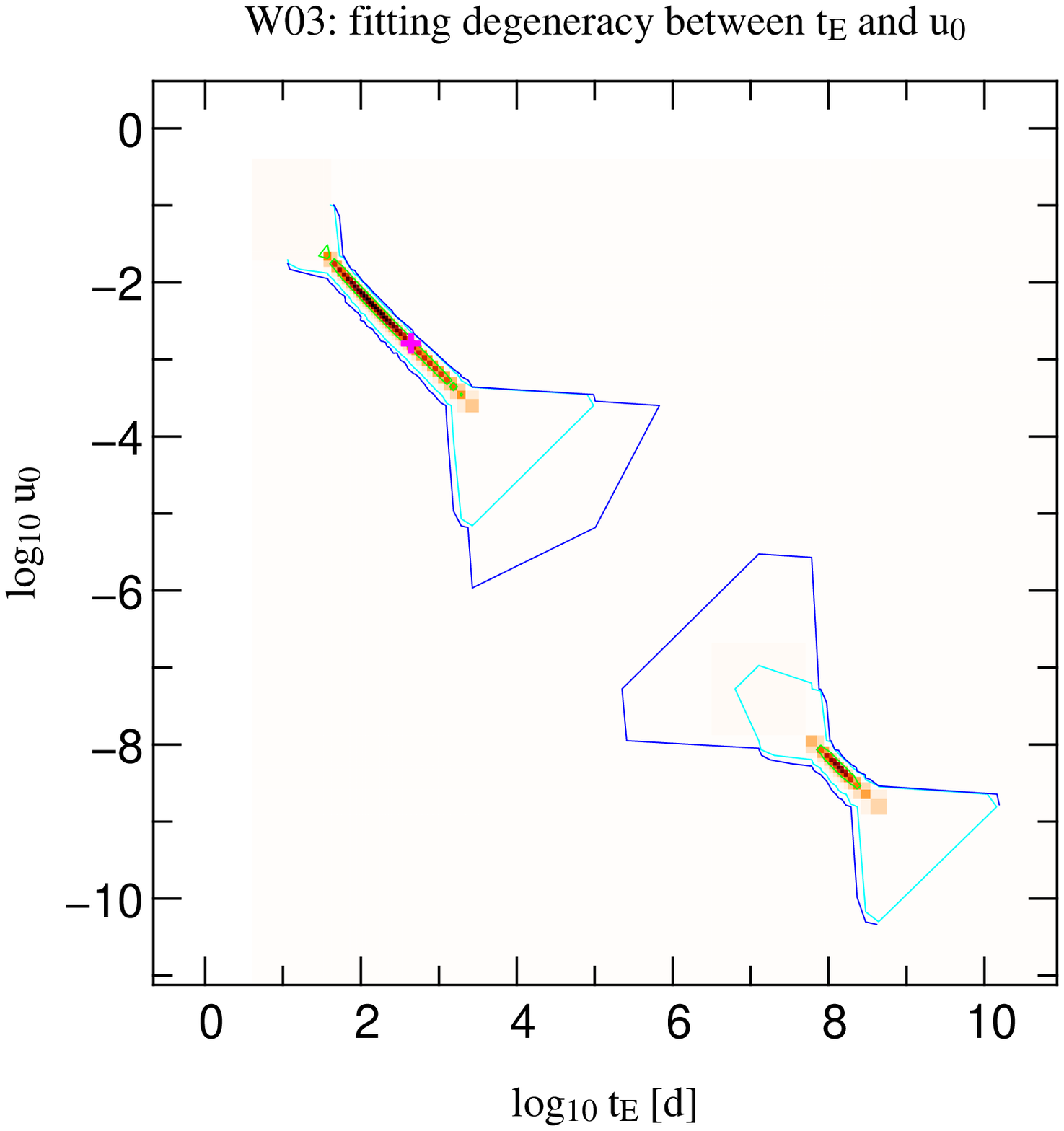}
  \includegraphics[width=3.5cm]{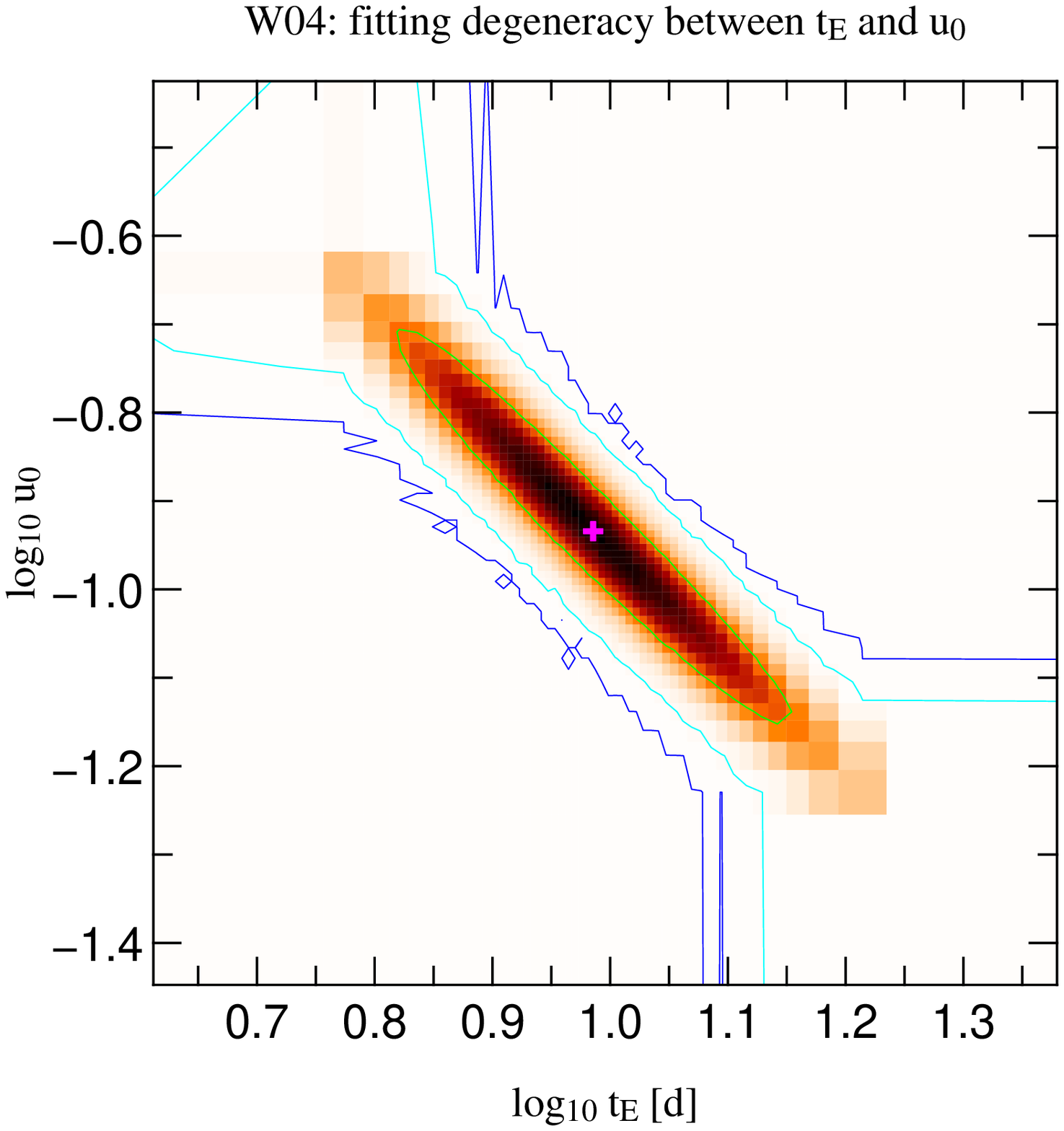}
  \includegraphics[width=3.5cm]{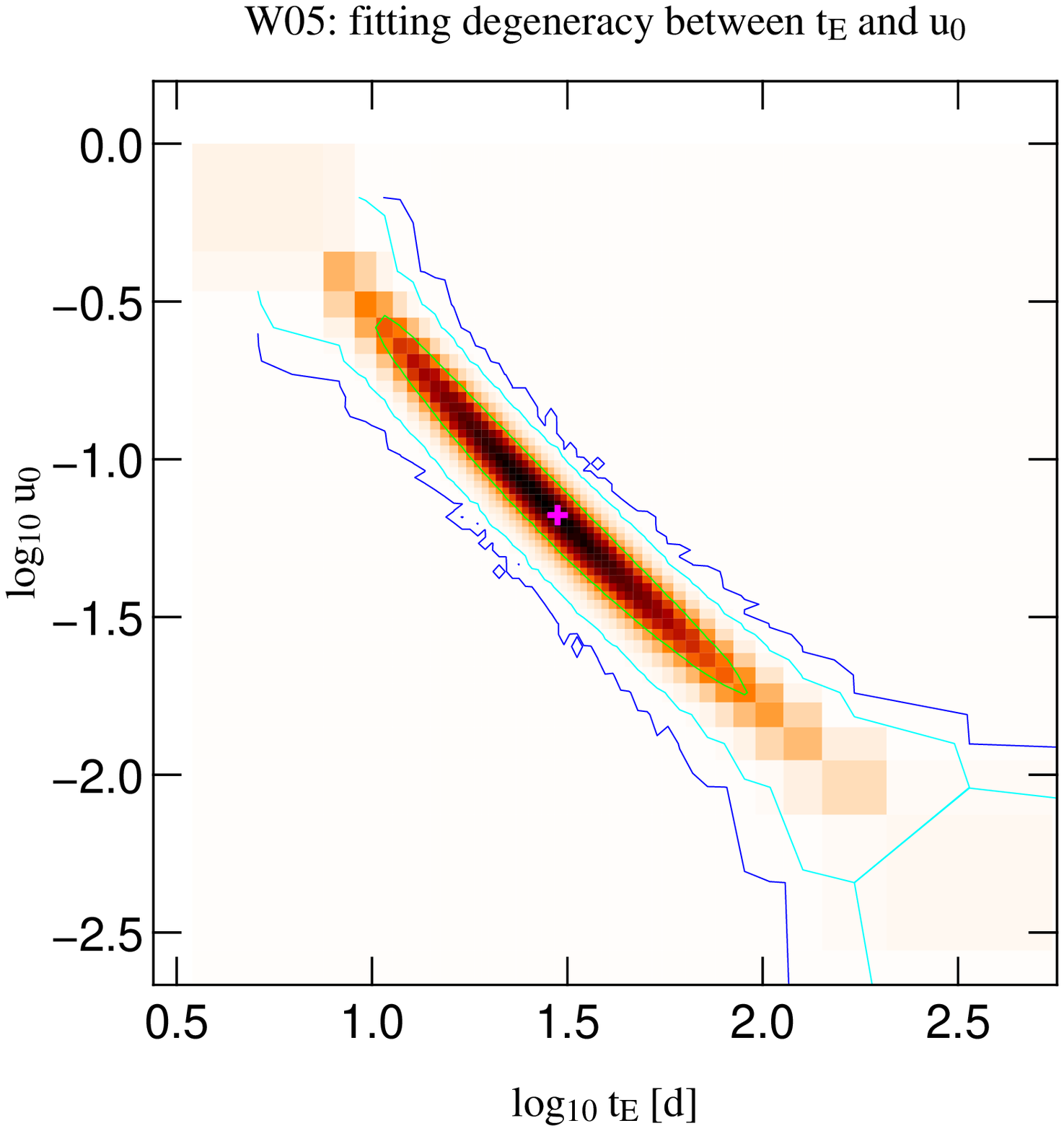}
  \includegraphics[width=3.5cm]{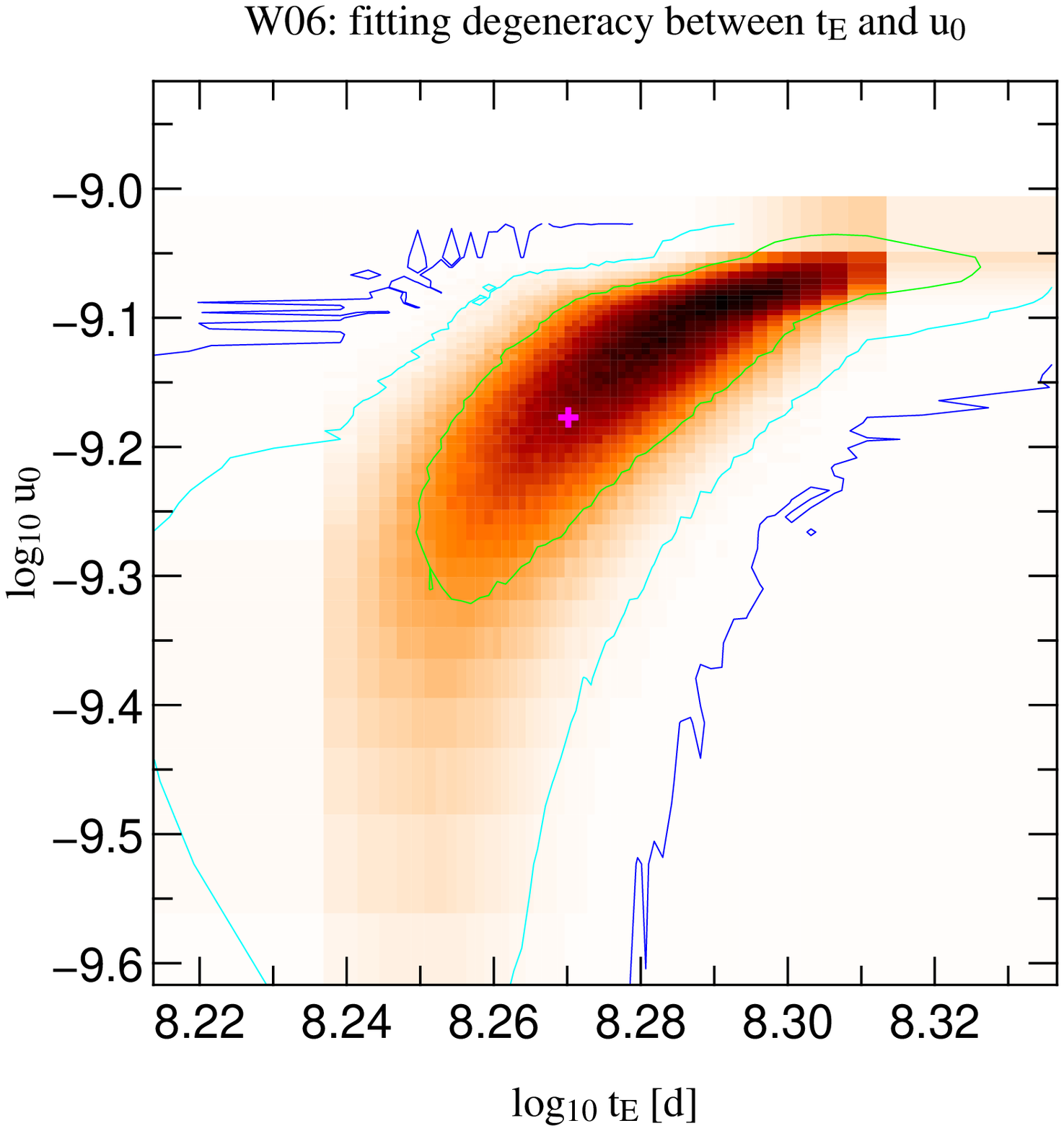}
  \includegraphics[width=3.5cm]{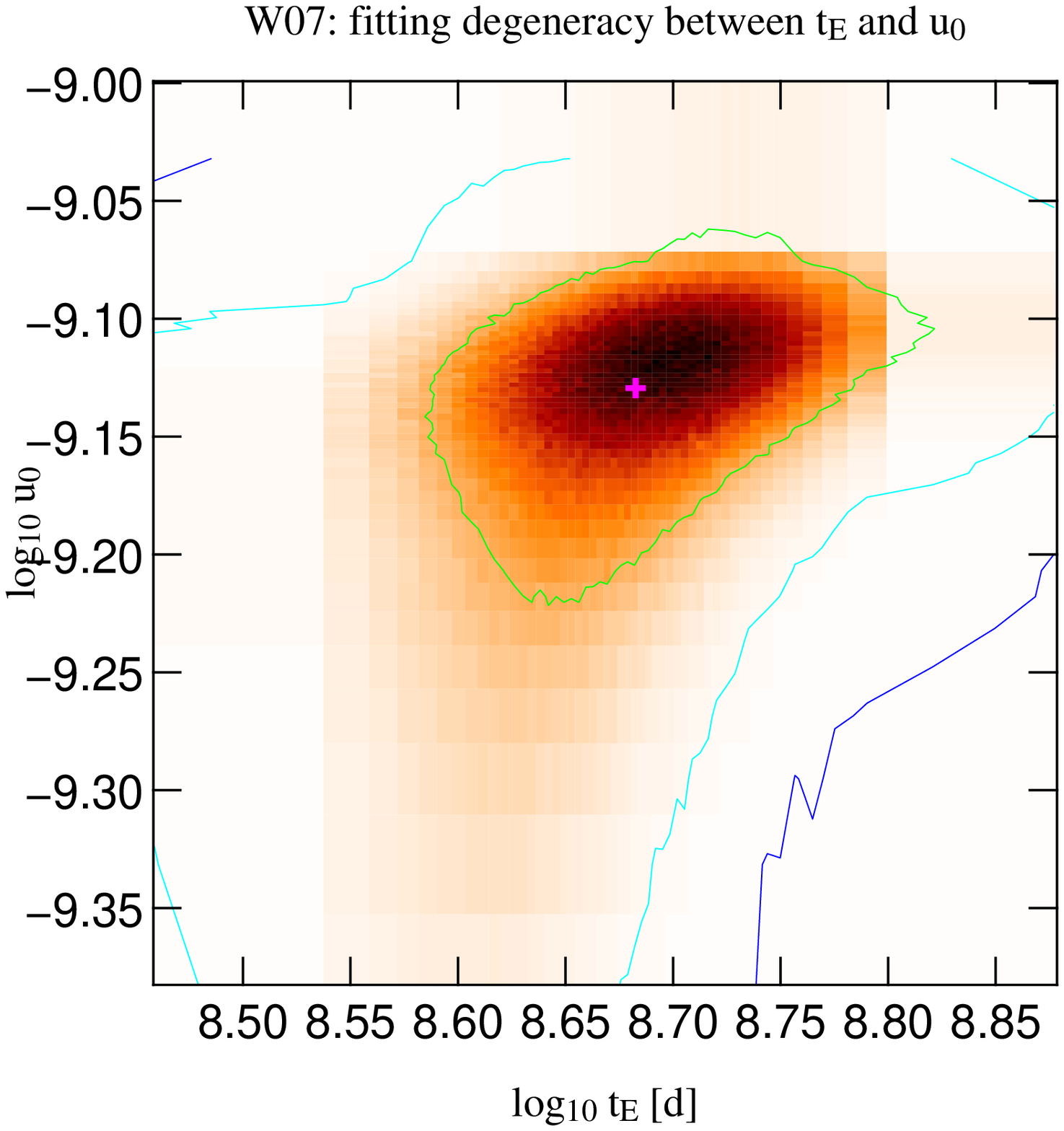}
  \includegraphics[width=3.5cm]{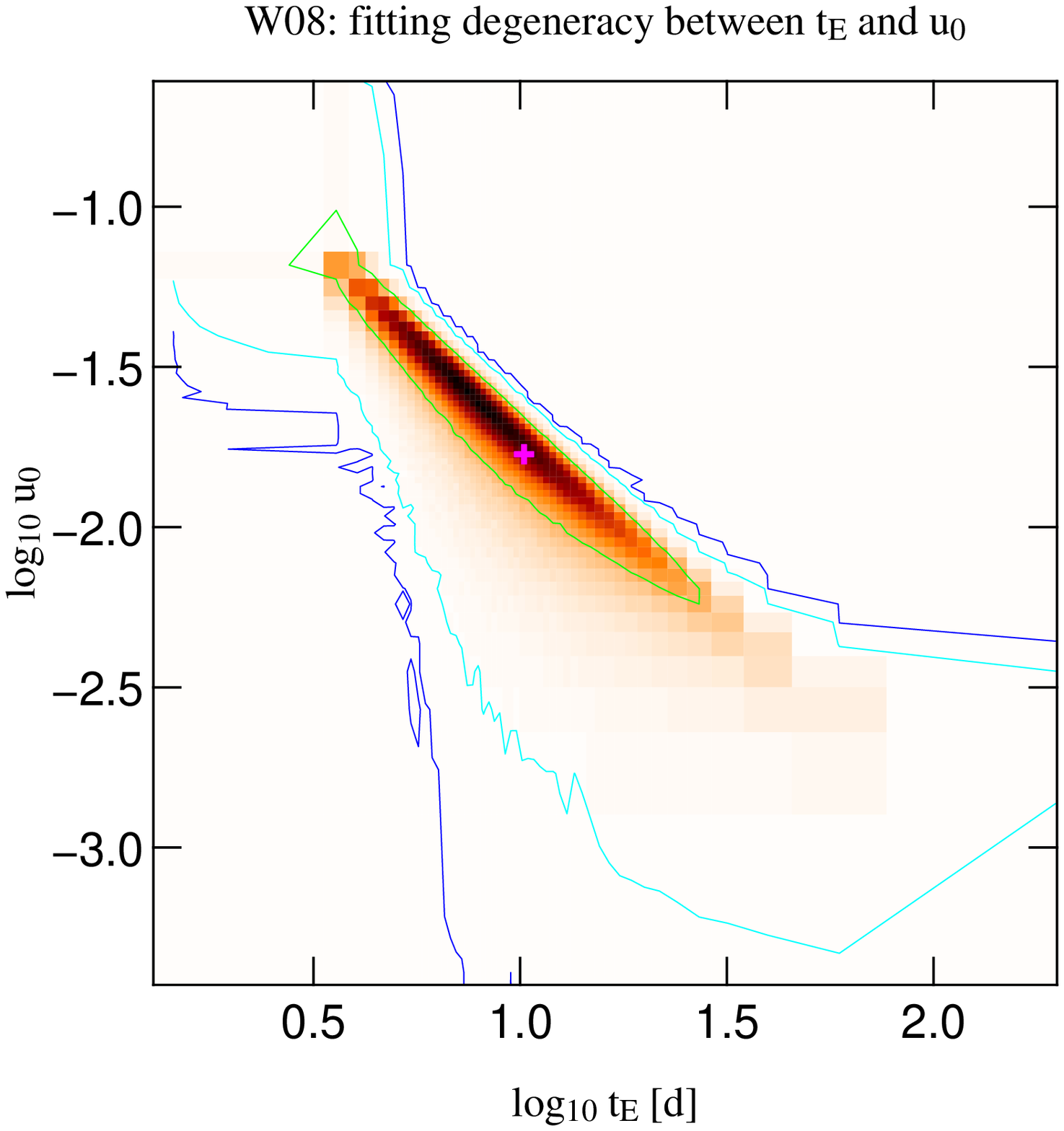}
  \includegraphics[width=3.5cm]{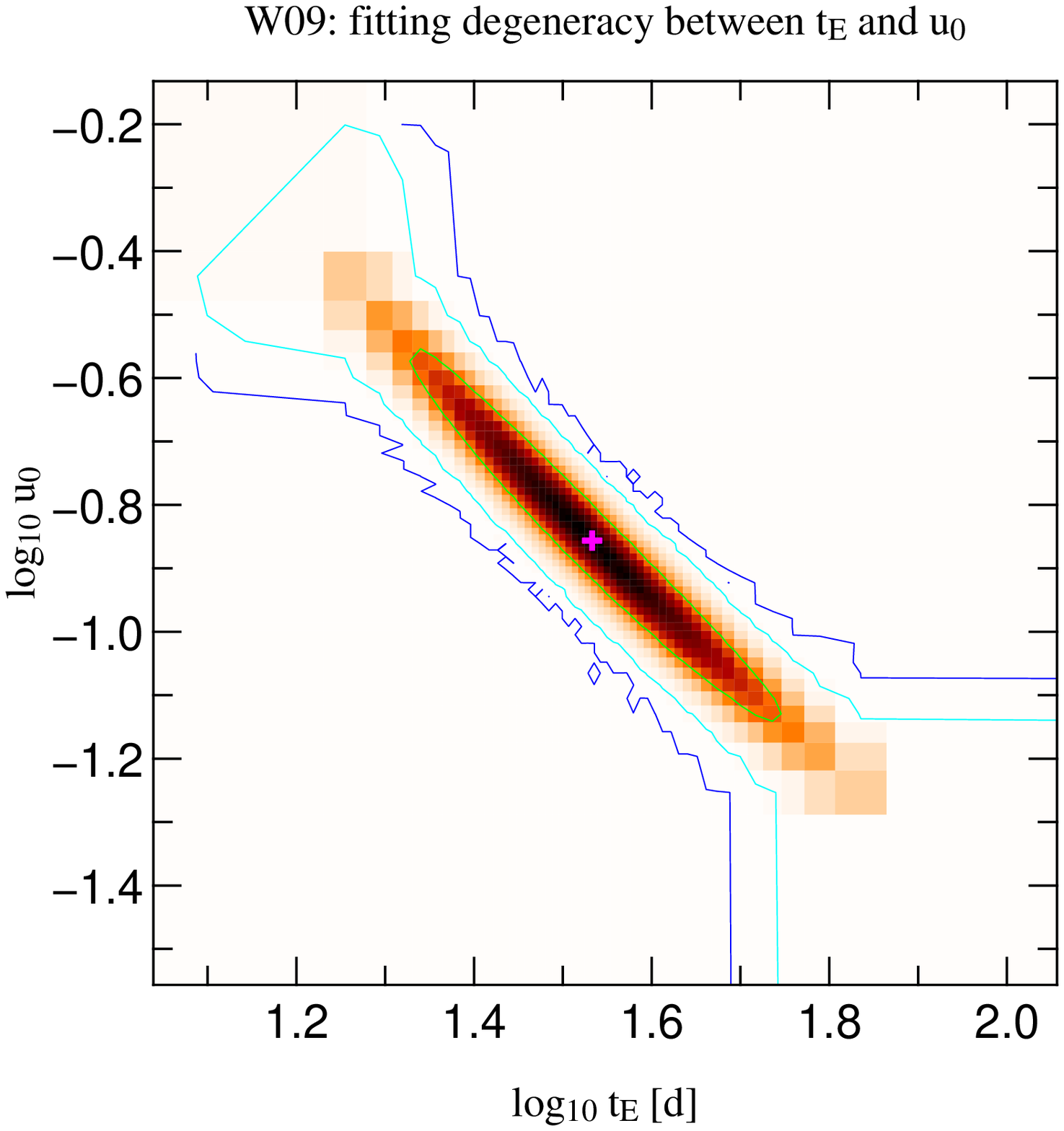}
  \includegraphics[width=3.5cm]{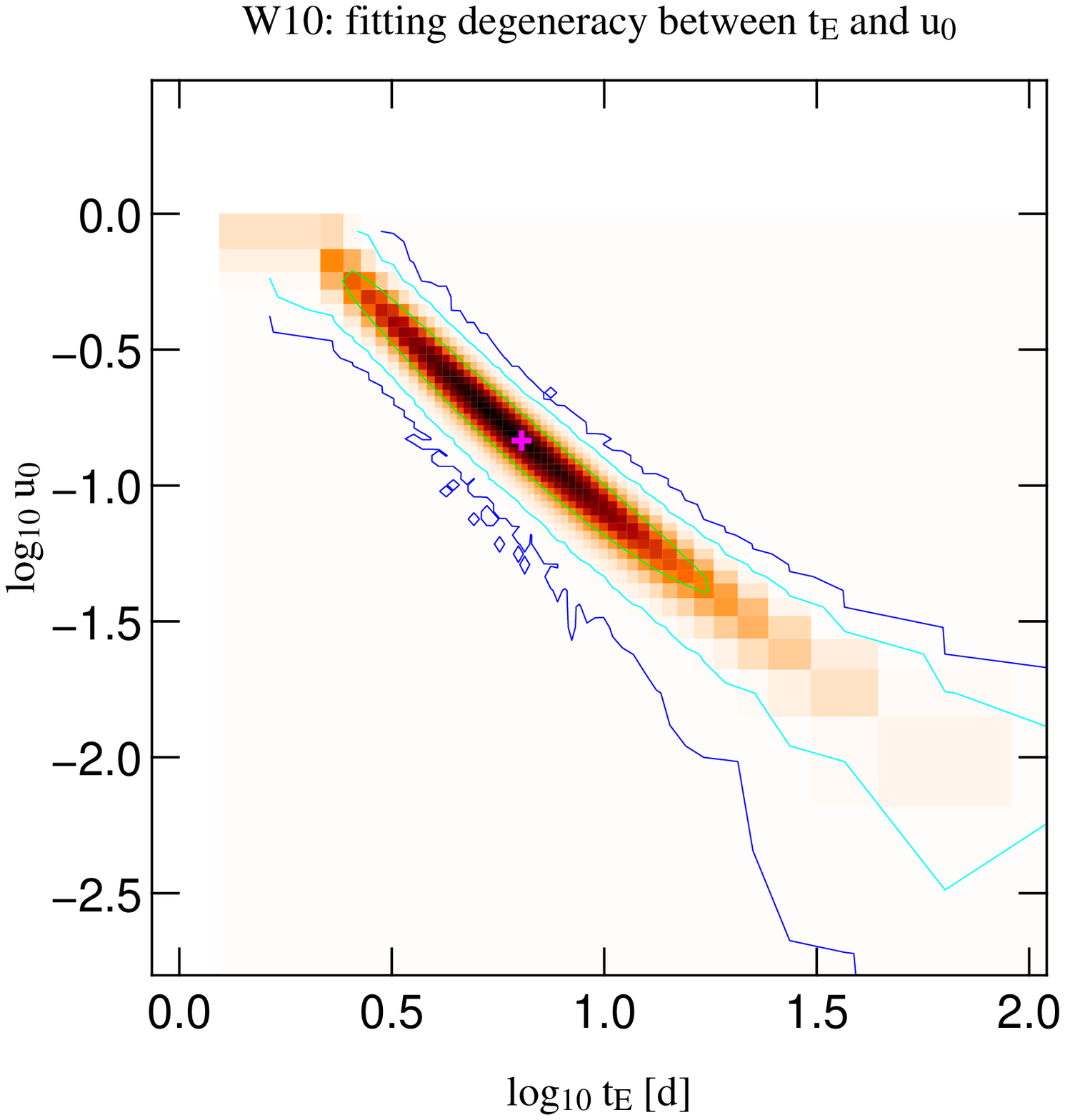}
  \includegraphics[width=3.5cm]{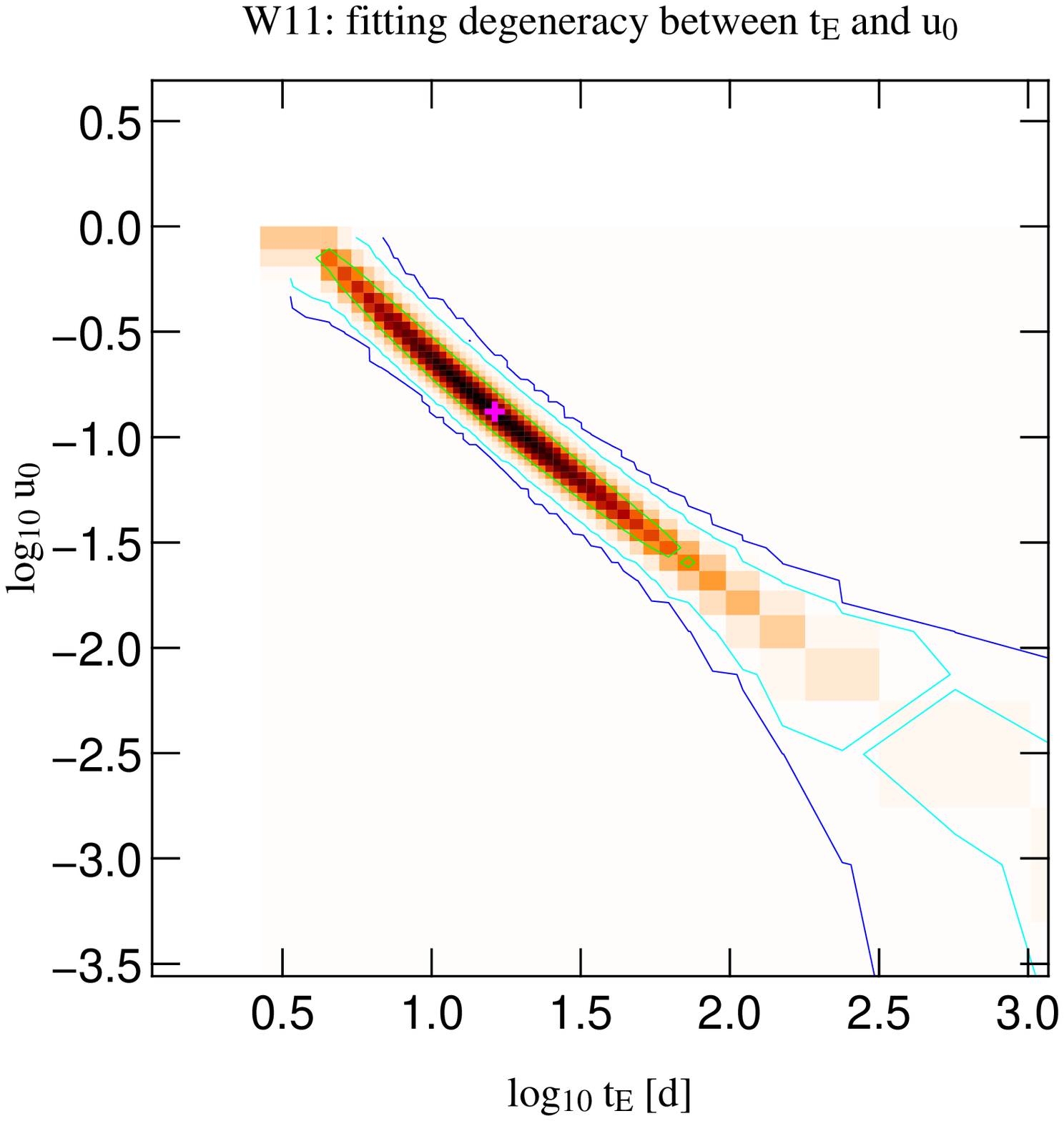}
  \includegraphics[width=3.5cm]{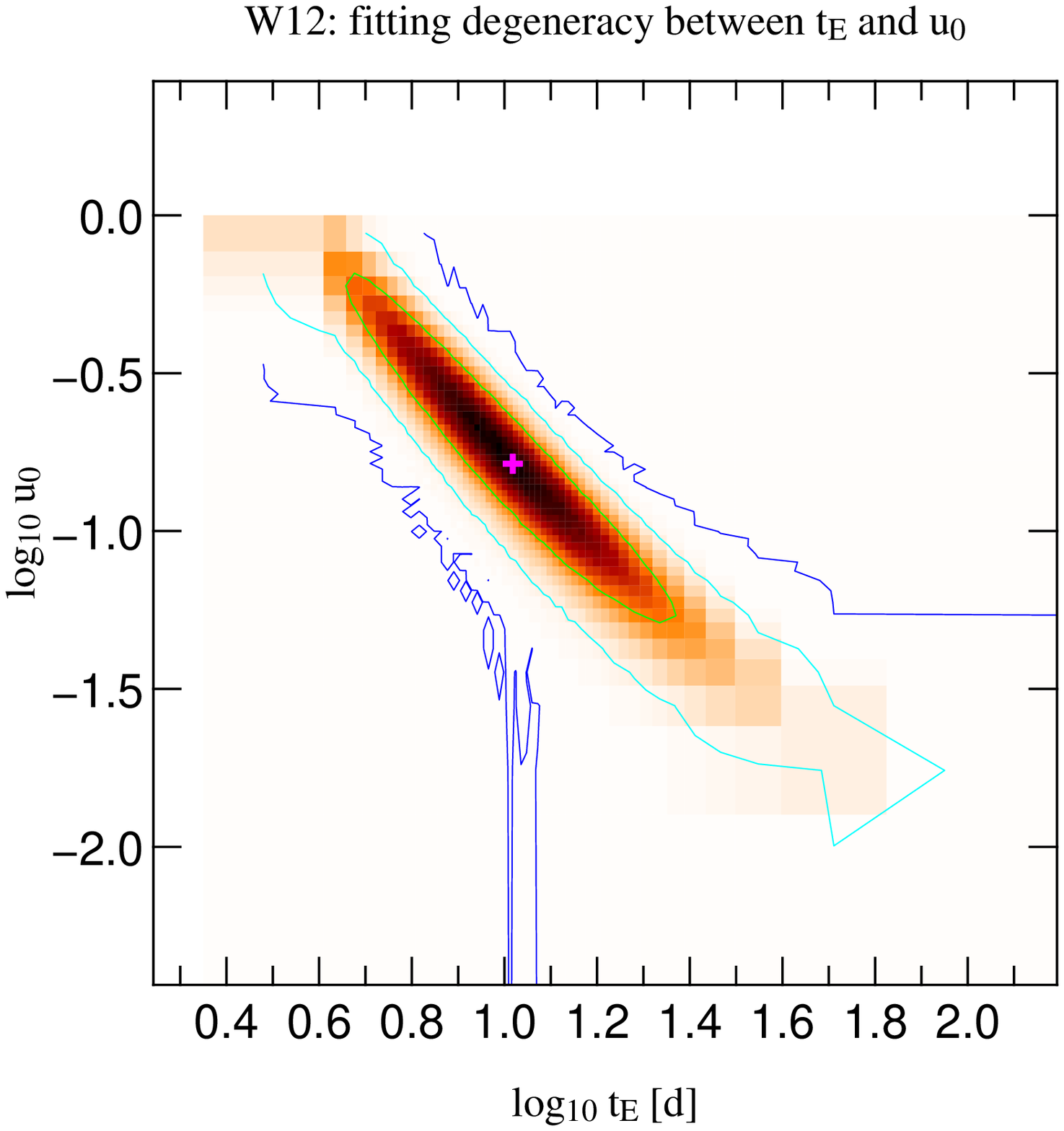}
  \caption{$t_{\mathrm{E}}$ vs. $u_0$ distribution in log scale of the 12 WeCAPP microlensing events. We note that the $t_{\mathrm{E}}$ and $u_0$-scale 
in the figures are different for each event. The 1, 2, and 3-$\sigma$ contours are marked in green, cyan and blue, respectively. The best-fit parameter is marked in magenta.}
  \label{fig.teu0}
\end{figure*}

We also show the degeneracy between $t_{\mathrm{E}}$ and $u_0$ for our 12 WeCAPP microlensing events in Fig. \ref{fig.teu0}. 
In the pixel-lensing regime, $t_{\mathrm{E}}$ and $u_0$ are highly degenerate, hence the $t_{\mathrm{E}}$-$u_0$ maps are much more degenerate than the $\tfwhm$-$\DF$ maps.

\end{document}